\documentclass[12pt]{article}
\usepackage[british]{babel}
\def\AnswerYes{y}
\def\pdflatex{y}                  
\def\ShowLabelsVersion{n}         
\def\ShowChangesVersion{n}        
\def\ShowAnnotationsVersion{n}    
\def\ShowFigures{y}               
\def\feynVersion{n}               
\def\MakeArXivLinksActive{y}      
\ifx\ShowFigures\AnswerYes
   \usepackage{graphicx}          
\else
   \usepackage[draft]{graphicx}   
   \renewcommand{\includegraphics}[2][]{\fbox{#2}}
\fi

\graphicspath{{figures/}{./}{"/home_th/judith/Dropbox/Compton-1N-asymmetries/figures/"}}

\ifx\pdflatex\AnswerYes
   \usepackage[update,prepend]{epstopdf} 
   \usepackage{grffile} 
\fi
\ifx\pdflatex\AnswerYes
    \usepackage{color}
\else
    \usepackage[dvips]{color}
\fi
\ifx\pdflatex\AnswerYes
    \usepackage{thumbpdf}    %
\else
    \usepackage[dvips]{thumbpdf}
\fi

\definecolor{green}{rgb}{0.,0.7,0.}
\definecolor{blue}{rgb}{0.,0.,1.0}
\definecolor{red}{rgb}{1,0,0}

\ifx\pdflatex\AnswerYes
    \usepackage[ 
        final,
        breaklinks=true,
        colorlinks=true,           
        pdfborder={0 0 1},    
        citecolor={blue},linkcolor={blue},urlcolor={red},
        citebordercolor={1 0 0},linkbordercolor={0 0 1},urlbordercolor={1 0 0},
        pdfpagemode=UseOutlines,
        bookmarks=true,bookmarksopenlevel=4
        ]{hyperref}         
\else
    \usepackage[dvips,              
        final,
        breaklinks=true,
        colorlinks=true,      
        pdfborder={0 0 1},    
        citecolor={blue},linkcolor={blue},urlcolor={red},
        citebordercolor={1 0 0},linkbordercolor={0 0 1},urlbordercolor={1 0 0},
        pdfpagemode=UseOutlines,
        bookmarks=true,bookmarksopenlevel=4
         ]{hyperref}         
\fi
\usepackage[numbers,sort&compress]{natbib}

\usepackage{multirow}                   
\usepackage{amssymb}
\usepackage{amsmath}
\usepackage{bbm} 
\usepackage{xspace}                    
\usepackage{dcolumn}                    
\usepackage{bm}                        
\usepackage{cancel}	                  

\usepackage{rotating}

\usepackage{floatpag}           
\floatpagestyle{plain}          

\usepackage{slashed}
\ifx\feynVersion\AnswerYes
   \usepackage{feynmp} 
\fi

\ifx\ShowLabelsVersion\AnswerYes
   \usepackage[color]{showkeys} 
   \definecolor{refkey}{gray}{.5}   
   \definecolor{labelkey}{gray}{.5} 
   
\fi

\ifx\ShowAnnotationsVersion\AnswerYes
   \newcommand{\comment}[1]{{\scriptsize\sffamily\bfseries{#1}}}
   \newcommand{\margin}[1]{\marginpar{\scriptsize\sffamily\bfseries{#1}}}
   \newcommand{\drafty}{\textbf{Draft version \today} \hfill}
\else
   \newcommand{\comment}[1]{}
   \newcommand{\margin}[1]{}
   \newcommand{\drafty}{}
\fi

\ifx\ShowChangesVersion\AnswerYes
   \usepackage{ulem}
    
   \newcommand{\delete}[1]{\sout{#1}}   
   \renewcommand{\emph}[1]{\textit{#1}} 
   
\else

   \newcommand{\sout}[1]{}
   \newcommand{\xout}[1]{}

   \newcommand{\delete}[1]{}
   
\fi

\ifx\MakeArXivLinksActive\AnswerYes
\usepackage{xparse}
\NewDocumentCommand{\arxiv} {r [: u{ [} u{]]} }{[\href{http://arxiv.org/abs/#2}{arXiv:#2}~[#3]]}
\NewDocumentCommand{\arxivold} {r[]}{[\href{http://arxiv.org/abs/#1}{#1}]}
\else
\newcommand{\arxiv}[1][]{[#1]}
\newcommand{\arxivold}[1][]{[#1]}
\fi

\newcommand{\eg}{\textit{e.g.}\xspace}
\newcommand{\ie}{\textit{i.e.}\xspace}
\newcommand{\etal}{\textit{et al.}\xspace}

\newcommand{\dis}{\displaystyle}

\newcommand{\phm}{\phantom{-}} 

\newcommand{\half}{\frac{1}{2}}

\newcommand{\ii}{\mathrm{i}}
\newcommand{\dd}{\mathrm{d}}
\newcommand{\tr}{\mathrm{tr}}

\newcommand{\kv}{\vec{k}}
\newcommand{\pv}{\vec{p}}

\renewcommand{\Re}{\mathrm{Re}}
\renewcommand{\Im}{\mathrm{Im}}

\newcommand{\mpi}{\ensuremath{m_\pi}}

\newcommand{\MeV}{\ensuremath{\mathrm{MeV}}}
\newcommand{\fm}{\ensuremath{\mathrm{fm}}}
\newcommand{\ChiEFT}{$\chi$EFT\xspace}

\newcommand{\NXLO}[1]{N\ensuremath{{}^{#1}}LO\xspace}

\newcommand{\HIGS}{HI$\gamma$S\xspace}

\newcommand{\fourHe}{${}^4$He\xspace}

\newcommand{\alphae}{\ensuremath{\alpha_{E1}}}
\newcommand{\betam}{\ensuremath{\beta_{M1}}}
\newcommand{\gammaee}{\ensuremath{\gamma_{E1E1}}}
\newcommand{\gammamm}{\ensuremath{\gamma_{M1M1}}}
\newcommand{\gammaem}{\ensuremath{\gamma_{E1M2}}}
\newcommand{\gammame}{\ensuremath{\gamma_{M1E2}}}
\newcommand{\gammazero}{\ensuremath{\gamma_{0}}}
\newcommand{\gammapi}{\ensuremath{\gamma_{\pi}}}
\newcommand{\gammamminus}{\ensuremath{\gamma_{M-}}} 
\newcommand{\gammaeminus}{\ensuremath{\gamma_{E-}}} 
\newcommand{\alphaep}{\ensuremath{\alpha_{E1}^{(\mathrm{p})}}}
\newcommand{\betamp}{\ensuremath{\beta_{M1}^{(\mathrm{p})}}}

\newcommand{\gammammp}{\ensuremath{\gamma_{M1M1}^{(\mathrm{p})}}}

\newcommand{\gammammn}{\ensuremath{\gamma_{M1M1}^{(\mathrm{n})}}}

\newcommand{\philin}{\ensuremath{\varphi_\text{lin}}}
\newcommand{\thetan}{\ensuremath{\vartheta_{\vec{n}}}}
\newcommand{\phin}{\ensuremath{\varphi_{\vec{n}}}}

\newcommand{\gammaN}{\gamma \mathrm{N}}

\newcommand{\MN}{\ensuremath{M_\mathrm{N}}} 
\newcommand{\DeltaM}{\ensuremath{\Delta_{\scriptscriptstyle M}}} 

\def\textfrac#1#2{{\textstyle \frac{#1}{#2}}}

\newcommand{\omegalab}{\ensuremath{\omega_\mathrm{lab}}}
\newcommand{\thetalab}{\ensuremath{\theta}}
\newcommand{\omegaprimelab}{\ensuremath{\omega'_\mathrm{lab}}}
\newcommand{\w}{\ensuremath{\omega}}

\newcommand{\ChPT}{\ensuremath{\chi \mathrm{PT}}~\xspace}

\newcommand{\noj}{\ensuremath{\bcancel{j}_0}}

\newcommand{\calM}{\mathcal{M}} 
\newcommand{\calO}{\mathcal{O}}

\newcommand{\mytitle}[1]{\begin{center}\LARGE{\textbf{#1}}\end{center}}
\newcommand{\myauthor}[1]{\textbf{#1}}
\newcommand{\myaddress}[1]{\textit{#1}}
\newcommand{\mypreprint}[1]{\begin{flushright}#1\end{flushright}}

\begin{document}
%

\begin{titlepage}
  \setcounter{page}{0} \mypreprint{
    \drafty
    30th November 2017, arXiv:1711.11546 \\
    Revised version 28th February 2018\\
    Accepted by Europ.~Phys.~J.~\textbf{A}.}
  
  
  \mytitle{Comprehensive Study of Observables in Compton Scattering on the
    Nucleon}

\begin{center}
  \myauthor{Harald W.\ Grie\3hammer$^{a}$}\footnote{Email:
    hgrie@gwu.edu}, 
  \myauthor{Judith
    A.~McGovern$^{b}$}\footnote{Email: judith.mcgovern@manchester.ac.uk} 
  \emph{and} 
  \myauthor{Daniel R.~Phillips$^{c}$}\footnote{Email: phillips@phy.ohiou.edu}
  
  \vspace*{0.5cm}
  
  \myaddress{$^a$ Institute for Nuclear Studies, Department of Physics, \\The
    George Washington University, Washington DC 20052, USA}
  \\[2ex]
  \myaddress{$^b$ School of Physics and Astronomy, The University of
    Manchester,\\ Manchester M13 9PL, UK}
  \\[2ex]
  \myaddress{$^c$ Department of Physics and Astronomy and Institute of Nuclear
    and Particle Physics, Ohio University, Athens, Ohio 45701, USA}

  \vspace*{0.2cm}

\end{center}


\begin{abstract}
  We present an analysis of $13$ observables in Compton scattering on the
  proton. Cross sections, asymmetries with polarised beam and/or targets, and
  polarisation-transfer observables are investigated for energies up to the
  $\Delta(1232)$ resonance to determine their sensitivity to the proton's
  dipole scalar and spin polarisabilities. The Chiral Effective Field Theory
  Compton amplitude we use is complete at \NXLO{4}, $\calO(e^2\delta^4)$, for
  photon energies $\omega\sim\mpi$, and so has an accuracy of a few per cent
  there. At photon energies in the resonance region, it is complete at NLO,
  $\calO(e^2\delta^0)$, and so its accuracy there is about $20$\%.  We find
  that for energies from pion-production threshold to about $250\;\MeV$,
  multiple asymmetries have significant sensitivity to presently
  ill-determined combinations of proton spin polarisabilities. We also argue
  that the broad outcomes of this analysis will be replicated in complementary
  theoretical approaches, \eg, dispersion relations.
  Finally, we show that below the pion-production threshold, $6$ observables
  suffice to reconstruct the Compton amplitude, and above it $11$ are
  required. Although not necessary for polarisability extractions, this opens
  the possibility to perform ``complete" Compton-scattering experiments.
  An interactive \emph{Mathematica} notebook, including results for the
  neutron, is available from \texttt{judith.mcgovern@manchester.ac.uk}.
\end{abstract}
\vskip 1.0cm
\noindent
\begin{tabular}{rl}
  Suggested Keywords: &\begin{minipage}[t]{10.7cm} Effective Field Theory,
    proton, neutron and nucleon
    polarisabilities, spin polarisabilities, Chiral Perturbation Theory,
    $\Delta(1232)$ resonance
                    \end{minipage}
\end{tabular}

\vskip 1.0cm

\end{titlepage}

\setcounter{footnote}{0}

\newpage

%

\section{Introduction}
\setcounter{equation}{0}
\label{sec:introduction}

Compton scattering from a spin-$\frac 1 2$ target is fully described by six
amplitudes, as first shown in ref.~\cite{Prange:1958}. These are functions of
the incoming photon energy $\omega$ and the photon scattering angle $\theta$;
they become complex above the pion-production threshold.  At sufficiently low
energy, gauge and Lorentz invariance require that the process reduces to
Thomson scattering, depending only on the charge and mass of the target.  At
somewhat higher energies, deviations become apparent. For nucleons, some
deviations are explained by treating the target as a Dirac particle with an
anomalous magnetic moment, and by including neutral pion exchange (the
pion-pole contribution). Others are due to internal hadronic structure and
excitations.  The principal such effects at low energy are the nucleon's
electric and magnetic dipole polarisabilities, $\alphae$ and $\betam$. They
reveal the extent to which charge and current distributions in the target
shift under the influence of external electromagnetic fields and parametrise
the strength of the induced radiation dipoles.  Then, in the amplitudes that
are sensitive to the target's spin, four ``spin polarisabilities'' $\gamma_i$
govern the departure from point-like scattering and parametrise the response
of the spin degrees of freedom.  Each of these six low-energy parameters is as
fundamental a property of the nucleon as its anomalous magnetic moment, and
they are equally useful summaries of its composition. However, even for the
proton, these $6$ numbers are not well determined: only one combination is
known with better than 2\% accuracy, while current uncertainties vary from
10\% to over 100\% for the rest.  A rigorous test of our understanding of QCD
will be provided by comparing lattice QCD computations of these parameters,
the first of which are emerging (see ref.~\cite{Griesshammer:2015ahu} for a
recent compilation), with extractions from Compton data on the proton and
light nuclear targets.  This will be a new benchmark of our ability to use QCD
to compute the proton's structure.

However, a significant problem is that in the low-energy expansion which
parametrises the amplitudes purely as ``Born plus pion-pole plus
polarisabilities", the polarisabilities, especially the spin ones, have little
effect below $\omegalab\approx 100$~MeV. And this expansion breaks down
altogether as the pronounced cusp due to photoproduction is approached; see
also fig.~\ref{fig:multipoles}. Thus, the link between the amplitudes and
their low-energy limits is lost entirely, unless one can resort to an
underlying theoretical description of the process. In this paper, we describe
the amplitudes using Chiral Effective Field Theory (\ChiEFT), the effective
field theory of nucleons and pions, in a variant which also includes the
$\Delta(1232)$. \ChiEFT is formulated as a perturbative expansion in the ratio
of the photon energy and other small scales over the mass of omitted degrees
of freedom such as the $\rho$ meson. \ChiEFT amplitudes thus have a
theoretical uncertainty that can be quantified from the truncation at a given
order.

Contributions from short-distance structure can have a significant effect on
\ChiEFT's predictions for polarisabilities. However, one can compare \ChiEFT's
amplitudes with those of the complementary, dispersion-relation,
approach~\cite{Drechsel:2002ar} when common values of the $6$ polarisabilities
are adopted. We make this comparison for moderate energies in
sect.~\ref{sec:multipoles} and show that the heavier degrees of freedom
omitted from the \ChiEFT calculation have very little impact on the residual
(beyond dipole polarisabilities) energy dependence of the Compton amplitudes
there. Thus, we argue it makes sense to trust the \ChiEFT energy dependence up
to $\omegalab \approx 250\;\MeV$, while fitting the polarisabilities to
Compton data. We then explore the sensitivity of observables to the
polarisability values, arguing that those sensitivities should be rather
independent of the theoretical approach we have adopted.

The world data set for unpolarised scattering on the proton between $50$ and
$170\;\MeV$ is quite rich and was used to extract the scalar polarisabilities
in the \ChiEFT framework in ref.~\cite{McGovern:2012ew}. Very similar results
were obtained in a related work~\cite{Lensky:2014efa} (see also
ref.~\cite{Pasquini:2017ehj}).
From $170$ to $250$~MeV, the data are sparse and not fully in agreement.
Low-energy data on the deuteron also exist and have been used to extract the
neutron's scalar polarisabilities in ref.~\cite{Myers:2014ace}, albeit with
larger statistical and theoretical uncertainties.  More recently, attention
has largely shifted to polarised scattering, in part with the goal of
extracting one or more of the spin polarisabilities. Two recent publications
from MAMI~\cite{Martel:2014pba, Sokhoyan:2016yrc} are a promising start,
though, for reasons we will explain in sect.~\ref{sec:multipoles}, from our
perspective, only the second presents data at energies where a polarisability
extraction is substantially free of theory
uncertainties~\cite{Griesshammer:2014xla}.

In a seminal paper, Babusci \etal~\cite{Babusci:1998ww} carried out a
comprehensive study of the relationship between the $6$ Compton amplitudes and
the $13$ observables which characterise scattering events in which at most one
photon and at most one nucleon, in either the in or out states, is
polarised. The primary observables are ratios of cross section differences for
different orientations of a particle's polarisation to sums of those same
cross sections. If all polarised particles are incoming, these are termed
asymmetries; if one is incoming and one outgoing, polarisation
transfers. While measuring the polarisation of outgoing photons at these
energies is impractical, recoil nucleon polarimetry is feasible: it has,
for example, been done for photoproduction at MAMI. But there has, to date,
been no study of the sensitivity of Compton polarisation-transfer reactions to
nucleon polarisabilities. As for asymmetries, after initial studies for linear
and circular photon polarisations by Hildebrandt
\etal~\cite{Hildebrandt:2003md,Hildebrandt:2005ix}, Pasquini
\etal~\cite{Pasquini:2007hf} explored their sensitivities to polarisabilities;
indeed, this study was instrumental in motivating the MAMI experiments
described in refs.~\cite{Martel:2012,Martel:2014pba,Collicott:2015,
  Huber:2015uza, Martel:2017pln}.  Other partial studies have since been
presented elsewhere~\cite{Griesshammer:2015wha,Mcgovern:2015mgf}. In this
paper, we present a comprehensive analysis of the sensitivities of all $13$
observables to the $6$ polarisabilities. Furthermore, since two combinations
of the polarisabilities can be deduced with considerable accuracy from total
photoproduction cross sections, we are particularly interested in sensitivity
to combinations that are orthogonal to those.

The observables and amplitudes presented in this paper are available via a
\emph{Mathematica} notebook from \texttt{judith.mcgovern@manchester.ac.uk}. It
contains routines for cross-sections, rates and asymmetries from zero to about
$340$~MeV in the lab frame and allows the scalar and spin polarisabilities to
be varied. It also gives all observables for scattering on a free neutron
target, for perusal in the context of few-nucleon experiments in quasi-free
neutron kinematics.

The paper is organised as follows. In sect.~\ref{sec:formalism}, we briefly
review the ingredients of our \ChiEFT calculation of Compton scattering and
list the pertinent observables. In sect.~\ref{sec:results}, we first 
discuss the magnitudes of the proton observables without adjusting the
polarisabilities, and then present the sensitivities of those observables to
changes in the polarisabilities. Finally, we provide selected results for the
free neutron. Our conclusions are given in sect.~\ref{sec:conclusions}. Our
reference values of polarisabilities are provided and briefly discussed in
appendix~\ref{app:polvalues}, technical details on the observables can be found in
appendix~\ref{app:matrices}, and comments on Babusci \etal~\cite{Babusci:1998ww}
in appendix~\ref{app:readbabusci}. This last appendix includes a proof that the
observables suffice to reconstruct the Compton amplitudes. A
supplement studying the sensitivities of neutron observables is available
online and as appendix~\ref{app:moreplots} of the arXiv version.

\section{Formalism}
\setcounter{equation}{0}
\label{sec:formalism}

\subsection{Definition of Polarisabilities}
\label{sec:pols}

As discussed above, the polarisabilities characterise the low-energy response
of the target to external electric and magnetic fields, $\vec E$ and $\vec B$.
This can be described by the following effective
Hamiltonian~\cite{Babusci:1998ww}, which includes all terms that contribute up
to order $\omega^4$ in the low-energy expansion:
\begin{align}
 \mathcal{H}_{\rm eff}^{(2)} =& -\frac12 \, 4\pi
    \left(\alphae \vec E^2 + \betam \vec B^2  \right. \nonumber\\
    &+
     \gammaee \vec\sigma \cdot \vec E \times \dot{\vec E}
   + \gammamm \vec\sigma \cdot \vec B \times \dot{\vec B}
   -2 \gammame E_{ij}\sigma_i B_j
   +2 \gammaem B_{ij}\sigma_i B_j \nonumber\\
&+\left.\alpha_{E1\nu} \dot{\vec E}^2 + \beta_{M1\nu} \dot{\vec B}^2
                                -\frac1{12}\, 4\pi (\alpha_{E2} E_{ij}^2 + \beta_{M2} B_{ij}^2)\right)+\dots\;\;,\label{eq:H-eff}
\end{align}
where $\vec{\sigma}$ is the nucleon spin operator and
$T_{ij} = \frac12 (\nabla_i T_j + \nabla_j T_i)$ with $T=E,B$. Here $\gamma_i$
are the four spin polarisabilities, labelled by the multipolarities of the
incident and outgoing photon, $Xl\to X^\prime l^\prime$ with
$l^\prime=l\pm\{0,1\}$. Accordingly, $\alphae$, $\betam$ and $\gamma_i$ are
called dipole polarisabilities, while $\alpha_{E2}$ and $\beta_{M2}$ are
quadrupole ones. The coefficients $\alpha_{E1\nu}$ and $\beta_{M1\nu}$ are
``dispersive polarisabilities": they contribute to the amplitudes in the same
places and with the same angular dependence as $\alphae$ and $\betam$, but
with an extra power of $\omega^2$. The expansion of
$\mathcal{H}_{\rm eff}$---and of the Compton amplitude---in photon energies
fails as the pion-production threshold is approached because of the
non-analyticities that introduces.  Another interpretation of
eq.~\eqref{eq:H-eff} is that it reproduces the first few terms of a multipole
expansion of those parts of the Compton amplitude which go beyond the Powell
amplitude, \ie beyond that for a point target with an anomalous magnetic
moment.  This last interpretation is discussed further in
sect.~\ref{sec:multipoles}, with details presented in
refs.~\cite{Hildebrandt:2003md,Hildebrandt:2003fm}. Finally, we mention that
we quote $\alphae$ and $\betam$ in units of $10^{-4}\;\fm^3$, and the spin
polarisabilities in units of $10^{-4}\;\fm^4$, throughout our presentation.

As stressed in refs.~\cite{Babusci:1998ww, Lensky:2015awa}, this
non-relativistic Hamiltonian is frame-dependent and hence so is the definition
of the polarisabilities; like those authors, we choose the Breit frame, which
has the advantage over the centre-of-mass (cm) frame (used by
ref.~\cite{Holstein:1999uu}) of being crossing-symmetric. The relation between
the polarisabilities and the low-energy expansion of the Breit- or cm-frame
amplitudes is given in eq.~\eqref{eq:polamp}.

\ChiEFT is summarised in the next section. It makes predictions based on
pion-cloud and Delta-resonance effects for all of these polarisabilities;
short-range effects enter the chiral expansion for a particular polarisability
at an order which increases with every derivative in the effective
Hamiltonian. This short-range nucleon structure---which is not resolved by
\ChiEFT---affects the $6$ polarisabilities under discussion here quite
significantly, but its effect on the remaining energy dependence of the
Compton amplitude occurs at a higher order in the chiral expansion.  The
dipole spin polarisabilities are, strictly speaking, predictions at the chiral
order to which we work, but we allow them to vary in order to assess the ways
in which they affect observables. Therefore, in
sects.~\ref{sec:crosssectionvar} and~\ref{sec:asymmetriesvar}, we discuss the
sensitivity of observables with respect to variations of the polarisabilities
around the \ChiEFT values quoted in appendix~\ref{app:polvalues}, assuming
that---at least for $\omegalab \lesssim 250$ MeV---the rest of the energy
dependence of the Compton amplitude is correctly captured in \ChiEFT.
Such variation is, in fact, the method by which we determined the static
polarisabilities $\alphae$ and $\betam$ (and gleaned some information on
$\gammamm$) in our earlier fit to the proton Compton
database~\cite{McGovern:2012ew}.  It also means that we vary only the $6$
dipole polarisabilities $\alphae$, $\betam$ and $\gamma_i$, and not any higher
polarisabilities. In fact, the entire contribution of $l\ge 2$ multipoles to
the asymmetries has been shown to be small at the energies we examine
here~\cite{Hildebrandt:2005ix}. In addition, the chance of extracting
coefficients from a sparse and noisy database in a statistically meaningful
way decreases dramatically with multipolarity. We note that our
polarisability-variation strategy concurrently provides an assessment of which
kinematics and observables give complementary tests of the $\chi$EFT
prediction for the Compton amplitude, since it shows places where that
prediction is robust against changes in the specific values chosen for the
dipole polarisabilities.

\subsection{Compton Scattering on One Nucleon in \texorpdfstring{\ChiEFT}{ChiEFT}}
\label{sec:formalism-chiEFT}

As the \ChiEFT Compton amplitudes have been described in great detail in
refs.~\cite{Griesshammer:2012we, McGovern:2012ew,Griesshammer:2015ahu}, we
refer to these for notation, the relevant parts of the chiral Lagrangian, and
the pertinent amplitudes and parameters. Here, we only briefly recapitulate
the main ingredients; full details are given in
ref.~\cite{McGovern:2012ew}. Three typical low-energy scales exist in \ChiEFT
with a dynamical $\Delta(1232)$ degree of freedom: the pion mass $\mpi$ as the
typical chiral scale; the Delta-nucleon mass splitting
$\DeltaM \approx 300\;\MeV$; and the photon energy $\omega$. Each provides a
small, dimensionless expansion parameter when measured in units of a natural
``high'' scale $\Lambda_\chi\gg\DeltaM,\mpi,\omega$ at which the theory is
expected to break down because new degrees of freedom enter. While a
three-parameter expansion is possible, it is more economical to follow
Pascalutsa and Phillips~\cite{PP03} and take advantage of a convenient
numerical coincidence by identifying
\begin{equation}
  \delta\equiv\frac{\Delta_M}{\Lambda_\chi}\approx
  \left(\frac{m_\pi}{\Lambda_\chi}\right)^{1/2}\approx 0.4\ll1\;\;.
\end{equation}
For simplicity, we count $\MN\sim\Lambda_\chi$ and employ one common breakdown
scale $\Lambda_\chi\approx650\;\MeV$, consistent with the masses of the
$\omega$ and $\rho$ as the next-lightest exchange mesons. We can then identify
two regimes of photon energy $\omega$ where the counting simplifies
further~\cite{Griesshammer:2012we, McGovern:2012ew}.  In \emph{regime I},
$\omega\lesssim\mpi$ counts like a chiral scale,
$\omega\sim\mpi\sim\delta^2\Lambda_\chi\ll\Delta_M$, and pion-cloud physics
dominates.  In contradistinction, in \emph{regime II},
$\omega\sim \Delta_M\sim\delta^1\Lambda_\chi\gg\mpi$, and the Delta
resonance may be excited so that it strongly dominates the relevant channels
and dwarfs contributions from the pion cloud through its large width and
strong $\gammaN\Delta$ coupling. Because of the increasing photon energy and
the reordering of contributions expected around the Delta resonance, the
level of theoretical uncertainty is actually different in these two
regimes. The ingredients of our calculation ensure that the Compton amplitude
contains in regime I all contributions at $\calO(e^2\delta^4)$ (\NXLO{4},
accuracy $\delta^5 \approx2\%$); and in regime II at $\calO(e^2\delta^0)$
(NLO, accuracy $\delta^2\approx 20\%$). For even
higher energies, the expansion parameter approaches one, and the EFT series
does not converge. 

As detailed in ref.~\cite{McGovern:2012ew}, the ingredients are covariant
nucleon-Born, pion-pole and Delta-pole graphs, and heavy-baryon $\pi N$ and
$\pi \Delta$ loop graphs to order $e^2\delta^4$ (\ie chiral order $p^4$ plus
leading Delta loops). The $\pi N$ loop corrections to the $\gamma N \Delta$
vertex are added to ensure Watson's theorem is satisfied in
photoproduction. These are higher order in regime I but NLO in regime II.
Most physical constants that enter are taken from the Review of Particle
Physics~\cite{Patrignani:2016xqp}; the $\gamma N\Delta$ coupling constants are
close to those determined from photoproduction in
ref.~\cite{Pascalutsa:2006up} but are adjusted slightly to fit the strength of
the unpolarised differential Compton cross section at the Delta
peak~\cite{McGovern:2012ew} (see also \cite{Cawthorne:2015orf}).  At
${\cal O}(e^2\delta^4)$, there are two Compton-specific low-energy constants
$\delta\alphae$ and $\delta\betam$ for each nucleon; these encode
short-distance contributions to the scalar polarisabilities and have been fit
to unpolarised Compton scattering data. To obtain a good description of this
data, it was also necessary to fit $\gammammp$, though strictly speaking it is
not a free parameter at this order; in this fit, the other spin
polarisabilities were left at their predicted values.

For reference, we repeat (from ref.~\cite{Griesshammer:2015ahu}) the \ChiEFT
values of the proton and neutron scalar and spin polarisabilities, together
with their theoretical and (for extracted quantities) statistical errors, in
appendix~\ref{app:polvalues}.  Table~1 of ref.~\cite{Griesshammer:2015ahu} shows
that, within the respective uncertainties, all values adopted here are
compatible with those of other approaches and with available experimental
information.  And, in fact, while the exact results for rates and
asymmetries depend on these particular values, we are more concerned here with
the sensitivities of observables to polarisabilities; those are not markedly
affected by the baseline polarisability values chosen. Those sensitivities should
also be less dependent on details of the calculational framework used to
compute them, as  discussed in the Introduction and in
sect.~\ref{sec:multipoles}.

\subsection{Comparing Theories via a Multipole Expansion}
\label{sec:multipoles}

The basic output of this \ChiEFT calculation is the $6$ energy- and
angle-dependent amplitudes of eq.~\eqref{eq:Tmatrix}, and all observables are
directly constructed from these.  Any extraction of polarisabilities from
experiment is only as good as the reliability of these amplitudes. We have
already mentioned the fact that the power-counting provides an \emph{a priori}
estimate of theoretical uncertainties and indicates that they are
intrinsically less reliable as the energy increases.  However, even at low
energies, other than the extent to which they have already been confronted
with data~\cite{McGovern:2012ew, Myers:2014ace,Sokhoyan:2016yrc}, external
validation is not easy.  One important test, therefore, is the
extent to which different approaches agree.

A useful summary is provided by the dynamical polarisabilities. They are the
coefficients of the multipole expansion of the non-Born amplitudes in the cm
frame at fixed energy, with a polynomial energy dependence divided out. At
fixed photon energy, they are distinguished by different angular dependencies,
and are single-variable functions of the photon energy. Dynamical
polarisabilities were defined in
refs.~\cite{Griesshammer:2001uw,Hildebrandt:2003fm} and further explained in
refs.~\cite{Hildebrandt:2003md,Hildebrandt:2005ix, Lensky:2015awa}, so we will
not repeat the definitions here. We do stress though that, as with any
multipole expansion, they do not contain more information than the amplitudes
themselves; it is just that the information is more readily
accessible\footnote{There is an approach, pioneered by
  Hildebrandt~\cite{Griesshammer:2004yn, Hildebrandt:2005ix}, of fitting only
  the $l = 1$ multipoles to cross-section data, exploiting the existence of
  such data at a range of angles for a particular energy. After the current
  paper was submitted, Krupina \etal~\cite{Krupina:2017pgr} presented an
  extended version of such an analysis; they supplement this information with
  low-energy theorems and sum-rule determinations of forward scattering
  amplitudes to extract both static and dynamical dipole polarisabilities.}.
\begin{figure}[!htbp]\thisfloatpagestyle{empty}
\begin{center}
    \includegraphics[width=0.93\linewidth]{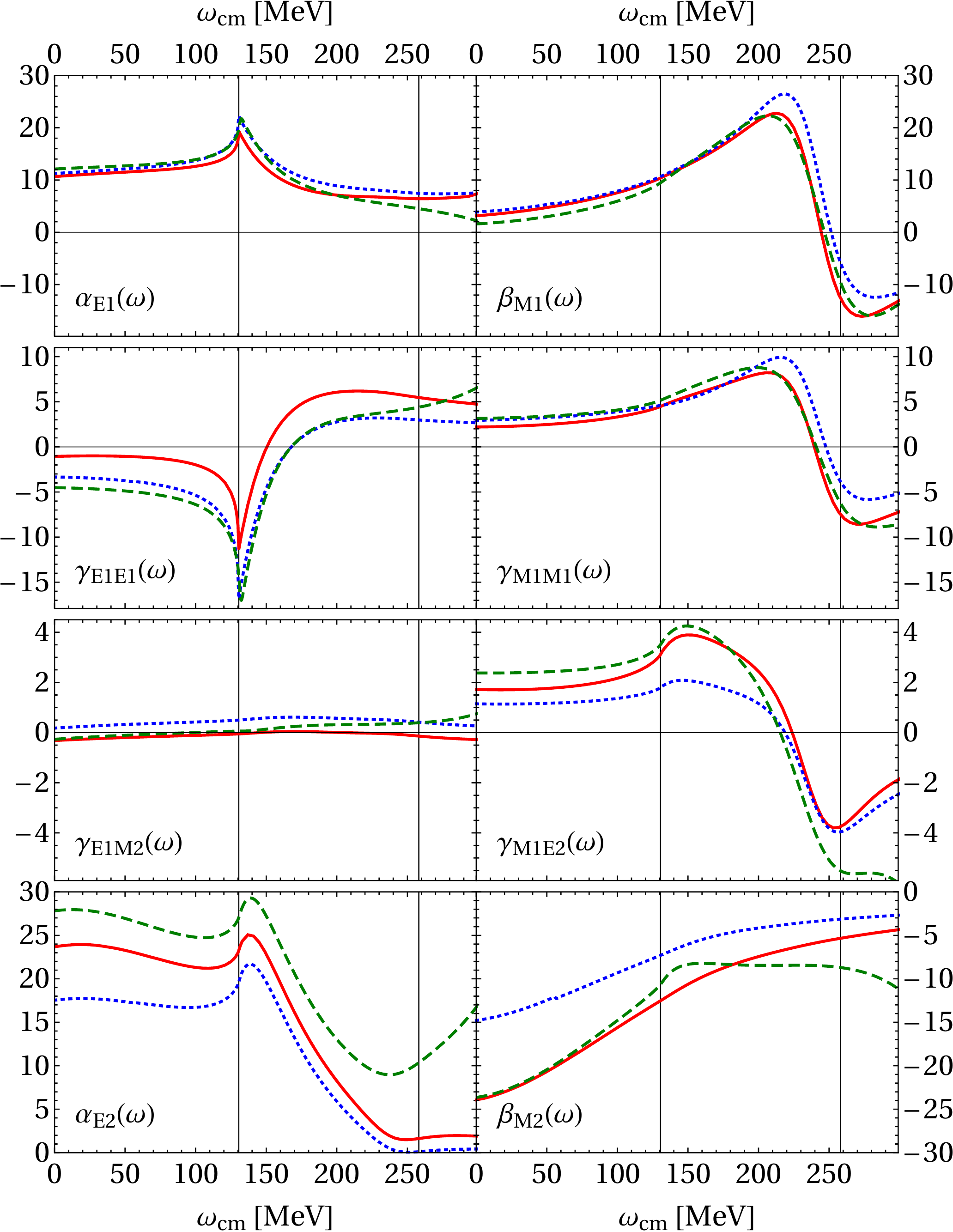}
    \caption{(Colour online) Real parts of the dominant dynamical
      polarisabilities for low-energy Compton scattering from the proton,
      plotted as a function of cm photon energy. The units are
      $10^{-4}~{\rm fm}^n$ where $n=3$ for $\alphae$ and $\betam$, $n=4$ for
      the $\gamma_i$, and $n=5$ for $\alpha_{E2}$ and $\beta_{E2}$. Red
      (solid): this work; green (dashed): DR-based by Pasquini
      \etal~\cite{Hildebrandt:2003fm}; blue (dotted) 3rd-order covariant
      \ChPT by Lensky \etal\ \cite{Lensky:2015awa}. Note that each row has its
      distinct plot scale.}
\label{fig:multipoles}
\end{center}
\end{figure}

For calculations and plots in the subsequent sections, we use the full
amplitudes, and not the multipoles.  But here, fig.~\ref{fig:multipoles} shows
the first $8$ multipoles in the present theory (after fitting as described
above), in the ${\cal O}(e^2\delta^3)$ covariant framework of Lensky
\etal~\cite{Lensky:2009uv,Lensky:2015awa} (without fitting to data), and in the
dispersion-relation framework of
Pasquini~\etal~\cite{Drechsel:2002ar,Hildebrandt:2003fm,barbaraprivate}, based
on integrals over pion-photoproduction multipoles (with $\alphae-\betam$ and
$\gamma_\pi$ fit to Compton scattering data)\footnote{As this article was
  being completed, Pasquini \etal~\cite{Pasquini:2017ehj} published a new analysis of the proton
  Compton database employing the dispersion-relation amplitudes. They used the
  bootstrap technique to obtain new static values of $\alphae$ and $\betam$,
  as well as the first results for the corresponding ``dispersive
  polarisabilities" (see eq.~(\ref{eq:H-eff}) and the subsequent
  discussion). The curves of fig.~\ref{fig:multipoles}
  do not reflect the new analysis, but the changes are quite small.}.
We will compare these approaches shortly.  The evolution of each dynamical
polarisability with $\omega$ displays the relevant physics in each channel
(cusps at the pion-production threshold, the Delta resonance, etc.).  The
value of a dynamical polarisability at $\omega=0$ is then the corresponding
static polarisability.  For the three sets of curves shown in
fig.~\ref{fig:multipoles}, these are the values tabulated in Table 1 of
ref.~\cite{Griesshammer:2015ahu} (see also \cite{Lensky:2015awa}). Varying a
static polarisability is identical to simply shifting the corresponding
dynamical polarisability up or down.

As first noted in ref.~\cite{Hildebrandt:2003fm}, based on more limited
results, there is a substantial degree of agreement on the \emph{shape} of the
polarisabilities up to around $250$~MeV lab energy ($200$~MeV cm energy in the
figure). The same pion-loop and Delta-resonance physics is encoded in all
three calculations.  Furthermore, after adjusting the static polarisabilities
to a common value, the results generically lie very close to one
another. Indeed, they agree more closely than the theoretical error estimates
shown in ref.~\cite{Lensky:2015awa} might suggest.  In the Delta-dominated
multipoles, this agreement continues up to surprisingly high energies, but
overall disagreement creeps in above $250$~MeV lab energy, where one also
expects polarisabilities beyond the dipole ones to play an increasing role.
(The imaginary parts of the multipoles are much smaller than the real parts in
this region, so we do not show them. They may actually be more easily
accessible via pion photoproduction, see ref.~\cite{Hildebrandt:2003fm}. For
the Delta-dominated multipoles, for which the real parts vanish near the Delta
pole, the imaginary parts agree well~\cite{Lensky:2015awa}.)

It is important to remember that the units of the polarisabilities are not all
the same. The large numerical factors in the definitions of $\alpha_{E2}$ and
$\beta_{E2}$, and the dividing out of $\omega^4$ from these, conspire to make
them look more important than they actually are, at least at low energies.
Conversely, discrepancies between the polarisabilities are multiplied in the
amplitudes by the appropriate powers of $\omega$, and hence are rather
down-played by this depiction at high energies.

On the theory side, two main differences exist between the ingredients of this
work and those of Lensky \etal~\cite{Lensky:2009uv,Lensky:2015awa}, which is also a \ChiEFT
calculation. First, in the latter, the pion loops are calculated in a
kinematically-covariant framework. Second, their work is at
${\cal O}(e^2\delta^3)$, \ie one order lower than in ours. The most important
physical consequence is that they, unlike us, omit the anomalous part of the
magnetic moment of the nucleons in the pion-nucleon loops.  Both differences
imply that deviations between their approach and ours are indicative of the
typical sizes of ${\cal O}(e^2\delta^4)$ corrections (which are fully included
in our approach).

A complementary approach is provided by dispersion
relations~\cite{Drechsel:2002ar} for the Compton amplitude, but these inherit
uncertainties from the photoproduction database (which deteriorates at higher
energies) and from the need to model the high-energy part of the dispersive
integrals. Neither \ChiEFT alone nor dispersion relations alone should be
taken as the gold standard for the theory of Compton scattering. Rather,
comparing the output of these two different approaches allows one to assess
the reliability of each; where they agree, the results can confidently be
described as framework-independent.

In ref.~\cite{Martel:2014pba}, data for $\Sigma_{2x}$ at a lab energy of
approximately $300\;\MeV$ have been analysed in the dispersion-relation
framework~\cite{Drechsel:2002ar}, and in Lensky's covariant
\ChiEFT~\cite{Lensky:2009uv}. The results extracted for spin polarisabilities
using the two theory approaches are consistent within the--somewhat
sizeable---statistical uncertainties. However, the appreciable differences of
their individual dynamical polarisabilities in fig.~\ref{fig:multipoles} at
this energy suggest the similarity may mask rather different physics in the
individual Compton amplitudes. From our perspective, therefore, this energy is
too high to reliably conclude that polarisability extractions are
framework-independent.

Thus, the message we take from fig.~\ref{fig:multipoles} is that there is a
concurrence of theory approaches for the energy dependence of all $6$
dynamical polarisabilities up to around $250$~MeV lab energy.  \ChiEFT
predictions of high intrinsic reliability only exist for energies up to a
point not far beyond the pion-production threshold. However, up to
$\omegalab \approx 250$~MeV, \ChiEFT and dispersion relations agree
quantitatively. We thus claim that there is an understanding of the energy
dependence of dynamical polarisabilities in this energy domain which does not
rely on the theoretical framework used. It is this range, therefore, on which
we focus our attention when varying the polarisabilities. Planning of
experiments then needs to balance the improved sensitivity to the
polarisabilities at higher energies with the decreasing reliability of any
extrapolation back to the zero-energy point.

\subsection{Observables}
\label{sec:Observables}

We follow Babusci \etal~\cite{Babusci:1998ww} closely. The interested reader
is directed to this invaluable resource for further details; some
elucidation of certain subtleties in ref.~\cite{Babusci:1998ww} is given in appendix~\ref{app:readbabusci}.

\begin{figure}[!htbp]
\begin{center}
\includegraphics[width=0.8\linewidth]{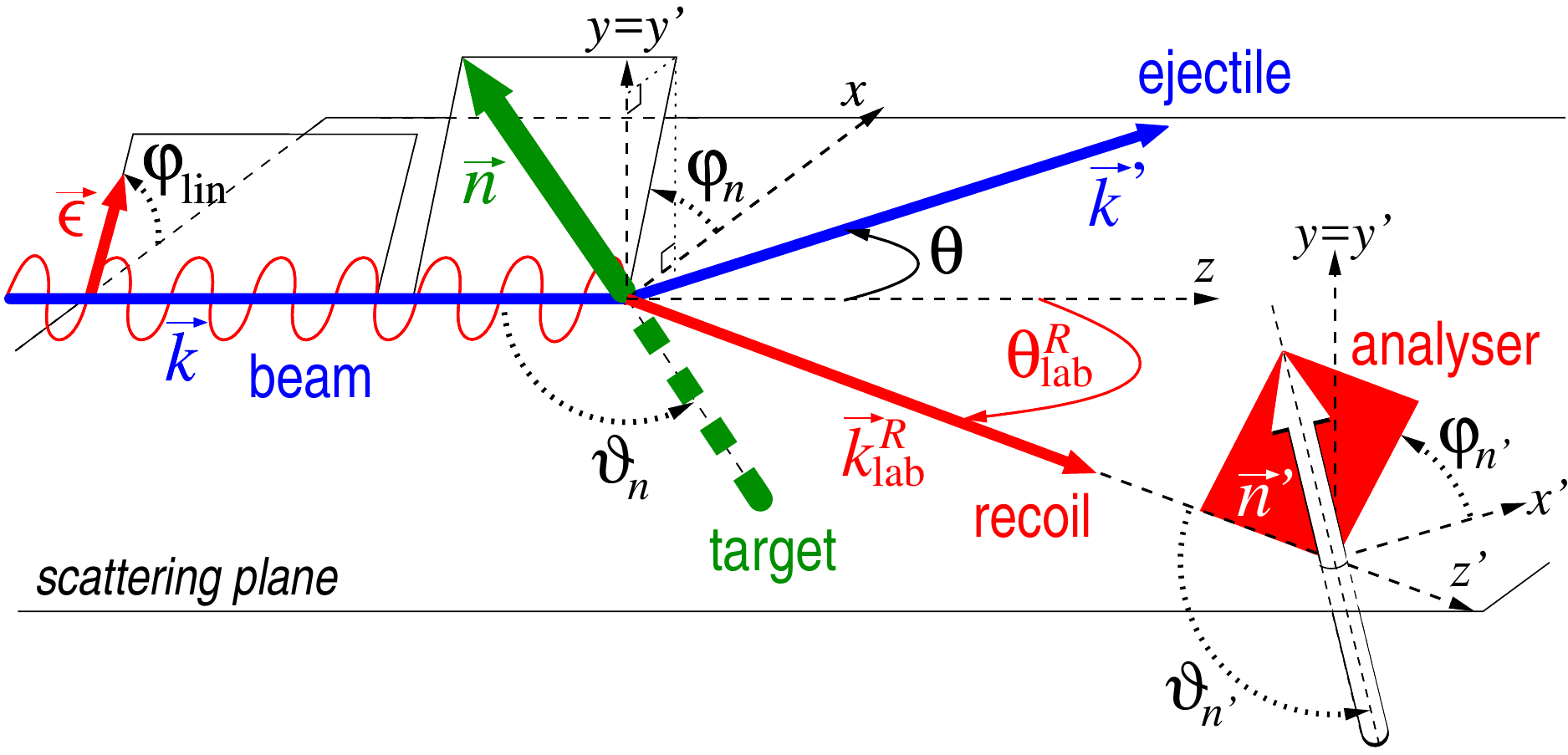}
\caption{(Colour online) Kinematics and variables for polarisation asymmetries
  and polarisation-transfer observables.}
\label{fig:spinkinematics}
\end{center}
\end{figure}

We first define the kinematics and coordinate system. Unless otherwise
specified, we work in the laboratory frame. The incident photon momentum is
$\kv$ ($|\kv|=\omegalab$); the outgoing one is $\kv'$
($|\kv'|=\omegaprimelab=\omegalab/(1+\omegalab(1-\cos\theta)/\MN)$); the
scattering angle $\theta$ is the angle between them, so
$\kv\cdot\kv'=\omegalab\,\omegaprimelab\cos\theta$.  As illustrated in
fig.~\ref{fig:spinkinematics}, the $z$-axis is defined as the incoming beam
direction, $\kv=k\;\vec e_z$; the scattering plane is the $xz$-plane, with the
$y$-axis perpendicular to it to form a right-handed triplet,
\ie $\kv\times\kv'= \omegalab\,\omegaprimelab\sin\theta\;\vec e_y$.  The
angle from the scattering plane to the polarisation axis of a
linearly-polarised photon is $\philin$.  The nucleon recoils in the scattering
plane; its momentum defines the $z^\prime$ axis; and a primed coordinate
system $\{x',y',z'\}$ is obtained from $\{x,y,z\}$ by a clockwise rotation
through the recoil angle $\theta_R$ about the $y$ axis (which is also the $y'$
axis).

For asymmetries with a polarised target, the target nucleon's polarisation density is
\begin{equation}
  \rho(P,\vec n)=\half\bigl(1+P\;\vec\sigma\cdot\vec n\bigr)
\end{equation}
where $\vec{n}=(\sin\thetan\cos\phin,\sin\thetan\sin\phin,\cos\thetan)$
($|\vec n|=1$) is the nucleon spin direction, and $P\in[0;1]$ its degree of
polarisation (Basel convention). We define the azimuthal angle $\thetan$ from
the $z$-axis to $\vec{n}$ and polar angle $\phin$ from the $x$-axis to the
projection of $\vec{n}$ onto the $xy$-plane; see
fig.~\ref{fig:spinkinematics}. For polarisation-transfer observables, it is
customary to resolve the recoil polarisation direction $\vec{n}^\prime$ in the
primed coordinate system, defining angles $\thetan^\prime$ and $\phin^\prime$
with respect to the $z'$ axis and the scattering plane.

``Primed'' indices are used to indicate the nucleon spin direction for
polarisation-transfer observables, while unprimed ones are used for polarised
targets. For example, $\Sigma_y$ is an asymmetry with the target nucleon
polarised along the $y$ axis, while $\Sigma_{y^\prime}$ is a polarisation
transfer observable from an unpolarised target to a recoil nucleon polarised
along the $y^\prime$ direction.

The photon beam polarisation is characterised by the three Stokes parameters
$\xi_{i}$, all of which satisfy $\xi_i\in[-1;1]$.  $\xi_2$ is the degree of
circular polarisation, with $\xi_2=\pm1$ describing a fully right/left circularly
polarised photon (positive/negative helicity); $\xi_1$ and $\xi_3$ describe
linear polarisation with degree $\sqrt{\xi_1^2+\xi_3^2}\in[0;1]$ and
polarisation angle specified by $\cos[2\philin]=\xi_3/\sqrt{\xi_1^2+\xi_3^2}$
and $\sin[2\philin]=\xi_1/\sqrt{\xi_1^2+\xi_3^2}$. Therefore, $\xi_3=\pm1$
with $\xi_1=0$ describes a beam which is linearly polarised
within/perpendicular to the scattering plane, and $\xi_1=\pm 1$ with $\xi_3=0$
one which is linearly polarised at angle $\philin=\pm\pi/4$ relative to the
scattering plane.

With these definitions, Babusci \etal~\cite{Babusci:1998ww} parametrise the
cross section with polarised beam and/or target and without detection of
final-state polarisation as
\begin{equation}
\begin{split}
  \label{eq:babcross}
  \frac{\dd\sigma}{\dd\Omega}=\left.\frac{\dd\sigma}{\dd\Omega}
  \right|_\text{unpol}\;\bigg[1&\;+\;\xi_3\;\Sigma_3(\omega,\theta)\;+
\;P\,n_y\;\Sigma_y(\omega,\theta)\;
  +\;P\,\xi_1\Big(n_x\;\Sigma_{1x}(\omega,\theta)+
  n_z\;\Sigma_{1z}(\omega,\theta)\Big)\\
  &\;+\;P\,\xi_2\Big(n_x\;\Sigma_{2x}(\omega,\theta)
  +n_z\;\Sigma_{2z}(\omega,\theta)\Big)
  \;+\;P\,\xi_3\;n_y\;\Sigma_{3y}(\omega,\theta)
  \bigg]\;\;,
\end{split}
\end{equation}
where $n_i$ are the components of the polarisation vector $\vec{n}$ of the
spin-$\half$ target in its rest frame\footnote{Babusci \etal\ denote
  them by $\zeta_i$~\cite{Babusci:1998ww}.}.
This parametrisation can also be related to a more generally applicable one
which uses spherical multipoles~\cite{3Hetobe,Griesshammer:2013vga}.

The $8$ linearly independent asymmetries are: 

\begin{itemize}

\item $1$ differential cross section
  $\dis\left.\frac{\dd\sigma}{\dd\Omega}\right|_\text{unpol}$ of unpolarised
  photons on an unpolarised target.

\item $1$ beam asymmetry of a linearly polarised beam on an unpolarised
  target:
\begin{equation} 
  \label{eq:asym3}
  \Sigma_3=\frac{\dd \sigma^{||}-\dd \sigma^{\perp}}{\dd \sigma^{||}+\dd \sigma^{\perp}}\;\;.
\end{equation}
Here and below, $\dd \sigma$ is shorthand for ${\dd\sigma}/{\dd\Omega}$;
superscripts refer to photon polarisations (``$\parallel$'' for polarisation
in the scattering plane, ``$\perp$'' for perpendicular to it); subscripts to
nucleon polarisations; and the absence of either means unpolarised.

\item $1$ target asymmetry for nucleons polarised out of the scattering plane
  along the $\pm y$ direction and an unpolarised beam:
\begin{equation} 
  \label{eq:asymy}
\Sigma_y=\frac{\dd \sigma_{y}-\dd \sigma_{-y}}{\dd \sigma_{y}+\dd \sigma_{-y}}\;\;.
\end{equation}

\item $2$ double asymmetries of right/left-circularly polarised photons on a
  target polarised along the $\pm x$ or $\pm z$ directions:
\begin{equation}\label{eq:asym2}
\Sigma_{2x}=\frac{\dd\sigma^{R}_{x}-\dd\sigma^{L}_{x}}{\dd\sigma^{R}_x+\dd\sigma^{L}_{x}}
\;\;,\;\;
\Sigma_{2z}=\frac{\dd\sigma^{R}_{z}-\dd\sigma^{L}_{z}}{\dd\sigma^{R}_z+\dd\sigma^{L}_{z}}\;\;.
\end{equation}

\item $3$ double asymmetries of linearly-polarised photons on a polarised target:
\begin{align}\label{eq:asym1}
\Sigma_{1x}&=\frac{\dd\sigma^{\pi/4}_{x}-\dd\sigma^{-\pi/4}_{x}}{\dd\sigma^{\pi/4}_x+\dd\sigma^{-\pi/4}_{x}}\;\;,\;\;
\Sigma_{1z}=\frac{\dd\sigma^{\pi/4}_{z}-\dd\sigma^{-\pi/4}_{z}}{\dd\sigma^{\pi/4}_z+\dd\sigma^{-\pi/4}_{z}}\\
\Sigma_{3y}&=\frac{(\dd \sigma^{||}_y-\dd \sigma^{\perp}_y)-(\dd \sigma^{||}_{-y}-\dd \sigma^{\perp}_{-y})}
{\dd \sigma^{||}_y+\dd \sigma^{\perp}_y+\dd \sigma^{||}_{-y}+\dd \sigma^{\perp}_{-y}}\;\;.\label{eq:asym3y}
\end{align}
\end{itemize}
The decomposition of eq.~\eqref{eq:babcross} holds in both the lab and
centre-of-mass frames, but the functions are frame-dependent. By time-reversal
invariance, the $8$ recoil polarisations $\Sigma_{1^\prime x^\prime}$
etc.~are related (but usually not identical) to the functions above.

Turning to polarisation-transfer observables, final-state photon polarisation
is very hard to detect in the energy range of interest. Thus, we concentrate
on those observables in which a polarised photon beam transfers its
polarisation to a recoil nucleon, with an unpolarised target and undetected
scattered-photon polarisation.  (Polarisation transfer from a polarised target
to a polarised scattered photon follows from time-reversal invariance.)  The
kinetic energy of the recoil nucleon increases near-linearly as a function of
$\cos\theta$ from zero at $\theta=0^\circ$ to a maximum at $\theta=180^\circ$.
The maximum recoil kinetic energy is $18\;\MeV$ for $\omegalab=100\;\MeV$ and
$62\;\MeV$ at $\omegalab=200\;\MeV$.  In polarisation transfer to the nucleon,
an ideal experiment places an analyser in front of the detector to allow only
certain recoil polarisations to be detected. Actual experiments use
polarisation-dependent scattering, \eg,~from \fourHe or
${}^{12}$C~\cite{Ohlsen:1972zz, Sikora:2013vfa}.

Polarisation-transfer observables are defined by a parametrisation of the cross section very similar to that 
of eq.~(\ref{eq:babcross}), but with the target polarisation axis
$(\vartheta_n,\varphi_n)$ replaced by the orientation of the axis of the
ideal analyser ($P^\prime\equiv1$), with $(\thetan^\prime,\phin^\prime)$
measured in the ``primed'' coordinate system specified above:
\begin{equation}
\begin{split}
  \label{eq:babrecoil}
\frac{\dd\sigma}{\dd\Omega}=\left.\frac 1 2\;\frac{\dd\sigma}{\dd\Omega}
  \right|_\text{unpol}\;\bigg[1&\;+\;n_{y^\prime}\;\Sigma_{y^\prime}(\omega,\theta)\;
  +\;\xi_3\;\Sigma_3(\omega,\theta)\;
  +\;\xi_1\Big(n_{x^\prime}\;\Sigma_{1x^\prime}(\omega,\theta)+
  n_{z^\prime}\;\Sigma_{1z^\prime}(\omega,\theta)\Big)\\
  &\;+\;\xi_2\Big(n_{x^\prime}\;\Sigma_{2x^\prime}(\omega,\theta)
  +n_{z^\prime}\;\Sigma_{2z^\prime}(\omega,\theta)\Big)
  \;+\;\xi_3\;n_{y^\prime}\;
  \Sigma_{3y^\prime}(\omega,\theta)\bigg]\;\;.
\end{split}
\end{equation}
The overall factor of $\frac 1 2$ arises from the fact that, as we specify the
final nucleon polarisation, we are no longer summing over final states.  Of
the $6$ extra functions introduced here, $\Sigma_{y^\prime}$ is equal to
$\Sigma_y$ by time-reversal invariance, so there are $5$ new ones:
\begin{itemize}
\item $\Sigma_{2x^\prime}$ and $\Sigma_{2z^\prime}$ are polarisation
  transfers from circularly-polarised photons to a polarised recoiling
  nucleon; 

\item $\Sigma_{1x^\prime}$, $\Sigma_{1z^\prime}$ and $\Sigma_{3y^\prime}$ are
  polarisation transfers from linearly-polarised photons to a polarised
  recoiling nucleon.
\end{itemize}
Their definitions exactly follow eqs.~\eqref{eq:asym1}, \eqref{eq:asym2} and
\eqref{eq:asym3y}, but with the target spin labels $\{x,y,z\}$ replaced by
$\{x',y',z'\}$ and now referring to the analyser orientation $\vec{n}^\prime$ in the
primed coordinate frame.

Only $6$ of the $13$ distinct observables are non-zero below the first
strong inelasticity, which is set by the pion-production threshold: the cross
section, the three asymmetries $\Sigma_3$, $\Sigma_{2x}$, $\Sigma_{2z}$, and
the two polarisation-transfer observables $\Sigma_{2x^\prime}$ and
$\Sigma_{2z^\prime}$. These suffice to determine the $6$ real Compton
amplitudes $A_i$. Above threshold, these combine with any $5$ of the
additional $7$ observables to provide all information on the $11$
independent real functions ($12$ minus an overall phase) of the complex
Compton amplitudes. We prove this in appendix~\ref{app:completeness}.

This reconstruction could potentially provide valuable information on Compton
ampli\-tudes---in the same way that complete photoproduction experiments enhance
our understanding of meson photoproduction. Complete low-energy Compton
experiments would allow the extraction of polarisabilities from experiments at
a single angle with no other theoretical biases, though the implementation of
this approach may be some way off. They may prove to be more interesting at
  high energies, where the multipole expansion breaks down and the full
  amplitude needs to be reconstructed.

But this is not the type of experiment we focus on here. Instead, in what
follows, we present the sensitivity of observables to the leading structure
effects in the proton Compton amplitude, as encoded by the static dipole
polarisabilities, with the goal of identifying a diverse set of observables,
over a range of energies and angles, that together can provide constraints on
those polarisabilities.

\section{Results}
\setcounter{equation}{0}
\label{sec:results}

We now present results for the $13$ observables defined in the previous
section. These are given as contour plots, presentational details of which we
explain in sects.~\ref{sec:noteonplots} and~\ref{sec:sensitivities}. The plots
in sects.~\ref{sec:crosssection} and \ref{sec:asymmetries} concern magnitudes
of observables, while sects.~\ref{sec:crosssectionvar} and
\ref{sec:asymmetriesvar} show sensitivities of observables to variations in
the dipole scalar and spin polarisabilities (``sensitivity plots'').  Unless
otherwise stated, the kinematics are for Compton scattering on a proton in the
lab frame (with the obvious exception being in sect.~\ref{sec:neutron}, where
results for the neutron are presented), and the baseline values for
polarisabilities are those given in appendix~\ref{app:polvalues}.  Before
continuing, we reiterate that in all subsequent figures, we use the
\emph{full} Compton amplitudes, and not a truncated multipole expansion.

\subsection{A Note on Contour Plots}
\label{sec:noteonplots}

We first discuss contour plots as a quick and intuitive way to assess both
magnitudes of observables and their sensitivity to polarisabilities. A
detailed analysis and extraction will of course need a more thorough
comparison.

In all plots, we use a ``heat scale'' colour gradient, from deep blue to deep
red. Except for the unpolarised cross section, deep blue indicates large and
negative, while deep red is large and positive.  The white region indicating
``numerical zero'' separates regions of small sensitivities, \ie slightly
blue and yellow tints. For asymmetries and polarisation-transfer observables,
the extreme values of $1$ and $-1$ set the range naturally. For sensitivities,
a common range was chosen by hand to maximise the information conveyed across
all plots.

\begin{figure}[!h]
\begin{center}    \includegraphics[width=\textwidth]{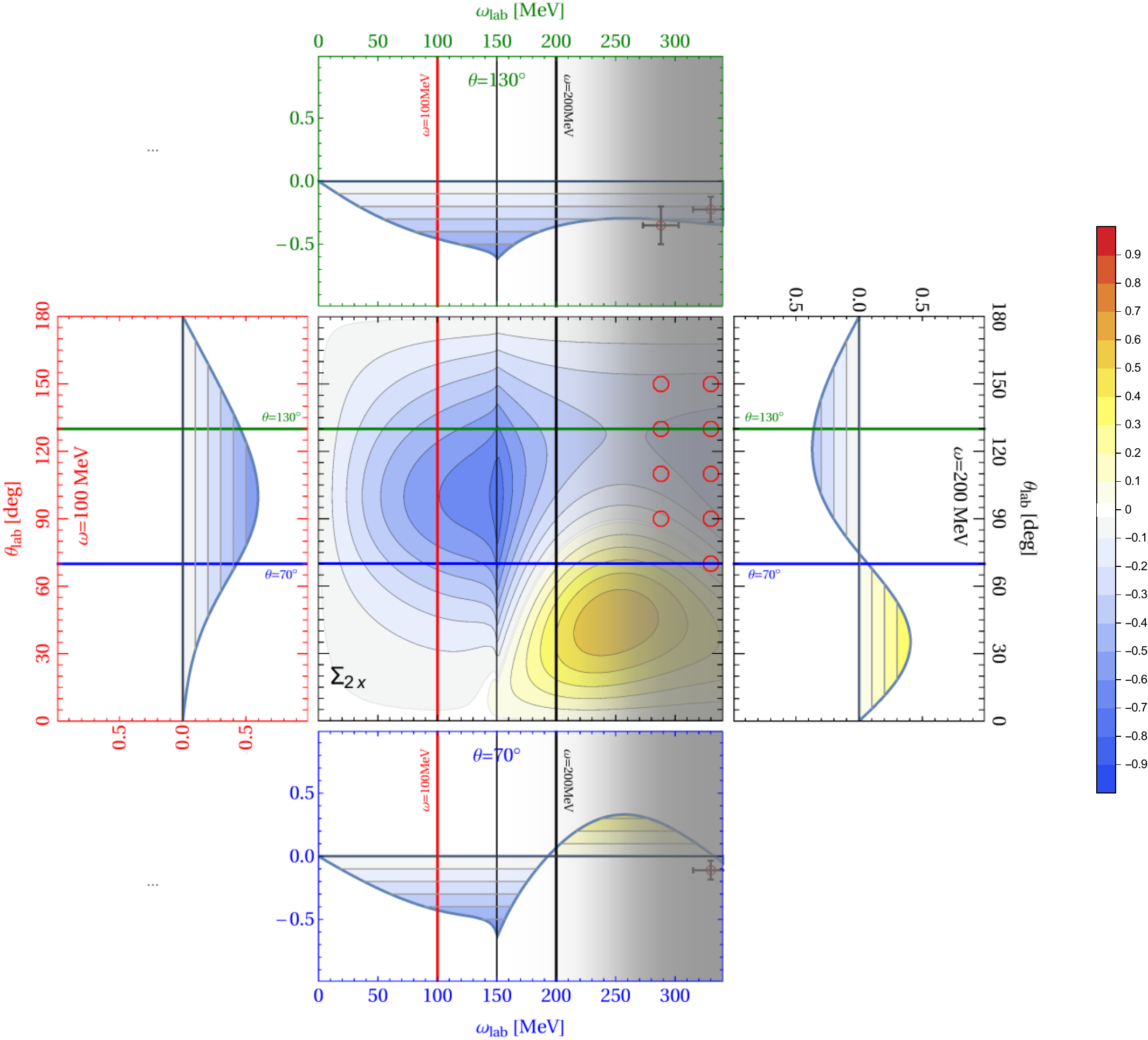}
     \caption{(Colour online) Illustration of a contour plot: the asymmetry
       $\Sigma_{2x}$, with transects at fixed $\omegalab=100\;\MeV$ (left,
       red), $\omegalab=200\;\MeV$ (right, black), $\thetalab= 70^\circ$
       (bottom, blue) and $\thetalab=110^\circ$ (top, green), and MAMI
       data~\cite{Martel:2012,Martel:2014pba} added as open (red) circles. In
       the central plot, the size of the circles reflects neither data
       uncertainties, nor the size of the energy or angle bins.  In the
       transects, statistical and systematic uncertainties, added in
       quadrature, are indicated, as is the width of the energy bins. Further
       comments in the text.}
\label{fig:clover}
\end{center}
\end{figure}

In fig.~\ref{fig:clover}, we use the asymmetry $\Sigma_{2x}$ as an example of
how a contour plot translates into transects along lines of constant photon
energy or scattering angle. In it, a transect at $\omegalab=100\;\MeV$ is red
and to the left, one at $\omegalab=200\;\MeV$ black and to the right. The
transect at $\thetalab=70^\circ$ is at the bottom of the contour plot and
colour-coded blue, and the one at $\thetalab=110^\circ$ is at the top and in
green. They are chosen to be two of the angles at which MAMI has published
data~\cite{Martel:2012,Martel:2014pba}; the width of the energy bin is
indicated as well. The coloured bands shown in these one-dimensional transects 
reproduce the heat scale used in the two-dimensional contour plot. 

In the constant-angle transects and contour plots, the one-pion production
threshold at $\omegalab^\pi=149.95\;\MeV$ is marked: the cusp there can also
usually be discerned. For observables which are zero below this first
inelasticity, the contour plot is shaded grey for $\omegalab < \omegalab^\pi$.
Plots also indicate the energies and angles where data is available (without
experimental uncertainties), so that regions which are already explored
experimentally can be quickly identified. We have, however, not attempted to
judge data quality. As it happens, the transects in fig.~\ref{fig:clover}
reveal that the agreement between the \ChiEFT prediction and the published
MAMI data~\cite{Martel:2012, Martel:2014pba} is quite good.

Finally, the figures also indicate the reliability of our predictions. As
discussed in sect.~\ref{sec:formalism-chiEFT}, the \ChiEFT expansion in
momenta can systematically be improved but always becomes gradually less
accurate with increasing photon energy.  A more extensive discussion of these
features and a less-handwaving, statistical interpretation using Bayesian
degrees of belief can be found in refs.~\cite{Griesshammer:2015ahu}
and~\cite{McGovern:2012ew}. Here, we pragmatically indicate the fact
demonstrated in sect.~\ref{sec:multipoles} that predictions from the two \ChiEFT
variants and from dispersion relations agree less well at higher energies by
putting a grey mist over the colours in the contour plots. The grey mist
begins to roll in at $\omegalab\gtrsim200\;\MeV$, and makes things opaque
above about $300\;\MeV$. In fig.~\ref{fig:clover}, this indicates the MAMI
data on $\Sigma_{2x}$ are in a region where polarisabilities cannot be
extracted with high confidence.

There could be an exception to this trend of lower accuracy at higher energy.
Since we tuned the parameters to reproduce the Delta peak, incorporated its
width and made sure Watson's theorem is approximately satisfied, we surmise
that observables which are Delta-dominated are somewhat more reliable than the
discussion in sect.~\ref{sec:multipoles} would suggest. The sensitivity of
Delta-dominated observables to polarisability variations may thus perhaps be
reliably predicted. However, there could be a similar degree of sensitivity to
omitted physics: that would render polarisability extractions problematic. We
also note that plots of most observables reveal quite a simple angular
dependence at high energies. Given the limited accuracy of data at higher
energies, different fits which each have just a few parameters but use
different theoretical descriptions of the amplitudes may compare equally well
with data there, but correspond to quite different static polarisability
values.

\subsection{Cross Section}
\label{sec:crosssection}

\begin{figure}[!htbp]
\begin{center}
     \includegraphics[width=0.6\textwidth]{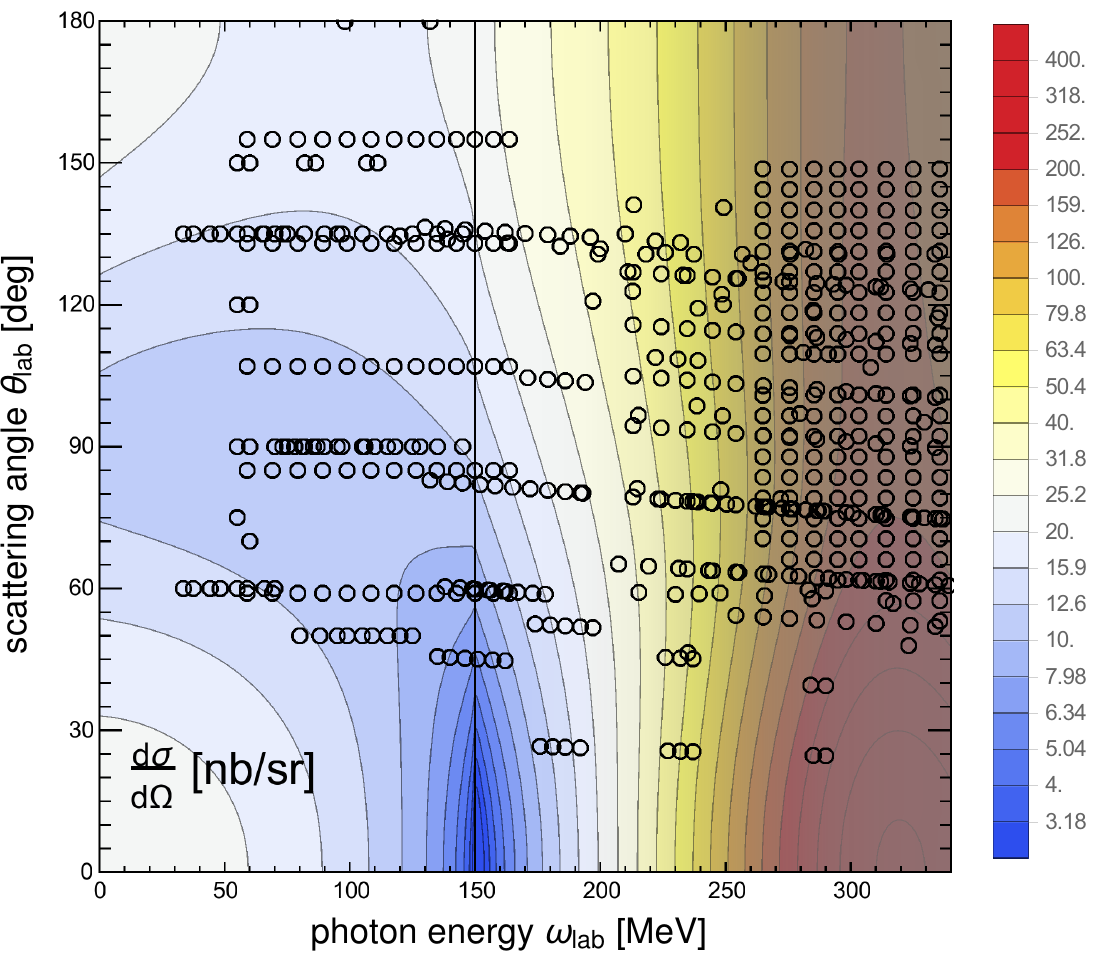}
     \caption{(Colour online) Contour plot of the unpolarised differential
       cross section as a function of $\omegalab$ and $\thetalab$, on a
       logarithmic scale, and with additional contours for very small and very
       large values. The colour coding is unique to this plot. The location of
       extant data is also indicated.}
\label{fig:crosssection}
\end{center}
\end{figure}

Figure~\ref{fig:crosssection} is a contour plot of the differential cross
section for Compton scattering on a proton, with contours on a logarithmic
scale. As is well-known, below $140\;\MeV$ the cross section is rather flat, staying between
 $10$ and $20$ $\mathrm{nb/sr}$ except at forward angles. The
cusp at the pion-production threshold is prominent for forward scattering,
where the cross section decreases to less than $2\;\mathrm{nb/sr}$ at the threshold, but is
hard to see for back angles. At higher energies, the broad width of the
$\Delta(1232)$ resonance leads to a rapid increase by about two orders of
magnitude. The maximum around $320\;\MeV$ is most pronounced at forward
angles, where the cross section exceeds $400\;\mathrm{nb/sr}$, but it still
reaches $170\;\mathrm{nb/sr}$ at backward angles.

In refs.~\cite{Griesshammer:2012we, McGovern:2012ew}, the world proton Compton
database is listed and discussed extensively. Below 170~MeV, a few very old
data sets as well as a small number of individual points must be discarded to
obtain a statistically-consistent database. Above about $150\;\MeV$, data sets
of purported high-precision experiments from different labs appear
incompatible, and one is forced to choose the more copious set over the others
in order to obtain a database for which the standard likelihood is a
meaningful objective function. The result is that between
$\omegalab=164$ MeV and $250\;\MeV$, the remaining data is
very sparse and confined to only a couple of angles. This data gap is not,
however, immediately apparent in plots such as fig.~\ref{fig:crosssection}
because there the data are shown as black circles, without discriminating
between experiments or indicating their quality. One can hence not discern all
regions of poor kinematic coverage, but only those where data is completely
absent. More details on the cross-section database, as well as comparisons
with our \ChiEFT results, can be found in ref.~\cite{McGovern:2012ew}.

Lastly, we note that data has recently been taken on the differential cross
section at $\omegalab\approx 85$ MeV at three different scattering angles at
\HIGS and is presently being analysed~\cite{Ahmed:2017}.

\subsection{Magnitudes of Asymmetries and Polarisation-Transfer Observables}
\label{sec:asymmetries}

\begin{figure}[!htbp]\thisfloatpagestyle{empty}
\begin{center}
 \includegraphics[width=\textwidth]{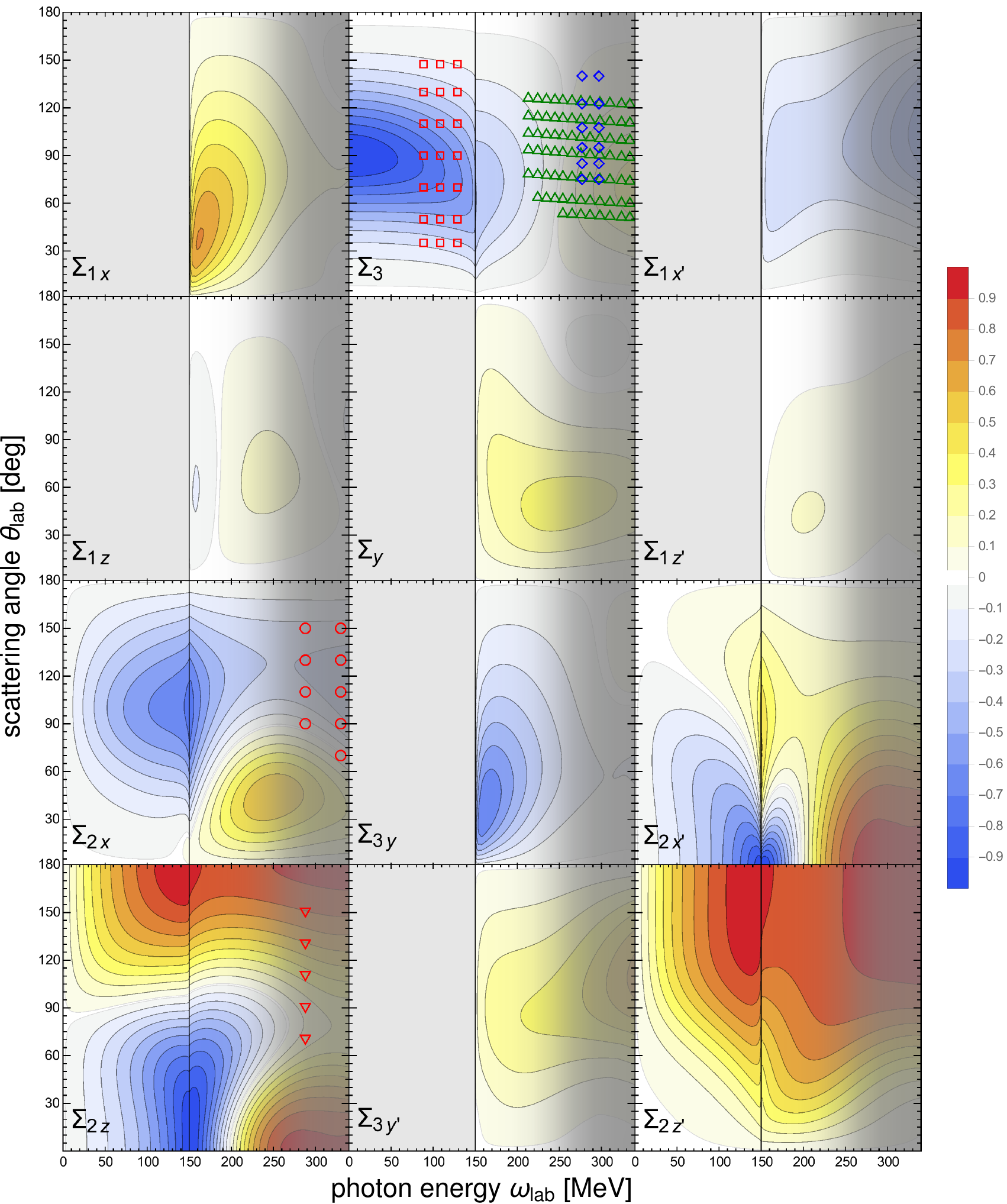}
     \caption{(Colour online) Contour plots of the asymmetries and
       polarisation-transfer observables; see text and
       sect.~\ref{sec:noteonplots} for details.  Data included as available,
       for $\Sigma_3$: open (green) triangles
       $\protect\textcolor{green}{\bm\bigtriangleup}$ from
       LEGS~\cite{Blanpied:2001ae}, open (red) squares
       $\protect\textcolor{red}{\bm\square}$ from
       MAMI~\cite{Sokhoyan:2016yrc}, open (blue) diamonds
       \protect\rotatebox{45}{$\protect\textcolor{blue}{\bm\square}$}
       preliminary from MAMI~\cite{Collicott:2015, Martel:2017pln}; for
       $\Sigma_{2x}$: open (red) circles $\protect\textcolor{red}{\bm\circ}$
       MAMI data from~\cite{Martel:2012,Martel:2014pba}; open (red) inverted triangle
       $\protect\textcolor{red}{\bm\bigtriangledown}$ preliminary from MAMI~\cite{Martel:2017pln}.   Symbol sizes do not reflect error bars, nor
       the size of energy or angle bins.}
\label{fig:asymmetries}
\end{center}
\end{figure}

In fig.~\ref{fig:asymmetries}, all $12$ asymmetries and polarisation-transfer
observables are shown, with symbols at the energies and angles of available
data: for $\Sigma_3$ from LEGS~\cite{Blanpied:2001ae},
MAMI~\cite{Sokhoyan:2016yrc} and MAMI (preliminary)~\cite{Collicott:2015,
  Martel:2017pln}; and from MAMI for $\Sigma_{2x}$~\cite{Martel:2012,
  Martel:2014pba} and $\Sigma_{2z}$ (preliminary)~\cite{Martel:2017pln}. As
in the case of the cross section, the symbols do not indicate data quality or statistical
consistency. We also mention that recent \HIGS data on
$\Sigma_3$ at $85\;\MeV$ is being analysed~\cite{Ahmed:2017}.

\begin{figure}[!htbp]
\begin{center}
     \includegraphics[width=\textwidth]{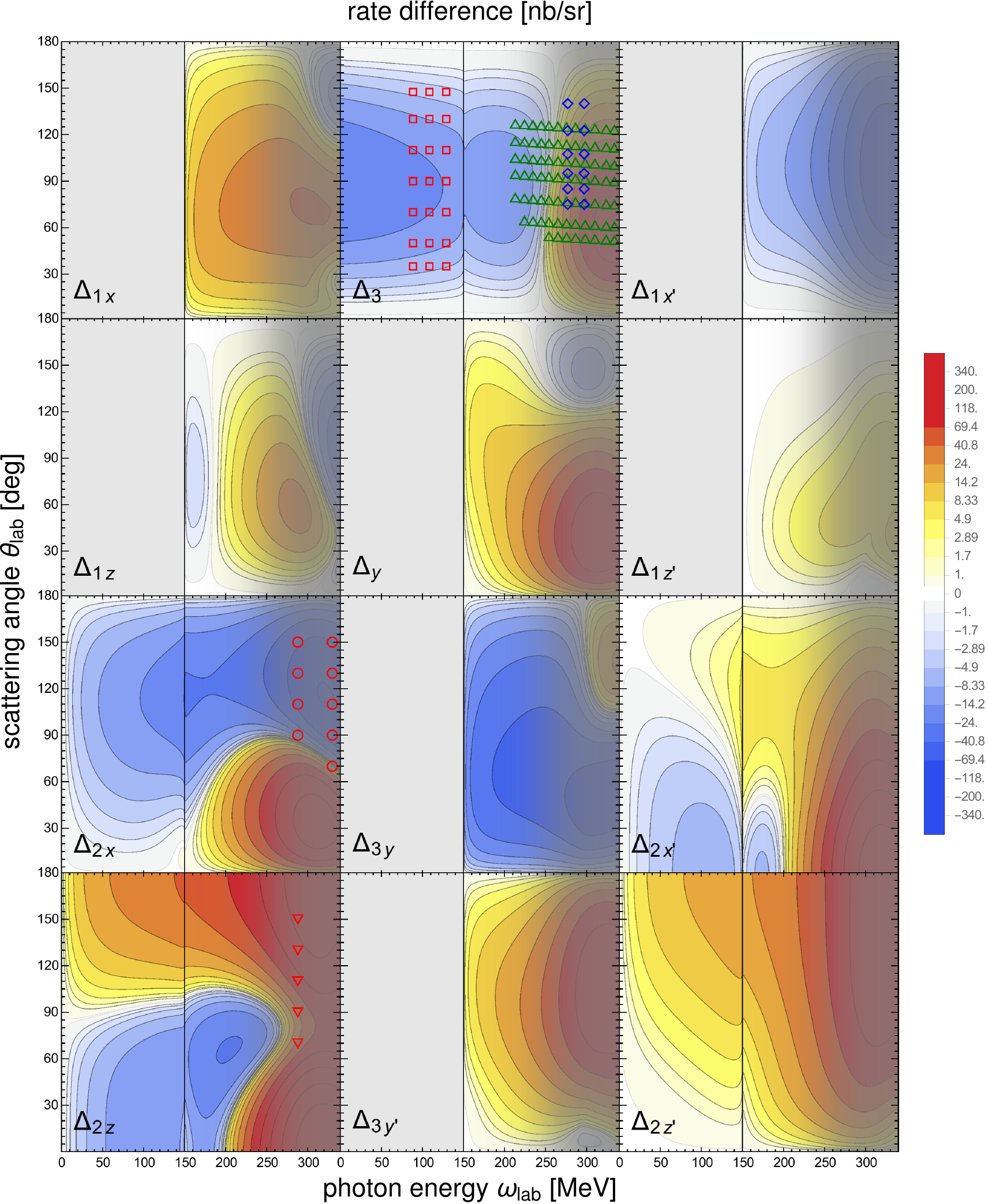}
     \caption{(Colour online) Contour plots of the rates associated with
       asymmetries and polarisation-transfer observables, with a unique colour
       coding on a logarithmic scale and additional contours for very small
       and very large values; see text for details. Data symbols as in
       fig.~\ref{fig:asymmetries}; their sizes do not reflect errors bars, nor
       the size of energy or angle bins.}
\label{fig:asymmetries-rates}
\end{center}
\end{figure}

It is worth pointing out that most of the observables are guaranteed to vanish
at $\theta=0^\circ$ and $180^\circ$; the exceptions are $\Sigma_{2z}$,
$\Sigma_{2x'}$ and $\Sigma_{2z'}$, with the last two vanishing at
$\theta=180^\circ$ and $0^\circ$, respectively. When these observables do not
vanish at $\theta=0^\circ$ or $180^\circ$, their contours meet the top and
bottom of the frame at right angles.  $\Sigma_{2z}$, $\Sigma_{2x'}$ and
$\Sigma_{2z'}$ also tend to have the largest magnitudes, reaching $0.7$ in
regions that are experimentally accessible.

The only asymmetry which does not vanish as the photon energy tends to zero is
the beam asymmetry $\Sigma_{3}$. Its exact shape at zero energy is dictated by
the Thomson term; this scattering on a point target without spin effects leads
to a $(1-\cos^2\theta)/(1+\cos^2\theta)$ shape which is well-known from
classical electrodynamics~\cite{jackson1975classical}. This behaviour
dominates at low energies, although its importance has decreased dramatically
by the pion-production threshold, with $\Sigma_3$ even changing sign above
about $240\;\MeV$. It will turn out that $\Sigma_3$ has limited sensitivity to
any polarisability at the energies where \ChiEFT can be trusted to
converge. Additional plots which compare to data can be found in
ref.~\cite{Sokhoyan:2016yrc}.

In general, all the observables which do not vanish below pion-production
threshold are strongly driven by the Born and pion-pole amplitude at low
energies, with small contributions from polarisabilities. At higher energies,
all asymmetries are driven to a large degree by structure, namely pion-cloud,
Delta-resonance and short-distance physics. Those observables which vanish
below the pion-production threshold increase to magnitudes of at least $0.2$
for some angles around $200\;\MeV$ and thus can provide reasonable count rates
there.

To facilitate run-time estimates, fig.~\ref{fig:asymmetries-rates} provides the
differences of the rates, $\Delta_\alpha$, for different orientations
associated with each asymmetry or polarisation-transfer observable, using
colour coding on a logarithmic scale. These are the numerators in
eqs.~\eqref{eq:asym3} to~\eqref{eq:asym3y} for the asymmetries, and their
analogues for polarisation transfers, for instance
$\Delta_{3}=2\,\frac{\dd\sigma}{\dd\Omega}\,\Sigma_{3}$; see
eq.~\eqref{eq:deltamatrix} and appendix~\ref{app:matrices} for details.

\clearpage
\subsection{General Comments on Sensitivities to Polarisabilities}
\label{sec:sensitivities}

After these plots of the $13$ observables with all values of the
polarisabilities fixed, figs.~\ref{fig:crosssection-polsvar} to
\ref{fig:polsvar-2Zp} display the sensitivity of each observable $\calO$ to
varying individual polarisabilities, measured by the derivative
\begin{equation}
  \frac{\dd\calO}{\dd\zeta}
\end{equation}
with respect to one of the $6$ polarisabilities, generically denoted
$\zeta$. This corresponds to the effect of a change of static polarisability
values (or, equivalently, of shifting a dynamical polarisability in
fig.~\ref{fig:multipoles} up or down by a constant amount). In all such plots,
we use the same heat scale as described in sect.~\ref{sec:noteonplots}. This allows one
to quickly identify large sensitivities. Where necessary, more contours are
added to display sensitivities beyond the extremes of the colour scale.
  
In what follows, our focus is on the region where three essential conditions
are met: there are significant sensitivities to spin polarisabilities (usually
$\omegalab\gtrsim100\;\MeV$); theoretical frameworks can extract
polarisabilities reliably and with high accuracy
($\omegalab\lesssim250\;\MeV$); and experiments are not overwhelmed by
backgrounds ($30^\circ\lesssim\theta\lesssim160^\circ$).

In a brave new world of high-accuracy experiments with well-controlled
systematic experimental uncertainties, high luminosities and 100\% beam and
target polarisations, an ideal observable should be very sensitive to one
polarisability or a simple combination, while being rather insensitive to all
others. Unfortunately, these plots show that one-nucleon Compton scattering
does not admit such a simple picture. Spin polarisabilities will need to be
extracted from a global analysis of a carefully selected set of
observables. Redundancies can be built into this process, so as to ameliorate
experimental and extraction uncertainties.

We therefore also explore sensitivities to particular linear combinations of
polarisabilities. So long as data is incomplete, it is likely that any fit of
polarisabilities to polarised cross sections and asymmetries will need to take
advantage of the two famous sum rules for $\alphae+\betam$ and $\gamma_0$
which are based on total photoabsorption cross sections.  The Baldin sum rule
constrains $\alphae+\betam$ and has been evaluated most recently to give
$14.0\pm0.2$~\cite{Gryniuk:2015eza}. The sum rule for the spin-dependent
amplitude constrains the combination $\gammazero$ of spin polarisabilities
encountered under forward angles, and the most recent evaluation gives
$0.93\pm0.1$ \cite{Gryniuk:2016gnm}. Both numbers are in good agreement with,
but more precise than, earlier
evaluations~\cite{OlmosdeLeon:2001zn,Pasquini:2010zr,Martel:2014pba}.

Our plots confirm that the sensitivity to $\alphae+\betam$ dominates forward
scattering, and that $\alphae-\betam$ can best be measured at back-angles.
The difference, unlike the sum which is known to $1.5\%$, carries a combined
theory and statistical uncertainty of greater than 10\%. Errors of the spin
polarisability combinations are greater than 20\%; see
ref.~\cite{Griesshammer:2015ahu}. Overall, the sensitivities to scalar
polarisabilities extend to lower energies than for the spin ones. In turn, an
extraction of spin polarisabilities must also address uncertainties induced by
the errors in the scalar ones. In particular at low energies, where the Born
amplitudes are large and the scalar polarisabilities are the dominant
deviation from Born, the sensitivity to $\alphae-\betam$ cannot be ignored
when discussing the potential for spin-polarisability extractions.  New
experimental information at $\omegalab\lesssim140\;\MeV$ can thus play a
valuable---if indirect---role in reducing uncertainties in spin
polarisabilities that are induced by the present error bars on $\alphae$ and
$\betam$.

Likewise, the two combinations
\begin{equation}
  \gammazero:=-\gammaee-\gammamm-\gammaem-\gammame\;\;,\;\;
  \gammapi:=-\gammaee+\gammamm-\gammaem+\gammame
\end{equation}
of the spin polarisabilities are  best measured at forward and
backward angles, respectively\footnote{$\gammapi$ is often quoted including
  the large contribution of about $46\times 10^{-4}$~fm$^4$ from the exchange
  of a neutral pion between the photon and the nucleon.  However this is
  normally excluded from the definition of ``structure" effects, and by
  convention is not included in the individual spin polarisabilities.}.
Finally, we include the two combinations
\begin{equation}
\gammaeminus:=\gammaee-\gammaem\;\;\mbox{ and } \;\;\gammamminus:=\gammamm-\gammame\;\;,
\end{equation}
which complement $\gammazero$ and $\gammapi$ to form an orthogonal basis for
the space of spin polarisabilities: an alternative to the multipole basis
$\gamma_i$.  We will see below, \eg,~in
fig.~\ref{fig:crosssection-polsvar}, that some observables display strong
sensitivities to many or all of the polarisabilities $\{\alphae$, $\betam$,
$\gammaee$, $\gammamm$, $\gammaem$, $\gammame\}$, but that looking at the set
$\{\alphae+\betam$, $\alphae-\betam$, $\gammazero$, $\gammapi$,
$\gammaeminus$, $\gammamminus\}$ reveals that a smaller number of linear
combinations actually accounts for the variability.

\subsection{Sensitivity of the Cross Section to Polarisabilities}
\label{sec:crosssectionvar}

\begin{figure}[!htbp]
\begin{center}
     \includegraphics[width=\textwidth]{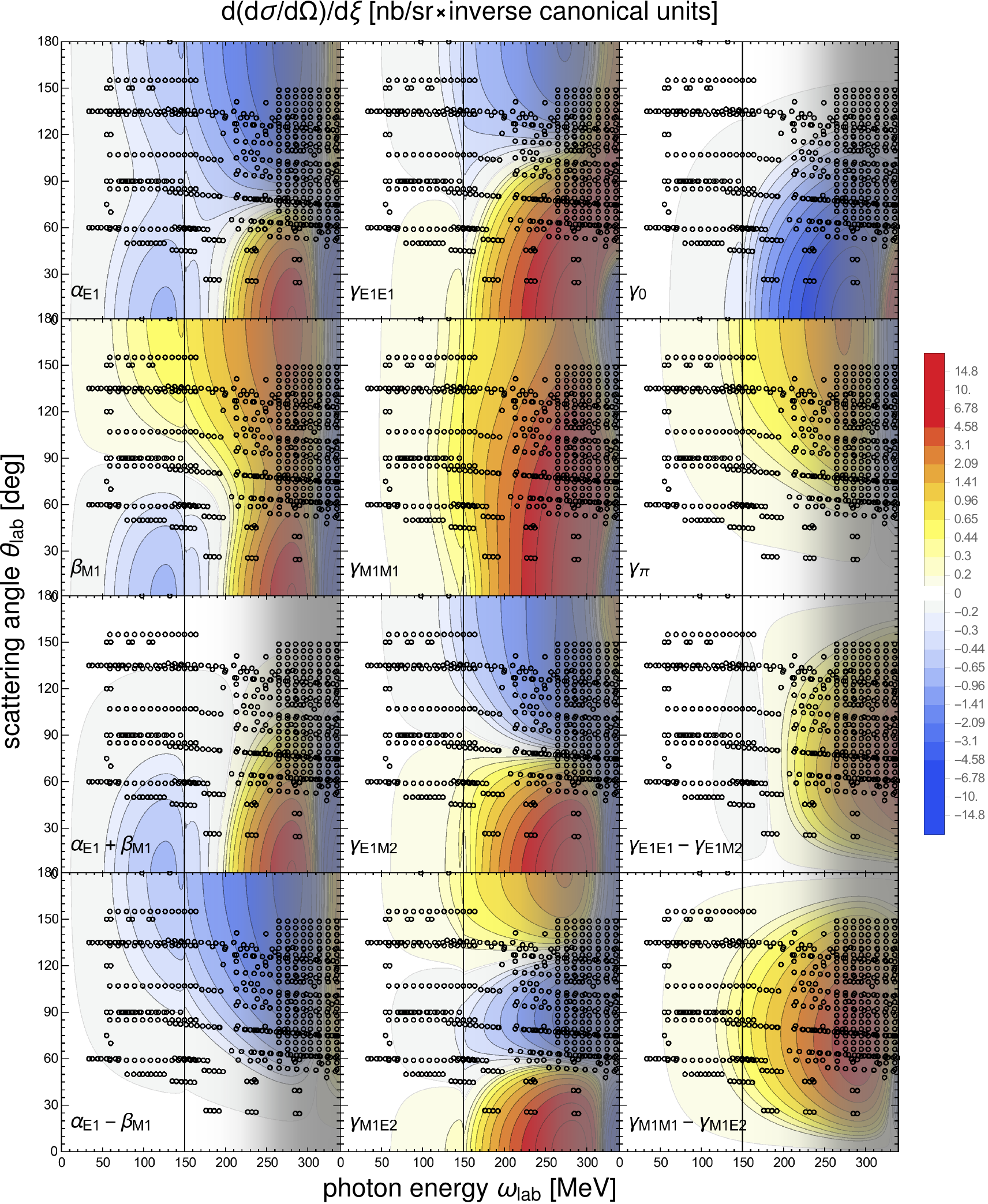}
     \caption{(Colour online) Sensitivity of the cross section to varying the
       polarisabilities, with a colour coding unique to this plot; see text
       for details.}
\label{fig:crosssection-polsvar}
\end{center}
\end{figure}

In fig.~\ref{fig:crosssection-polsvar}, we present the derivative
\begin{equation}
  \frac{\dd}{\dd\zeta}\left(\frac{\dd\sigma}{\dd\Omega}\right)
\end{equation}
in units of $\mathrm{nb\;sr}^{-1}\times 10^4\;\fm^{-3}$ for the scalar
polarisabilities, and $\mathrm{nb\;sr}^{-1}\times 10^4\;\fm^{-4}$ for the spin
polarisabilities, on a logarithmic scale unique to this figure. As the cross
sections are small below the pion-production threshold, the resulting
sensitivities are more pronounced at higher energies.

As an example of our assertion in sect.~\ref{sec:sensitivities} that using
different bases for the polarisabilities can reduce correlations, consider
scattering around the Delta peak. As the red areas in the central column of
fig.~\ref{fig:crosssection-polsvar} indicate, there appear to be large
sensitivities to all spin polarisabilities $\gamma_i$. However, when we
instead look at the right-hand column, we see that at forward angles the
sensitivity is indeed (and not surprisingly) only to $\gammazero$. More
interestingly, even non-forward angles show little sensitivity to $\gammapi$
or $\gammaeminus$, and only some limited sensitivity to $\gammamminus$.

The plot also provides a good example of correlations between variations of
different polarisabilities, some of which are not captured by the alternative
basis. At all energies and angles, the dependencies on changing $\alphae$,
$\gammaee$ and $\gammaem$ are near-identical, and especially so at
$\omegalab\approx200\;\MeV$ where one expects the biggest signals. The three
polarisabilities are thus extremely hard to disentangle in cross-section
data. In the alternative basis, the correlation is weaker for $\alphae-\betam$
and $\gammapi$, and for the anti-correlation between $\alphae+\betam$ and
$\gammazero$. Still, such degeneracies remain. A global analysis of a data
base containing several high-accuracy measurements of a diverse set of
observables will ultimately be the best way to pin down the proton's static
dipole polarisabilities.

\begin{figure}[!htbp]
\begin{center}
     \includegraphics[width=\textwidth]{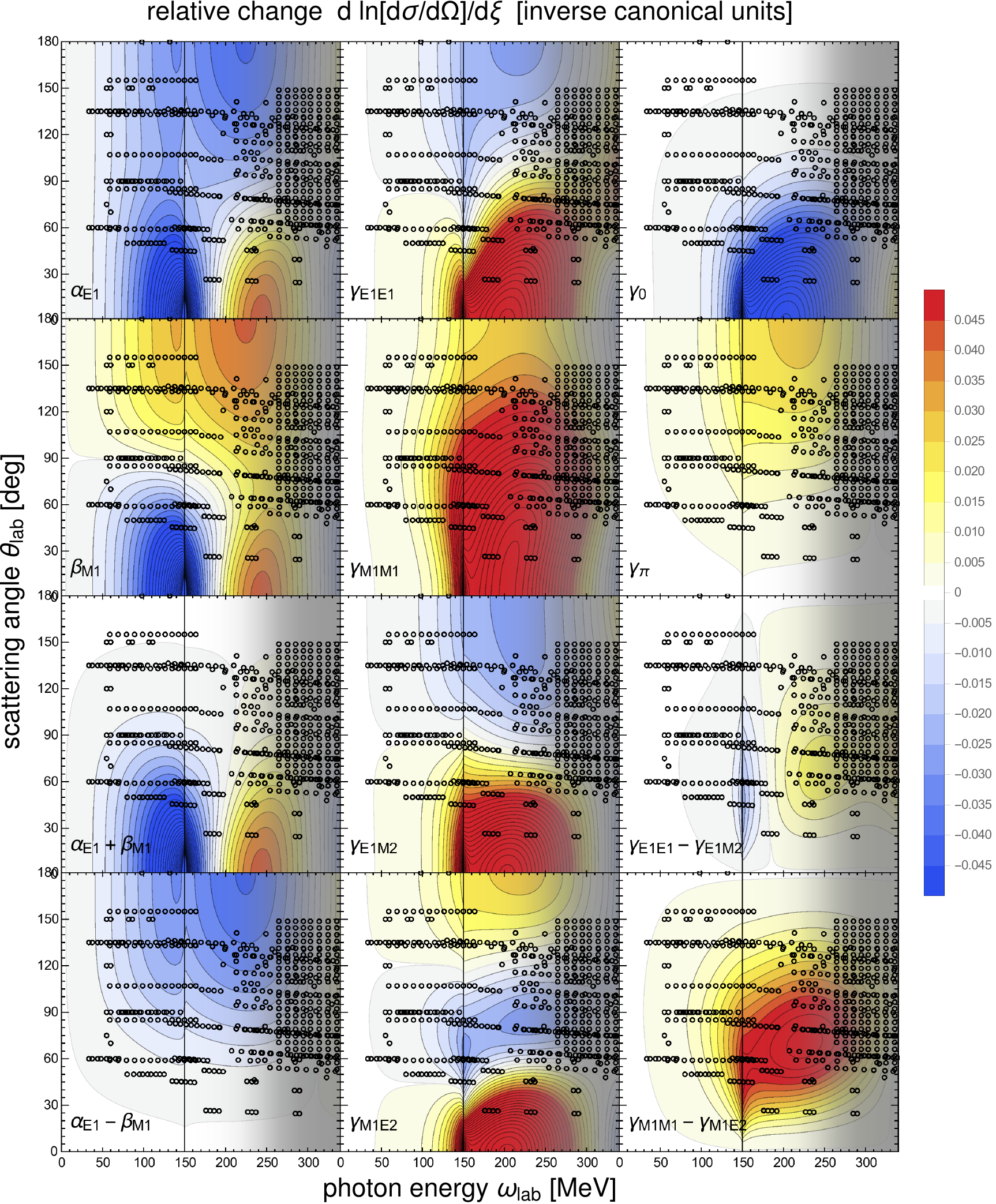}
     \caption{(Colour online) Sensitivity of the cross
       section to varying the polarisabilities, normalised to the cross
       section; see text for details.}
     \label{fig:crosssection-relative-change-polsvar}
\end{center}
\end{figure}

Since the cross section increases by two orders of magnitude, this plot
provides only an incomplete picture of the sensitivities at lower energies,
where extractions and predictions are naturally more
reliable. Figure~\ref{fig:crosssection-relative-change-polsvar} shows thus the
relative change of the cross section, \ie its logarithmic derivative
\begin{equation}
\left[\frac{\dd}{\dd\zeta}\left(\frac{\dd\sigma}{\dd\Omega}\right)\right]
\bigg/\left(\frac{\dd\sigma}{\dd\Omega}\right)\;\;,
\end{equation}
in inverse canonical units of the polarisabilities. It provides a measure
which is more akin to variations in the subsequent sensitivity studies of
asymmetries and polarisation-transfer observables, for which we use the same
colour coding and linear scale.

However, these relative changes are often very pronounced when the cross
section itself is small, \ie around the pion-production threshold, especially
at forward angles. Therefore, the combination of absolute and
relative sensitivities in figs.~\ref{fig:crosssection-polsvar}
and~\ref{fig:crosssection-relative-change-polsvar} allows those planning
experiments to balance appreciable rates with significant sensitivities, as
well as taking into account the diminished reliability of extractions at
higher photon energies, indicated in our plots by the fading colours.

Which polarisabilities can reliably be extracted from cross sections?
Sensitivities to $\alphae$ and $\betam$ occur in the same kinematic regions,
largely because the Baldin-constrained combination $\alphae+\betam$ influences
forward angles, while its counterpart $\alphae-\betam$ dominates at backward
angles. However, around $\omegalab\approx180$~MeV, there is almost no
sensitivity to $\alphae+\betam$ at any angle. Similarly, the plots show that
sensitivities to $\gammaee$, $\gammame$ and, to a lesser degree, $\gammaem$
are quite similar and not simple to disentangle. Below about 170~MeV, for
scattering angles that are neither forward not backward and hence where most
of the current low-energy data is, the greatest sensitivity is to $\gammamm$,
which enabled it to be fit in ref.~\cite{McGovern:2012ew}.  However,
significant and simpler dependence to our alternative combinations of spin
polarisabilities exists between the pion-production threshold and about
$250\;\MeV$:
\begin{itemize}
\item Extremely strong sensitivity to $\gammazero$ for
  $\thetalab\lesssim90^\circ$.
\item Strong sensitivity to $\gammamminus$ for
  $30^\circ\lesssim\thetalab\lesssim130^\circ$.
\item Some sensitivity to $\gammapi$ for $\thetalab\gtrsim90^\circ$.
\item Very little sensitivity to $\gammaeminus$ below $200\;\MeV$.
\end{itemize}
Complementary cross-section experiments, which use the well-established
sum-rule value of $\gammazero$ as input, may therefore have an opportunity to
disentangle the dipole spin polarisabilities from precise cross-section
measurements alone.  The plots also confirm that little data exists between
about $170$ and $200\;\MeV$, where an extraction would be quite reliable.  We
note, though, that measurements of cross sections usually carry larger
systematic uncertainties than those of asymmetries and polarisation-transfer observables,
where many experimental systematic uncertainties cancel. We therefore now
explore these other observables.

\clearpage

\clearpage

\subsection{Sensitivity of  Asymmetries and Polarisation-Transfer Observables to Polarisabilities}
\label{sec:asymmetriesvar}

\begin{figure}[!htbp]
\begin{center}
     \includegraphics[width=\textwidth]{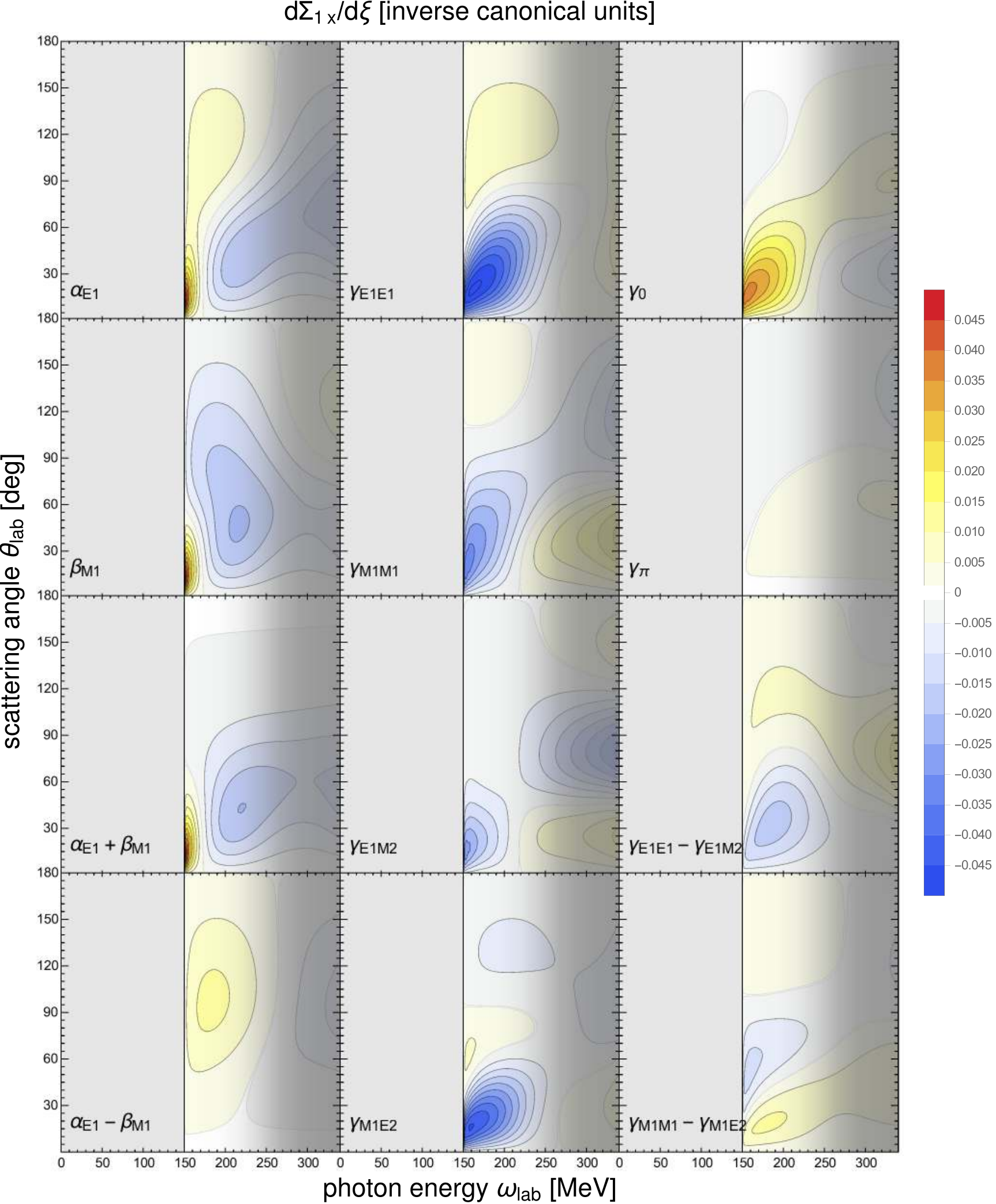}
     \caption{(Colour online) Sensitivity of the double asymmetry $\Sigma_{1x}$
       (linearly polarised photons on a proton target polarised along the
       $x$ axis) to varying the polarisabilities; see text for details.}
     \label{fig:polsvar-1X}
\end{center}
\end{figure}

\begin{figure}[!htbp]
\begin{center}
     \includegraphics[width=\textwidth]{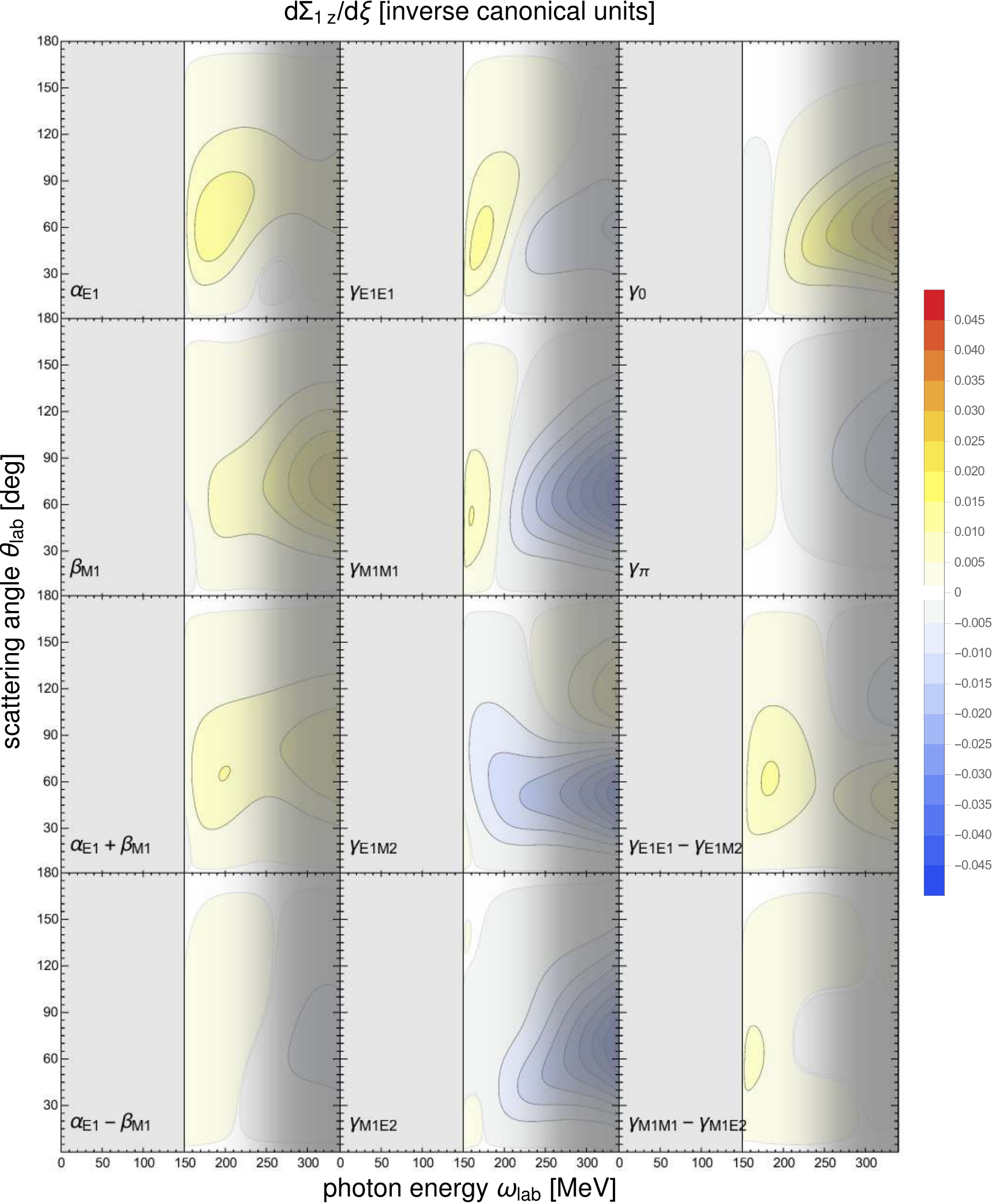}
     \caption{(Colour online) Sensitivity of the double asymmetry $\Sigma_{1z}$
       (linearly polarised photons on a proton target polarised along the
       $z$ axis) to varying the polarisabilities; see text for details.}
     \label{fig:polsvar-1Z}
\end{center}
\end{figure}

\begin{figure}[!htbp]
\begin{center}
     \includegraphics[width=\textwidth]{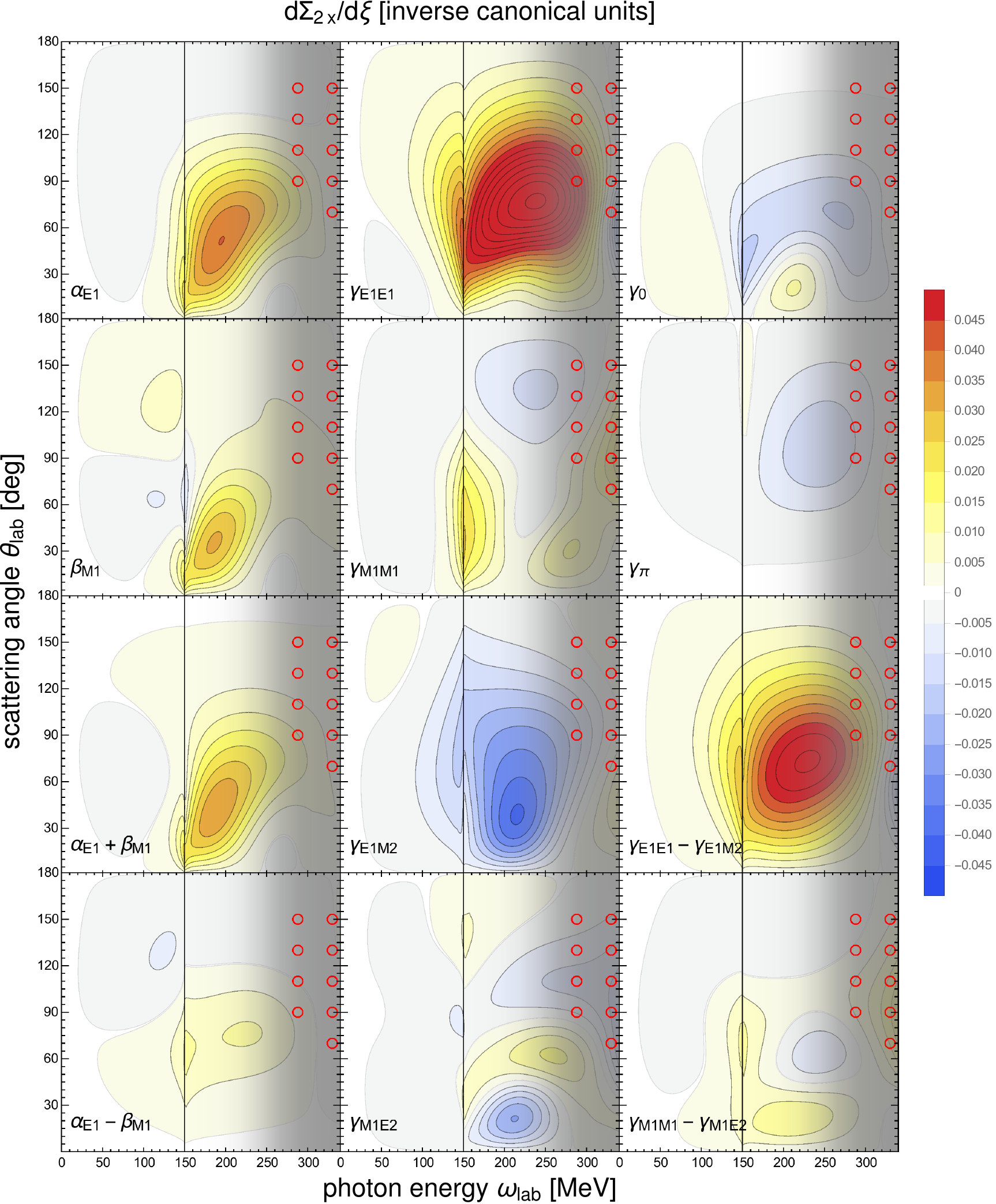}
     \caption{(Colour online) Sensitivity of the double asymmetry
       $\Sigma_{2x}$ (circularly polarised photons on a proton target
       polarised along the $x$ axis) to varying the polarisabilities; see text
       for details.  Data symbols as in fig.~\ref{fig:asymmetries}; their size
       does not reflect the error bars, nor the size of energy or
       angle bins.}
     \label{fig:polsvar-2X}
\end{center}
\end{figure}

\begin{figure}[!htbp]
\begin{center}
     \includegraphics[width=\textwidth]{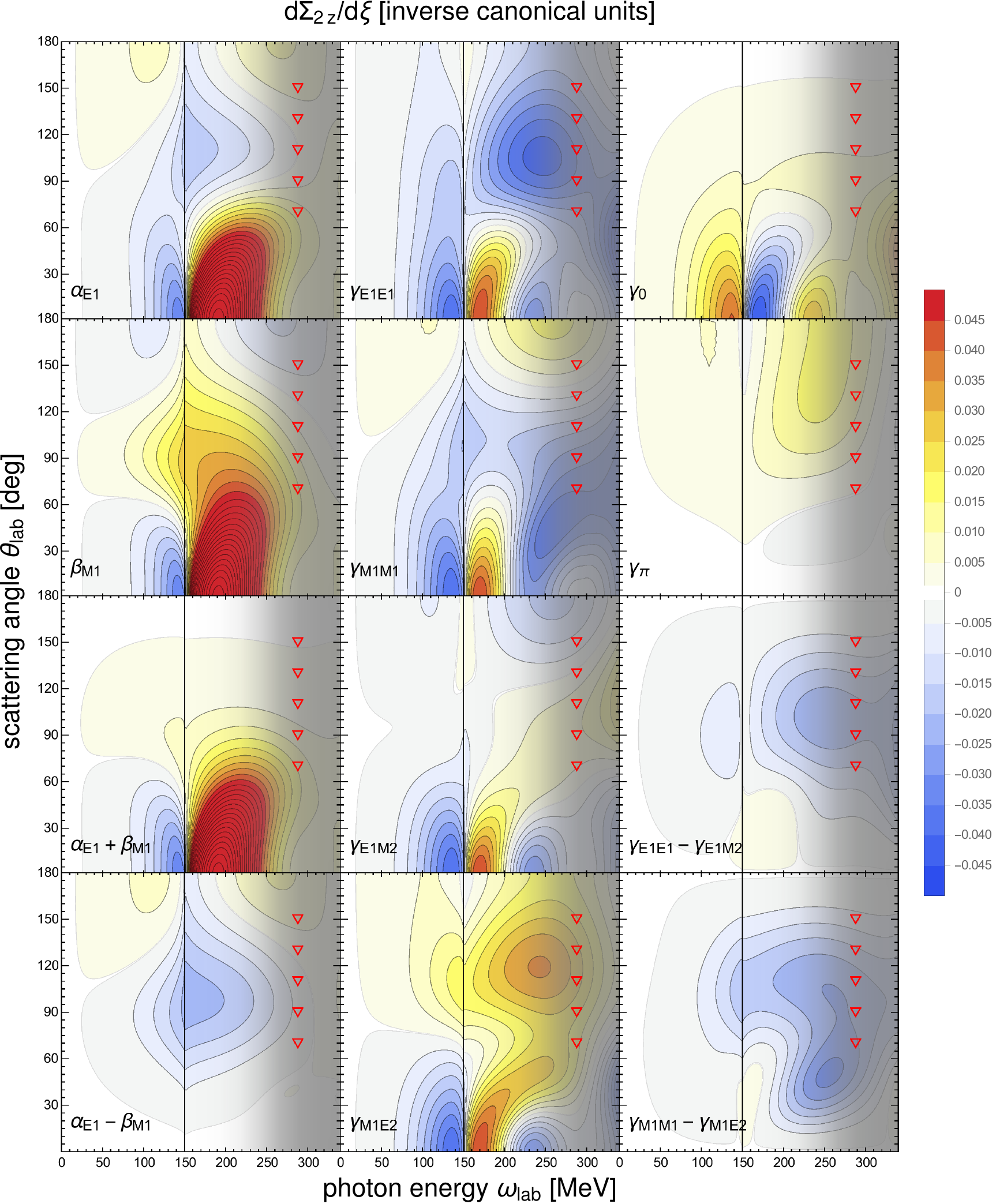}
     \caption{(Colour online) Sensitivity of the double asymmetry
       $\Sigma_{2z}$ (circularly polarised photons on a proton target
       polarised along the $z$ axis) to varying the polarisabilities; see text
       for details.  Data symbols as in fig.~\ref{fig:asymmetries}; their size
       does not reflect the error bars, nor the size of energy or
       angle bins.}
     \label{fig:polsvar-2Z}
\end{center}
\end{figure}

\begin{figure}[!htbp]
\begin{center}
     \includegraphics[width=\textwidth]{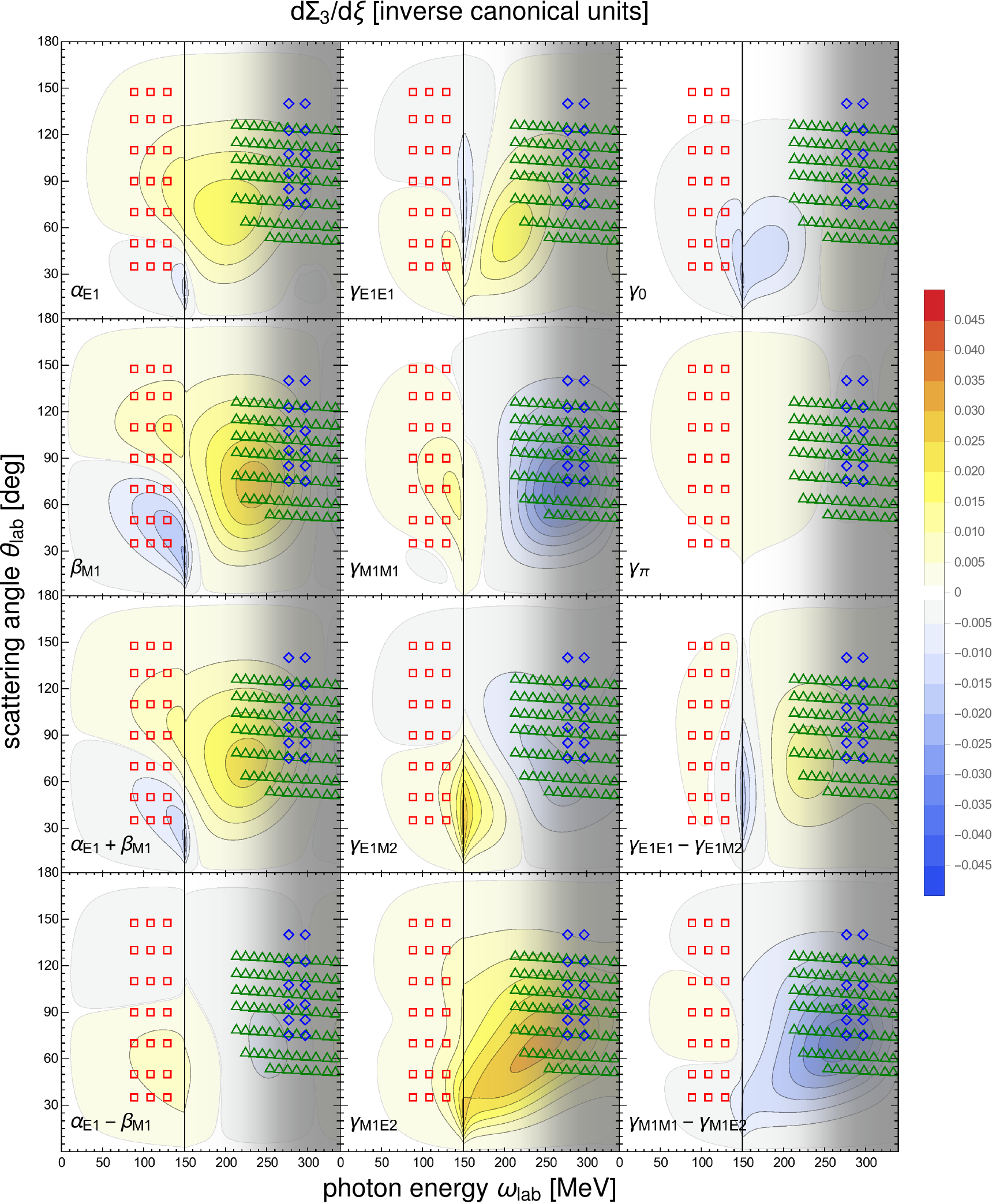}
     \caption{(Colour online) Sensitivity of the beam asymmetry $\Sigma_{3}$
       (linearly polarised photons on an unpolarised proton target) to varying
       the polarisabilities; see text for details. Data symbols as in
       fig.~\ref{fig:asymmetries}.}
     \label{fig:polsvar-3}
\end{center}
\end{figure}

\begin{figure}[!htbp]
\begin{center}
     \includegraphics[width=\textwidth]{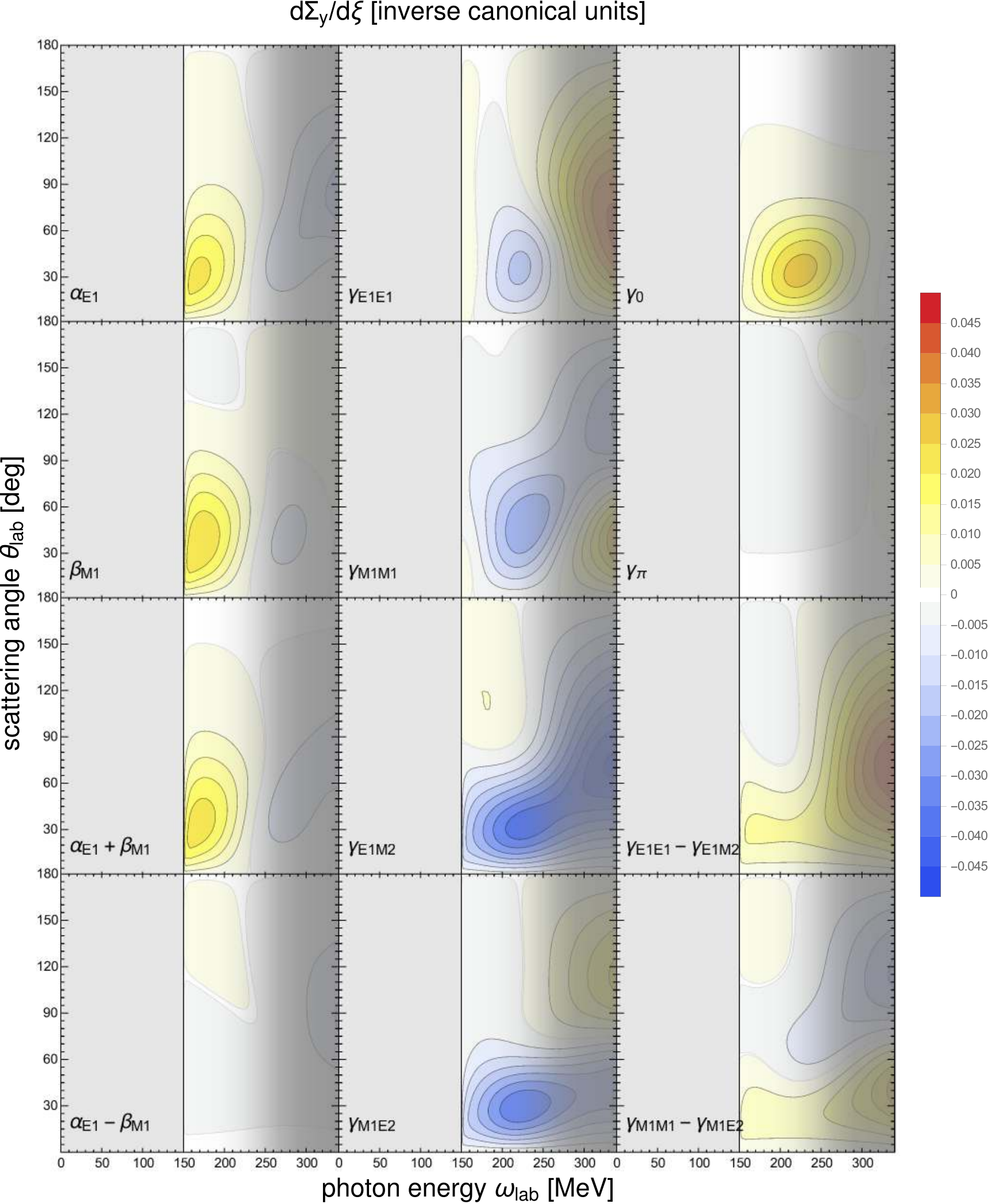}
     \caption{(Colour online) Sensitivity of the target asymmetry $\Sigma_{y}$
       (unpolarised photons on a proton target along the $y$ axis) to varying
       the polarisabilities; see text for details.}
     \label{fig:polsvar-Y}
\end{center}
\end{figure}

\begin{figure}[!htbp]
\begin{center}
     \includegraphics[width=\textwidth]{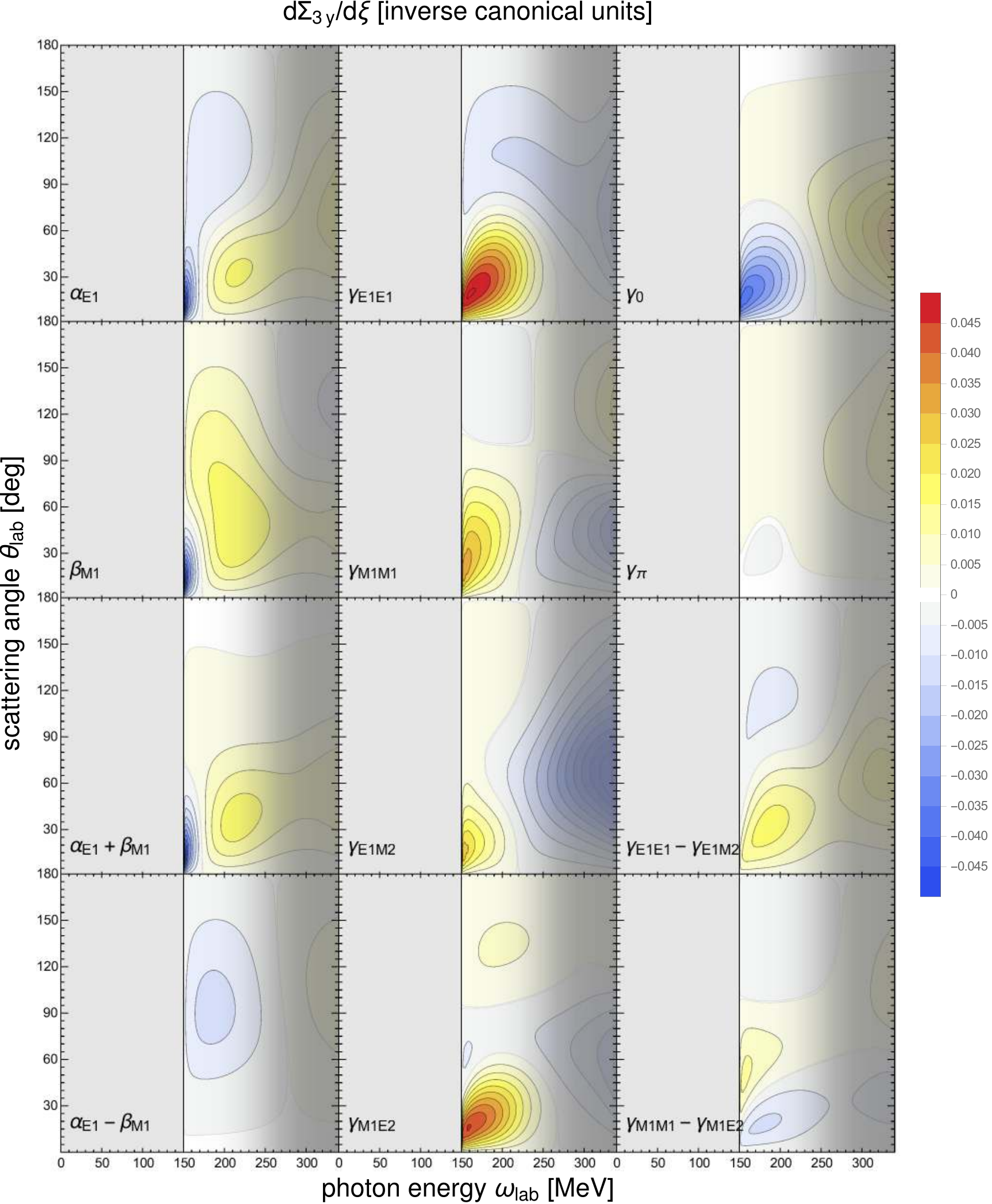}
     \caption{(Colour online) Sensitivity of the double asymmetry $\Sigma_{3y}$
       (linearly polarised photons on a proton target polarised along the
       $y$ axis) to varying the polarisabilities; see text for details.}
     \label{fig:polsvar-3Y}
\end{center}
\end{figure}

\begin{figure}[!htbp]
\begin{center}
     \includegraphics[width=\textwidth]{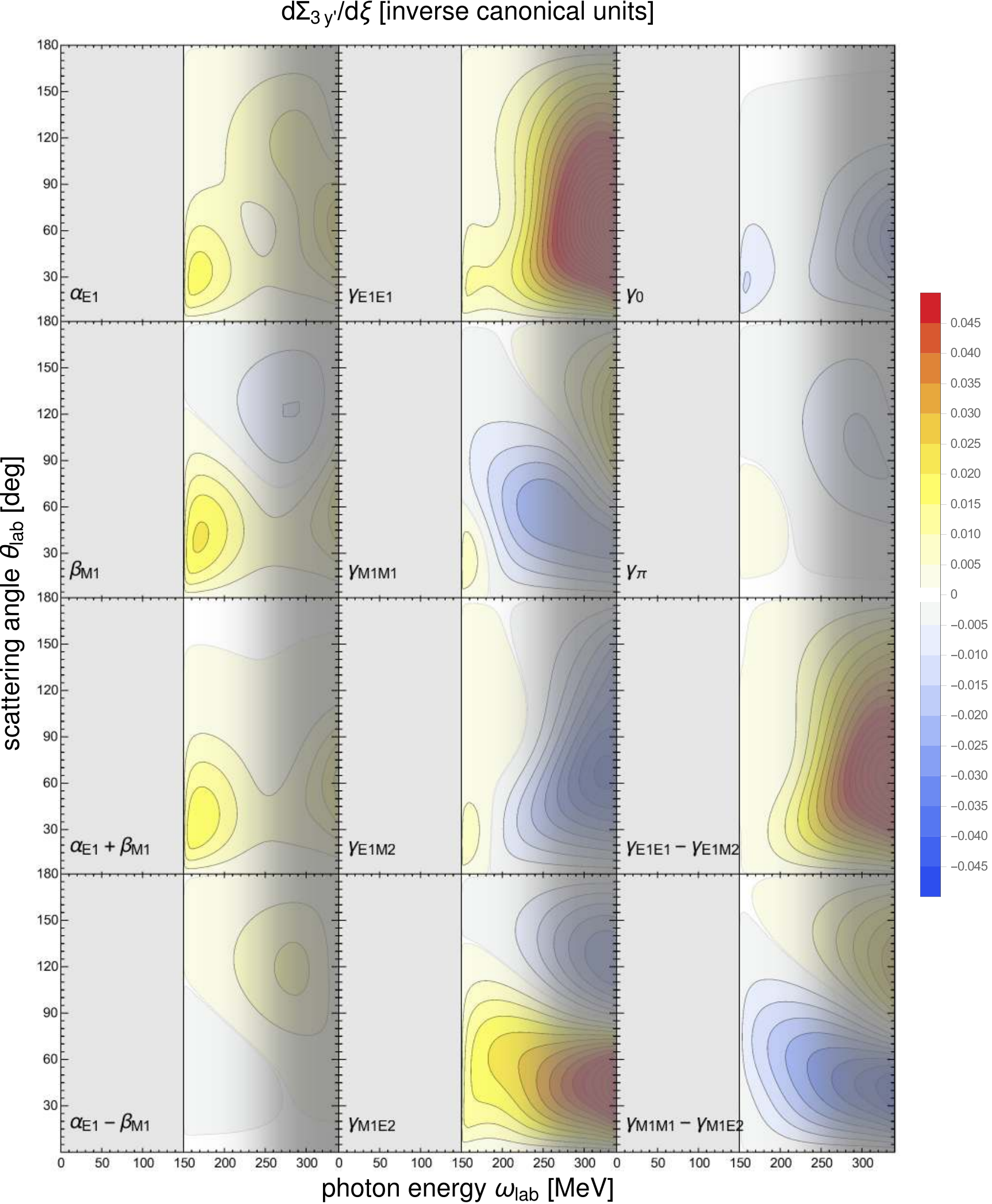}
     \caption{(Colour online) Sensitivity of the polarisation-transfer
       observable $\Sigma_{3y^\prime}$
       (linearly polarised photons on an unpolarised proton target, recoil polarised along the $y^\prime$ axis) to varying the polarisabilities; see text for details.}
     \label{fig:polsvar-3Yp}
\end{center}
\end{figure}

\begin{figure}[!htbp]
\begin{center}
     \includegraphics[width=\textwidth]{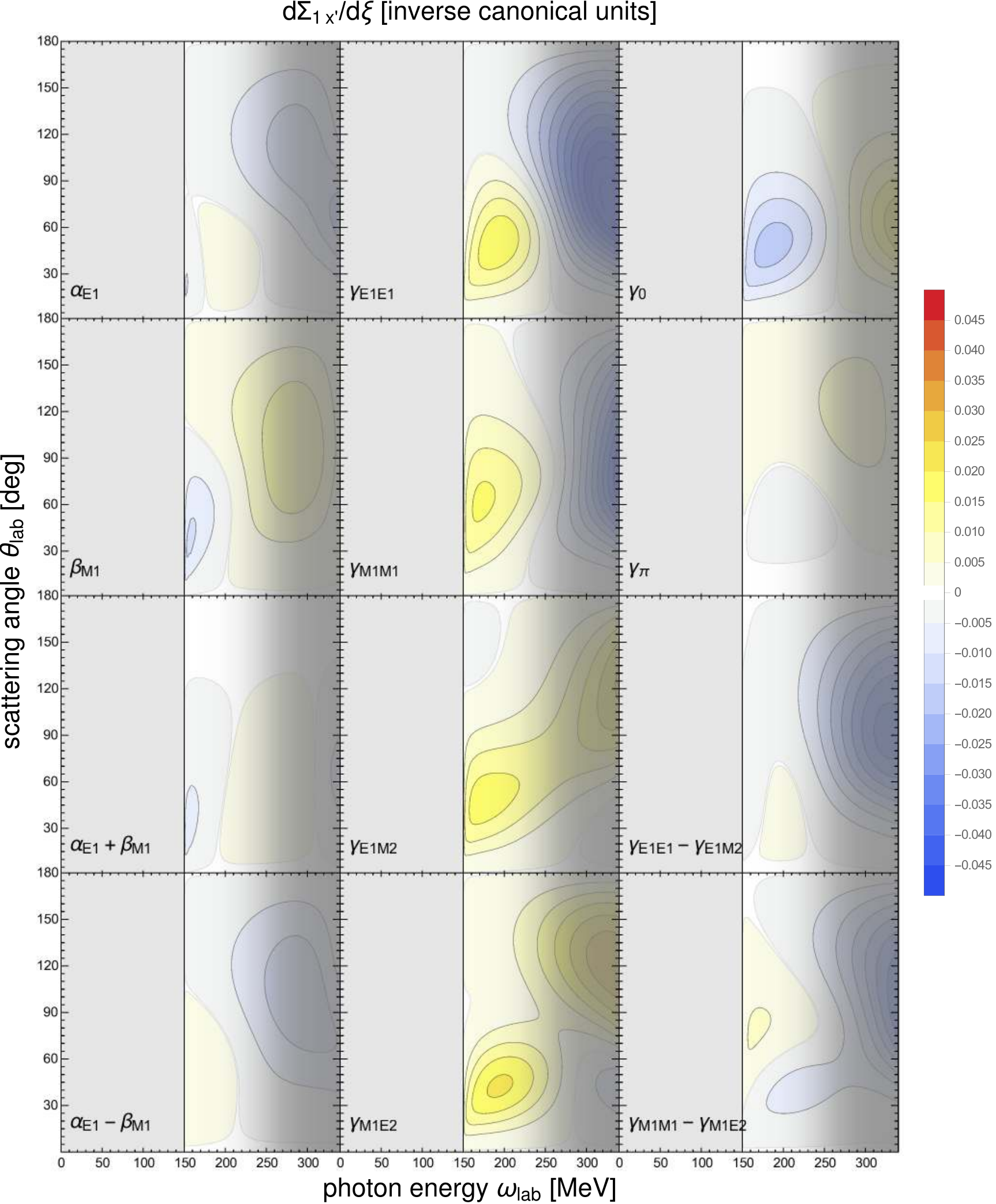}
     \caption{(Colour online) Sensitivity of the polarisation-transfer
       observable $\Sigma_{1x^\prime}$ (linearly polarised photons on an
       unpolarised proton target, recoil polarised along the $x^\prime$ axis) to
       varying the polarisabilities; see text for details.}
     \label{fig:polsvar-1Xp}
\end{center}
\end{figure}

\begin{figure}[!htbp]
\begin{center}
     \includegraphics[width=\textwidth]{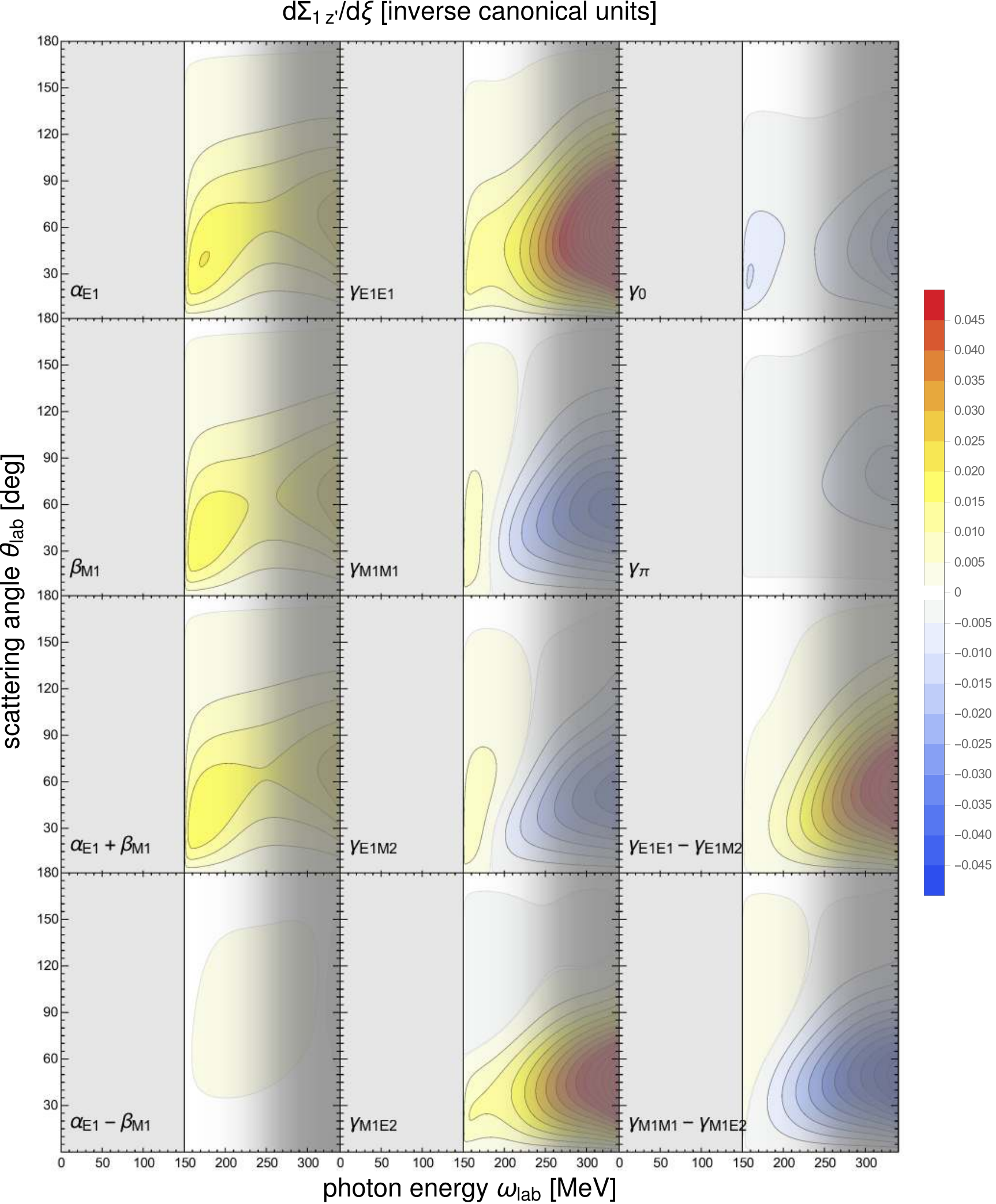}
     \caption{(Colour online) Sensitivity of the polarisation-transfer
       observable $\Sigma_{1z^\prime}$ (linearly polarised photons to an
       unpolarised proton target, recoil polarised along the $z^\prime$ axis) to
       varying the polarisabilities; see text for details.}
     \label{fig:polsvar-1Zp}
\end{center}
\end{figure}

\begin{figure}[!htbp]
\begin{center}
     \includegraphics[width=\textwidth]{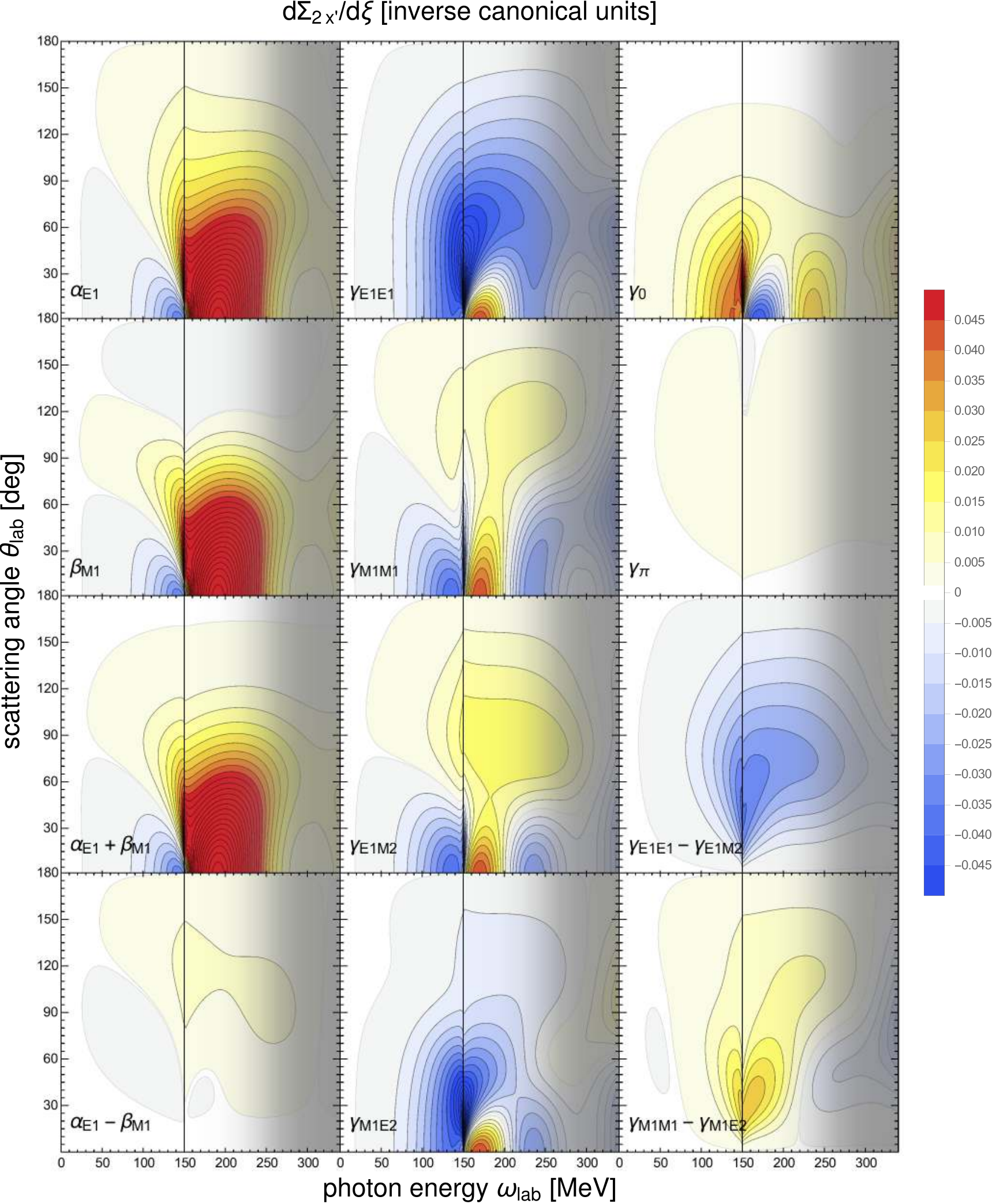}
     \caption{(Colour online) Sensitivity of the polarisation-transfer
       observable $\Sigma_{2x^\prime}$ (circularly polarised photons on an
       unpolarised proton target, recoil polarised along the $x^\prime$ axis) to
       varying the polarisabilities; see text for details.}
     \label{fig:polsvar-2Xp}
\end{center}
\end{figure}

\begin{figure}[!htbp]
\begin{center}
     \includegraphics[width=\textwidth]{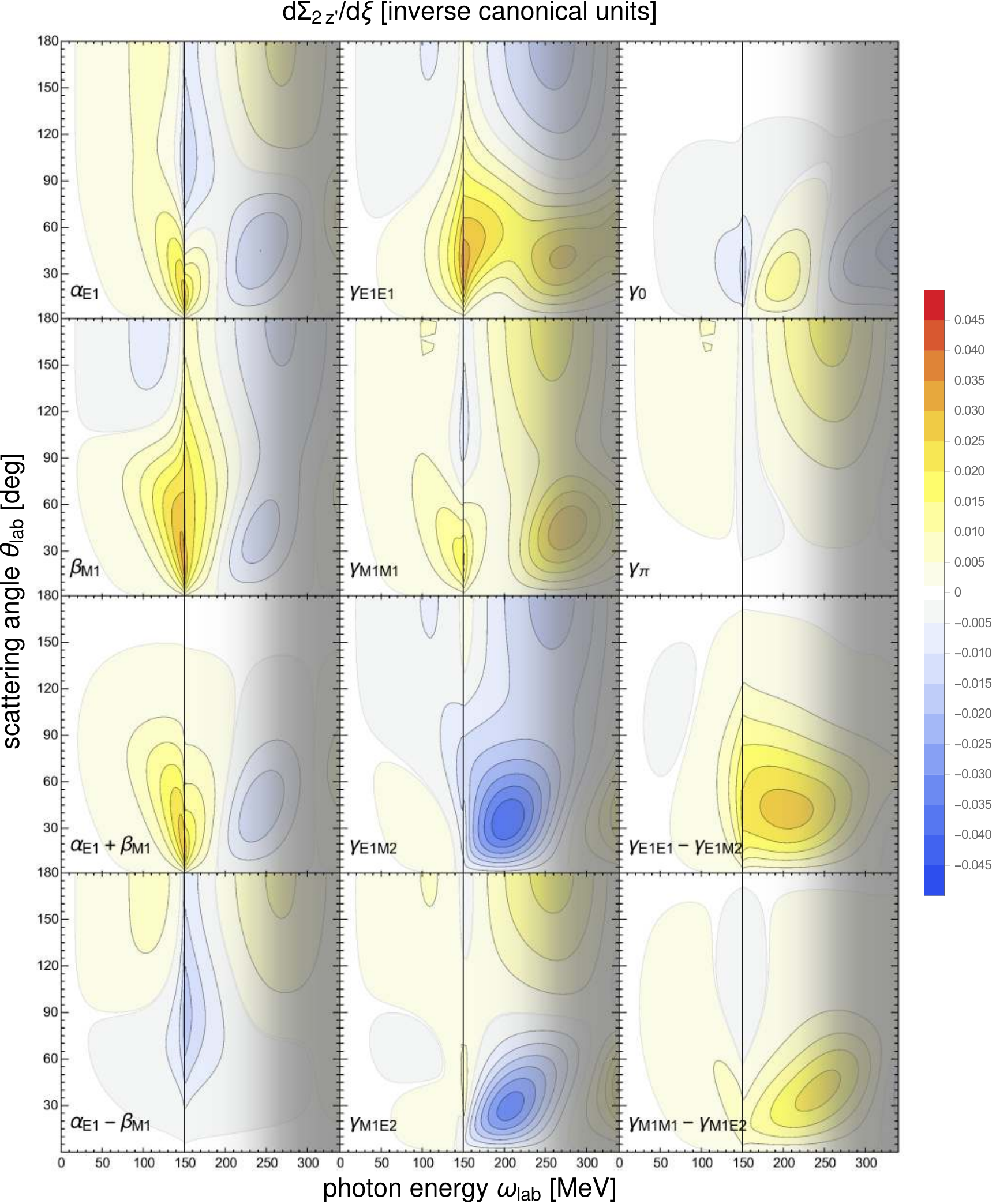}
     \caption{(Colour online) Sensitivity of the polarisation-transfer
       observable $\Sigma_{2z^\prime}$ (circularly polarised photons to an
       unpolarised proton target, recoil polarised along the $z^\prime$ axis) to
       varying the polarisabilities; see text for details.}
     \label{fig:polsvar-2Zp}
\end{center}
\end{figure}

Figures \ref{fig:polsvar-1X} to \ref{fig:polsvar-2Zp} all use the same colour
coding to indicate sensitivities to the polarisabilities as in
fig.~\ref{fig:crosssection-relative-change-polsvar}, now for the variation
\begin{equation}
   \frac{\dd\Sigma_\alpha}{\dd\zeta}
\end{equation}
of the asymmetry or polarisation-transfer observable $\Sigma_\alpha$ with
respect to one of the polarisabilities $\zeta$.  It varies between deep red
($\ge+0.045$ inverse canonical units) and deep blue ($\le-0.045$), with
additional contours when the extreme values of the colour scale 
are exceeded. All contours are separated by $0.005$
inverse canonical units.

The variations are in inverse units of $10^{-4}\;\fm^3$ for the scalar
polarisabilities, and of $10^{-4}\;\fm^4$ for the spin ones. By a lucky
coincidence, on these different scales the individual uncertainties are
numerically roughly comparable (with all errors combined in quadrature):
$\pm0.9$ units for $\alphae-\betam$; $\pm0.5$ for the individual scalar
polarisabilities; and between $\pm0.5$ and $\pm0.8$ (different) units for the
spin ones, with the exception of $\pm1.9$ for
$\gammaee$~\cite{Griesshammer:2015ahu}.  Therefore, similar colours in the
sensitivity plots also loosely correlate to similar impacts on improving the
absolute size of polarisability values, even between scalar and spin
polarisabilities, which are measured in different units.

It is important to reiterate the need for caution over the strong signals that
exist for some variables around the Delta resonance. In this region, the
theory is complete only at NLO, while the spin polarisabilities enter at a
higher order than other neglected physics.  The sensitivity that we see is
genuine, but without an accurate description of the amplitudes, an extraction
of the actual value of the polarisabilities should not be made.

We now list some observations. First, we consider the observables and make
general comments on sensitivities, largely concentrating again on energies
below 250~MeV, where the three essential conditions of
sect.~\ref{sec:sensitivities} are met.  At this point, we are more concerned
with the strength of signals than with their uniqueness, and indeed some of
the most sensitive observables are affected by several of the
polarisabilities. We will consider possible ``sweet spots" that may allow the
targetting of particular polarisabilities or combinations next.  A cursory
glance at the plots shows that there is almost no place with sensitivity to
only one of the multipole-basis spin polarisabilities, whereas sensitivities
to the combinations defined above are in general better separated. We will
therefore confine our comments to those combinations.

\begin{itemize}
\item $\Sigma_{2x}$, $\Sigma_{2z}$, $\Sigma_{2x^\prime}$, and to a lesser
  extent $\Sigma_{2z^\prime}$ and $\Sigma_{3y}$, show the largest
  sensitivities. This is particularly notable, as it occurs not very far above
  the pion-production threshold. The sensitivity of $\Sigma_{3y^\prime}$,
  while sizeable, is largely at higher energies.  For the other five, in most
  cases there is marked sensitivity to $\alphae + \betam$ and $\gammazero$,
  and not exclusively at very forward angles. None has strong dependence on
  $\alphae - \betam$, but $\Sigma_{2z}$ is something of an exception, with at
  least some sensitivity even below threshold. $\gammapi$ does not feature,
  either. The possible exceptions are again $\Sigma_{2z}$ and $\Sigma_{2z'}$,
  which are also the only ones that do not vanish at backward angles.
  $\Sigma_{2x}$ and $\Sigma_{2x'}$ display a strong dependence on
  $\gammaeminus$ around $\theta\approx60^\circ$ and
  $\omegalab\approx200\;\MeV$, and particularly for $\Sigma_{2x}$ this is
  paired with little sensitivity to $\alphae-\betam$ and the other spin
  polarisability combinations.  $\Sigma_{2z}$ at mid angles and $\Sigma_{3y'}$
  at more forward angles have some sensitivity to $\gammamminus$, and to some
  degree to $\gammaeminus$ as well.  $\Sigma_{3y}$ has some very strong
  sensitivities at forward angles very close to threshold, but both its
  absolute magnitude and the cross section there are small.
 
  MAMI data, preliminary and published, exists for both $\Sigma_{2x}$, and
  $\Sigma_{2z}$ \cite{Martel:2012,Martel:2014pba,Martel:2017pln}, all of which
  lies above the ideal energy region.  Taking the sensitivity plots at face
  value, though, we see that the 288~MeV $\Sigma_{2x}$ data, though at a
  higher energy than peak sensitivity, is still sensitive primarily to
  $\gammaeminus$ and, to a lesser extent, to $\gammapi$ and $\alphae+\betam$,
  with almost no sensitivity to the other combinations.  The $\Sigma_{2z}$
  data probes all the spin polarisabilities except $\gammazero$, with little
  dependence on $\alphae-\betam$.

\item $\Sigma_3$: While the beam asymmetry has a reliably large magnitude
  below threshold, its sensitivity to the polarisabilities is limited there,
  see fig.~\ref{fig:polsvar-3}. This is expected from the discussion in
  sect.~\ref{sec:asymmetries}, as the Thomson term provides a low-energy
  theorem for this observable.  As noted in Ref.~\cite{Krupina:2013dya}, at
  very low energy there is a linear dependence on $\betam$, which however
  vanishes at $90^\circ$.  It can easily be seen, though, that at most angles
  above 80~MeV, the dependence on $\alphae$, though formally higher-order in
  the low-energy expansion, is actually as important as the one to $\betam$,
  and that the sensitivity to $\alphae-\betam$ is particularly weak. Of the
  spin polarisabilities, though, only $\gamma_0$ has any influence below
  pion-threshold.  This implies that below-pion-threshold, measurements of
  $\Sigma_3$, like those of ref.~\cite{Sokhoyan:2016yrc}, or the $\Sigma_3$
  data recently taken at \HIGS~\cite{Ahmed:2017}, will require high
  precision---and tight control of systematics---if they are to provide useful
  information on polarisabilities.

  Above 200~MeV, at mid angles, dependence on $\alphae+\betam$ and
  $\gammamminus$ sets in, peaking around 250~MeV.  There is no sensitivity to
  $\gammapi$, almost none to $\alphae-\betam$, and even $\gamma_0$ has little
  effect in this region. If we exploit the Baldin sum rule, $\Sigma_3$
  provides an opportunity to fix $\gammamminus$ with only a little sensitivity
  to $\gammaeminus$.

\item $\Sigma_y$: The region somewhat above the pion-production threshold and
  at moderate forward angles is sensitive to $\alphae+\betam$, with marked
  dependence on $\gammazero$ and some on $\gammamminus$ and $\gammaeminus$
  developing at slightly higher energies.  In view of the information already
  given by sum rules, this observable does not look very promising.

\item $\Sigma_{1x}$ and, to a lesser extent, $\Sigma_{1x^\prime}$ show a strong
  pocket of sensitivity to $\gammazero$ just above the pion-production
  threshold and at moderate forward angles.  In $\Sigma_{1x^\prime}$, the
  sensitivity region is somewhat larger, shifted to less acute angles, and
  shows little dependence on the scalar polarisabilities. However, we also
  note that the cross section is quite small at forward angles around the pion
  threshold.

\item $\Sigma_{1z}$ and $\Sigma_{1z^\prime}$: At moderate forward angles and
  at energies above 200~MeV, the polarisation transfer has increasing
  sensitivity to $\gammaeminus$ and $\gammamminus$, with no contamination at
  all from $\alphae-\betam$ or $\gammapi$. The best sensitivity, though, is at
  energies where 
  results of different theoretical approaches may differ.  Below 240~MeV, the
  asymmetry $\Sigma_{1z}$ has the smallest overall sensitivities of any
  observable. This would be a good place to test how well different
  theoretical approaches describe the non-structure part. Around
  $\theta\approx60^\circ$ and $\omegalab\approx200\;\MeV$, there is some mild
  sensitivity to $\gammaeminus$, with only a slight contamination from
  $\alphae+\betam$ and $\gammazero$.

\end{itemize}

Turning this around, we now list, in no particular order, regions of marked
sensitivity to various polarisability combinations, within the kinematic
domain where we trust the theoretical
  extraction and cross sections are not exceedingly small.  We are now
particularly interested in isolated sensitivities, or at least those that
become so when the forward sum rules are used to fix $\alphae+\betam$ and
$\gammazero$.

\begin{itemize}

\item $\gammaeminus$ in $\Sigma_{2x}$ for $30^\circ \lesssim \theta \lesssim 60^\circ$ and
  $200~{\rm MeV} \lesssim \omegalab \lesssim 250$ MeV. Though the sensitivity
  decreases with energy, it is still present in the region of some of the data
  of ref.~\cite{Martel:2012,Martel:2014pba}. However, those data are at high
  enough energies that we begin to significantly mistrust theoretical
  extractions. In fact, our prediction is that the peak (absolute) sensitivity
  to $\gammaeminus$ in this observable is at lower energies and more forward
  angles than these data. At higher angles, $\gammapi$ starts to play a role.

\item $\gammamminus$ in $\Sigma_3$ for $60^\circ \lesssim \theta \lesssim 90^\circ$ and
  $200~{\rm MeV} \lesssim \omegalab \lesssim 250~{\rm MeV}$, though
  sensitivity to $\gammaeminus$ is not absent. The LEGS
  data~\cite{Blanpied:2001ae} overlaps this region, and re-examination of it
  in \ChiEFT is likely worthwhile (cf.~ref.~\cite{Pascalutsa:2003zk,
    Mcgovern:2015mgf, Lensky:2015awa}), especially in light of the preliminary
  MAMI results~\cite{Collicott:2015, Martel:2017pln} at kinematics which
  overlap the data of ref.~\cite{Blanpied:2001ae}.  $\theta\lesssim 60^\circ$ and
  energies of around 230~MeV in $\Sigma_{2z}$ are also promising for this
  polarisability.
  
\item $\gammaeminus$ and $\gammamminus$ affect almost every observable to some
  degree, and there is rarely contamination from $\alphae-\betam$ and
  $\gammapi$ for angles less than $90^\circ$.  For $\Sigma_{2x'}$ and
  $\Sigma_{2z'}$ and, to a lesser extent, for $\Sigma_{3y'}$, $\Sigma_{3y}$
  and $\Sigma_{1z'}$, sensitivity to both combinations is present below
  200~MeV. There is even a noticeable effect below pion-production threshold
  in $\Sigma_{2x'}$.  Strong dependence on both $\gammaeminus$ and
  $\gammamminus$ is also found for $60^\circ\lesssim \theta\lesssim 90^\circ$ and
  energies around 230~MeV in $\Sigma_{2z}$. At lower energies,
  $\alphae-\betam$ is a confounding variable.

\item $\gammapi$ 
  is hard to isolate. The vast majority of observables are almost completely
  insensitive to it below 240~MeV.  Exceptions are $\Sigma_{2z}$ at around
  $150^\circ$ and 240~MeV, where fortunately sensitivity to $\alphae-\betam$
  happens to be small, and $\Sigma_{2z'}$ in a somewhat similar
  region. However, in the latter observable, sensitivity to $\alphae-\betam$
  is not negligible.

\item $\alphae-\betam$ is another combination with very limited opportunities
  for determinative measurements in asymmetries and polarisation-transfer
  observables. At central angles around the pion-production threshold, there is
  good sensitivity in $\Sigma_{2z}$ and $\Sigma_{2z'}$, but not independent of
  $\gammaeminus$ and $\gammamminus$.

\item $\alphae + \betam$ affects many observables at forward angles, most
  notably $\Sigma_{2x'}$ and $\Sigma_{2z}$, though usually in conjunction with
  $\gammazero$. The strongest isolated sensitivity is well below threshold at
  moderate forward and backward angles in $\Sigma_{3}$ and $\Sigma_{2z'}$, but
  it is hard to see experiments here reaching an accuracy that could rival the
  error bars of the Baldin sum rule.

\item $\gammazero$ dominates the sensitivities for a reasonable-sized region
  in $\Sigma_{1x'}$: angles around $30^\circ$ and $\omegalab \approx 200$ MeV,
  and in $\Sigma_y$ in a rather narrower energy range at similar angles.
  Again, it is hard to see the sum rule being rigorously tested in any
  practical experiment here.
\end{itemize}

The bottom line is encouraging. If $\alphae+\betam$ and $\gammazero$ are
regarded as well-constrained by the forward sum rules, there are multiple
opportunities to extract the spin-polarisability combinations $\gammaeminus$
and $\gammamminus$ in regions that are experimentally accessible,
theoretically clean and largely independent of the values of $\alphae-\betam$
and $\gammapi$.  Access to the scalar polarisabilities, to $\gammazero$ and to
$\gammapi$ is best achieved through cross-section measurements, with
coverage of markedly forward and backward angles being very useful in that
regard.


\clearpage

\subsection{Neutron Observables}
\label{sec:neutron}

\begin{figure}[!htbp]
\begin{center}
     \includegraphics[width=0.6\textwidth]{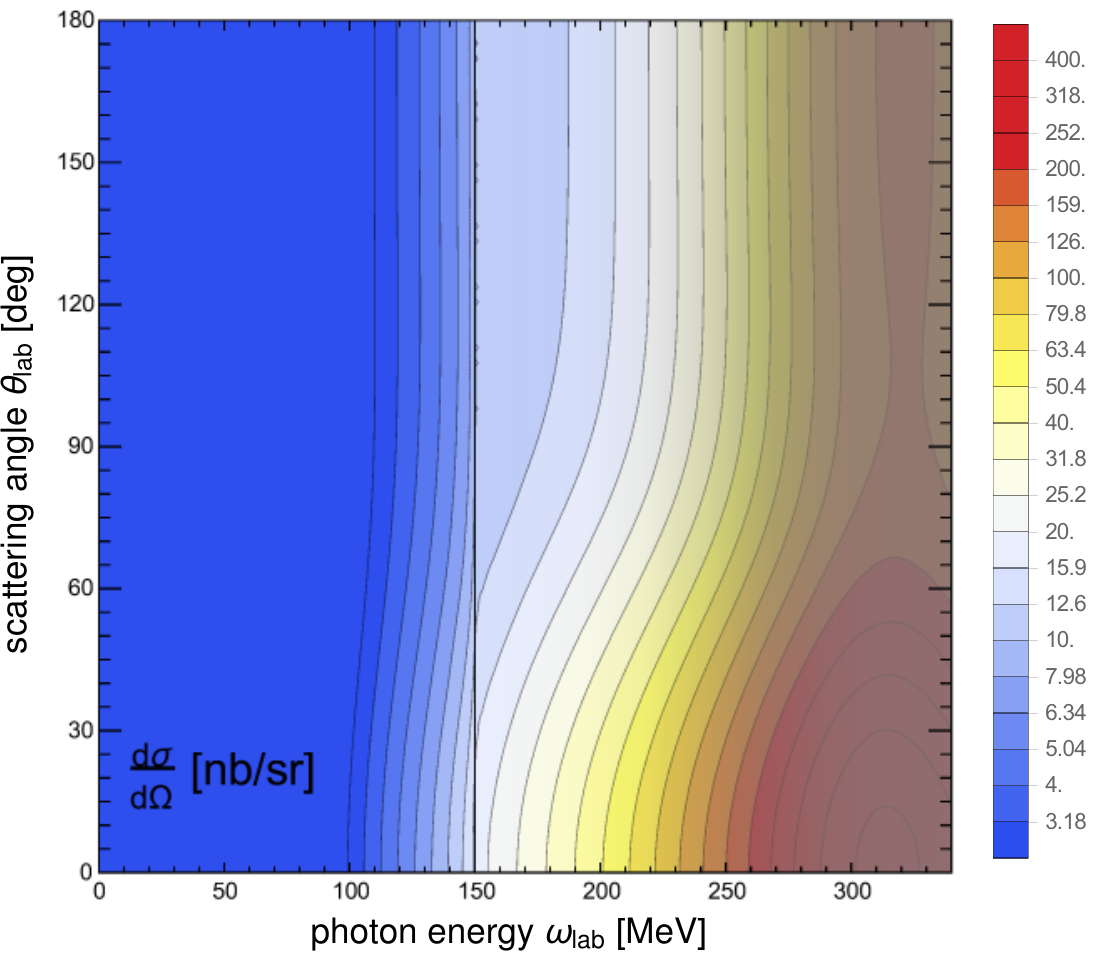}
     \caption{(Colour online) Contour plot of the unpolarised differential
       cross section for the neutron as a function of $\omegalab$ and
       $\thetalab$, on a logarithmic scale, and with additional contours for
       very small and very large values. The colour coding is identical to
       that of the corresponding fig.~\ref{fig:crosssection} for the
       proton.}
\label{fig:crosssection-neutron}
\end{center}
\end{figure}

\begin{figure}[!htbp]
\begin{center}
     \includegraphics[width=\textwidth]{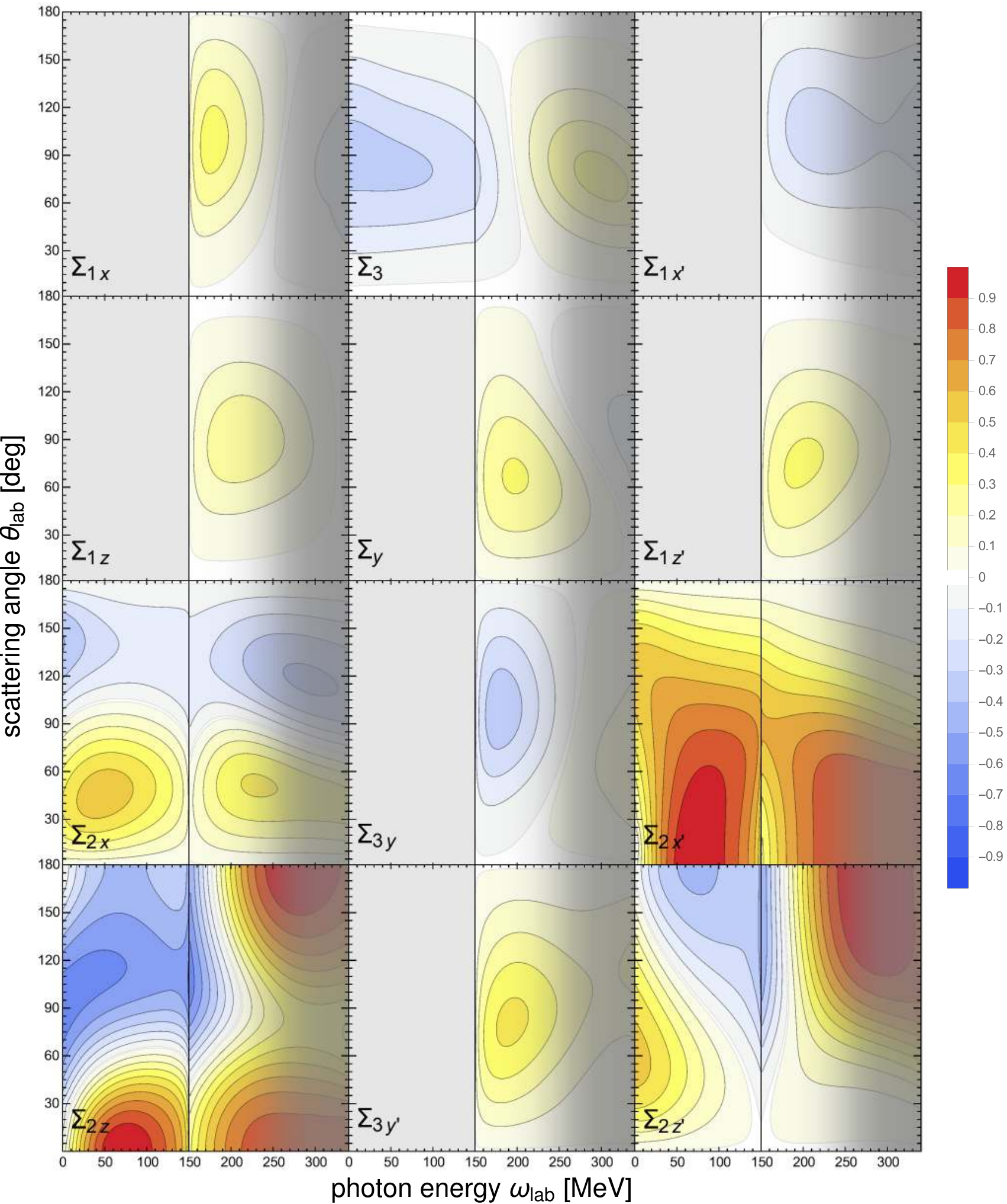}
     \caption{(Colour online) Contour plots of the asymmetries and
       polarisation-transfer observables for the neutron; see text for
       details.}
\label{fig:asymmetries-neutron}
\end{center}
\end{figure}

\begin{figure}[!htbp]
\begin{center}
     \includegraphics[width=\textwidth]{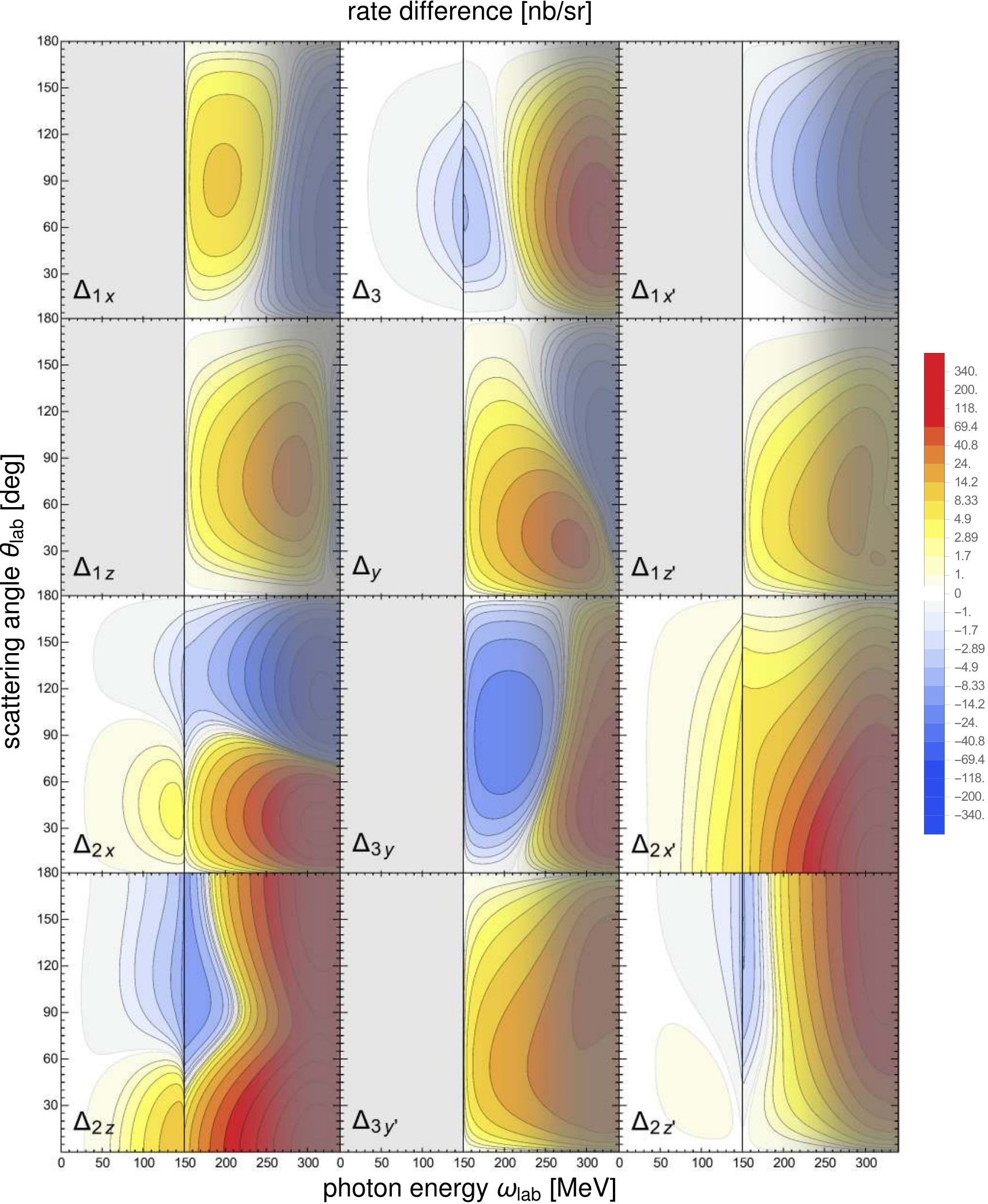}
     \caption{(Colour online) Contour plots of the rates associated with
       asymmetries and polarisation-transfer observables for the neutron, with
       additional contours for very small and very large values; see text for
       details. The colour coding is identical to that of the corresponding
       fig.~\ref{fig:asymmetries-rates} for the proton. }
\label{fig:asymmetries-rates-neutron}
\end{center}
\end{figure}

There are no free neutron targets stable and dense enough for meaningful
Compton scattering experiments. But inelastic processes in quasi-free neutron
kinematics can be approximated as free-neutron processes multiplied by the
momentum distribution of the neutron inside a nucleus; see
refs.~\cite{Levchuk:1994, Levchuk:2000, Demissie:2016ktr} and references
therein for descriptions of the unpolarised process on the deuteron. However,
such an ``Impulse Approximation" description is accurate only for energies
well above the pion-production threshold. With these caveats in mind, in this
section we present an overview of free-neutron Compton scattering
observables in figs.~\ref{fig:crosssection-neutron} to~\ref{fig:asymmetries-rates-neutron}. The atlas of sensitivities of observables to variations of the
neutron polarisabilities is available in the online supplement
(appendix~\ref{app:moreplots} of the arXiv version), and also covered by the
\emph{Mathematica} notebook. For observables corresponding to $\Sigma_{2z}$,
$\Sigma_{2x}$, $\Sigma_{2z}$, $\Sigma_{1x}$, $\Sigma_{1z}$ and $\Sigma_3$,
results to order $e^2\delta^3$ for $\omega\lesssim170\;\MeV$ were already
presented in refs.~\cite{Hildebrandt:2003md, Hildebrandt:2005ix}, and the
effects of higher-multipole polarisabilities were again shown to be
small. Here, we add one order, extend to higher energies, and provide a more
comprehensive sensitivity study.

The neutron cross section is of course much smaller than the proton one at low
energies due to the absence of the charged contributions to the Born
amplitudes. Only pion-pole, magnetic-moment and structure effects are
non-zero. Interestingly, it exceeds the proton cross section around the
pion-production threshold, where charge-contributions are overwhelmed by the
other Born and structure effects. Isospin symmetry guarantees that the two
cross sections are very similar closer to the Delta resonance.

The sensitivity of all asymmetries and polarisation transfers to the spin
polarisabilities is also much bigger than for the proton
analogues. Quasielastic kinematics for $\omegalab\lesssim250\;\MeV$ may
therefore be an interesting venue to extract neutron spin polarisabilities: it
is possible there are even better signals there than for the proton case.

\clearpage

\section{Summary and Conclusions}
\setcounter{equation}{0}
\label{sec:conclusions}

In this paper, we have provided a comprehensive study of asymmetries and
polarisation-transfer observables for polarised photon scattering from a
nucleon, including both polarised-target and recoil-polarisation-detection
observables. Our main goal is to provide guidance to experimentalists planning
the next generation of polarised Compton scattering experiments. We have used
Chiral Effective Field Theory to calculate the magnitudes of the observables
over the full range of angles and energies up to the peak of the
$\Delta(1232)$ resonance.  We have investigated the effects of varying the
scalar and spin polarisabilities, with a view to identifying regions of
particular sensitivity. The results presented here, supplemented by additional
ones for the neutron, are also available as an interactive \emph{Mathematica}
notebook from \texttt{judith.mcgovern@manchester.ac.uk}.

We have also shown that, at a given energy and angle, the $6$ observables that
do not vanish below the pion-production threshold suffice to determine the
Compton amplitudes there. They form a ``complete set of experiments" and in
principle could be used to obtain the energy-dependence of the scalar and spin
dipole polarisabilities from purely sub-threshold data---without the use of
any sum-rule constraints. Above pion threshold, $5$ additional observables
are needed if the Compton multipoles are to be completely determined from
experiment. Polarisation-transfer observables are essential to such a
programme. This is however not necessary to determine the static
polarisabilities which were our focus.

The Baldin and forward-spin-polarisability sum rules constrain the
polarisability combinations $\alphae + \betam$ and $\gamma_0$ quite
tightly. While achieving concurrence with the sum-rule values is a worthy
goal, it will be challenging for the experiments discussed here to improve
upon them. Because of this, we placed particular emphasis on opportunities to
pin down individual polarisabilities, as well as combinations such as
$\alphae - \betam$, $\gammaeminus:=\gammaee-\gammaem$ and
$\gammamminus:=\gammamm-\gammame$. This shows that a concerted effort to
measure several asymmetries and polarisation transfers, in the various
kinematics where we have identified sensitivities, is the optimum overall
strategy. Definitive values of the spin polarisabilities will ultimately be
established through the same procedure by which $\alphae - \betam$ has been
determined: a comprehensive fit to a statistically consistent experimental
database, with carefully formulated correlated and point-to-point systematic
errors, that has been compiled at several labs around the world.

By comparing different approaches to calculating Compton amplitudes and by
considering where the particular \ChiEFT amplitude we used here is reliable,
we have identified two distinct energy regions where such a campaign could be
carried out with a minimum of theoretical bias.

In the low-energy region, \ie below the pion-production threshold, we are
confident in the \ChiEFT predictions, with almost the only residual
uncertainty being the precise value of the polarisabilities themselves. At
these energies, plenty of high-quality cross section data is available, but
there is only limited information on the asymmetries---only one is measured at
all.  Accurate experiments with good energy and angular range that explore the
$6$ observables which do not vanish in this region are required to pin down
all $6$ polarisabilities. However, though asymmetries can be large, the
absolute rates are not huge and the sensitivities to the spin polarisabilities
are quite modest in the main. Demands on beam time and control of systematics
will certainly be high.

In the intermediate-energy region, namely between the pion threshold and
$\omegalab\approx250$~MeV, the agreement between \ChiEFT and
dispersion-relation calculations remains rather good, and both the cross
section and sensitivities tend to rise rapidly. This is a region for which
very limited data exist. It is therefore a promising avenue for further
exploration. Once again, only one asymmetry has been explored, with limited
coverage and accuracy. Indeed, significant gaps exist even in the $\gamma$p
cross section data base. Measurements motivated by sensitivities to spin
polarisabilities in this energy regime would thus likely have the added benefit of also
delivering high-quality unpolarised data. This could help to resolve the
disagreement between those proton Compton scattering experiments that accrued
significant cross section data in this energy range (see Table 3.1 of
ref.~\cite{Griesshammer:2012we} and discussion in
ref.~\cite{McGovern:2012ew}). Indeed, we argued that further information on
$\alphae$, $\betam$, $\gammazero$ and $\gammapi$ will be most easily and
reliably obtained from cross-section measurements---perhaps especially from
cross-section measurements at markedly forward or backward angles.

At even higher energies, the cross sections and certain sensitivities become
quite sizeable; however, the theoretical predictions of the underlying
amplitudes in \ChiEFT and the dispersion-relation approach start to
diverge. Though high-statistics measurements in this region will challenge and
discriminate between these two approaches, the comparisons may do little to
shed light on the specific value of the polarisabilities. Accurate connection
to the static polarisabilities, which are defined in the low-energy limit, is
likely to be very challenging here.

Therefore, we propose that future explorations focus on a region in which
three essential conditions are met: there are significant sensitivities to
spin polarisabilities (usually $\omegalab\gtrsim100\;\MeV$); theory can
extract polarisabilities reliably and with high accuracy
($\omegalab\lesssim250\;\MeV$); and experiments are not overwhelmed by
backgrounds ($30^\circ\lesssim\theta\lesssim160^\circ$). Within these
constraints, we identified several kinematic regions where various asymmetries
and polarisation-transfer observables are significantly affected by the
spin-polarisability combinations $\gammaeminus$ and $\gammamminus$, while
being rather insensitive to $\alphae-\betam$ and $\gammapi$.

We hope this presentation has provided indications of which proton
Compton-scattering data are most promising for polarisability extractions.
Nevertheless, complex sensitivities and the experimental challenges of small
rates paired with polarised targets and/or recoils mean that no single
experiment will suffice to determine a polarisability, or a simple combination
thereof, in particular when it comes to the spin polarisabilities. Instead,
what is needed is a comprehensive programme of mutually complementary
experiments---ideally conducted at different facilities.
As theorists, we recognise that there are many experimental realities that can
prevent a promising observable or kinematic region from being the best place
to do an experiment, and look forward to a lively exchange regarding the
planning and analysis of the experiments at the new generation of experiments
at high-luminosity facilities with polarised beams and targets such as \HIGS
and MAMI.


\section*{Acknowledgements}

We gratefully acknowledge discussions with J.~R.~M.~Annand, M.~W.~Ahmed,
E.~J.~Downie, D.~Hornidge, V.~Lensky and B.~Pasquini. We are particularly
grateful to the organisers and participants of the workshop \textsc{Lattice
  Nuclei, Nuclear Physics and QCD - Bridging the Gap} at the ECT* (Trento),
and of the US DOE-supported \textsc{Workshop on Next Generation Laser-Compton
  Gamma-Ray Source}, and for hospitality at KITP (Santa Barbara; supported in
part by the US National Science Foundation under grant NSF PHY-1125915) and at
KPH (Mainz). HWG is indebted to the kind hospitality and financial support of
the Institut f\"ur Theoretische Physik (T39) of the Physik-Department at TU
M\"unchen and of the Physics Department of the University of Manchester, where
much of this work was completed, and to both for the quick and unbureaucratic
help when hardware issues threatened to derail carefully crafted plans.
This work was supported in part by UK Science and Technology Facilities
Council grants ST/L005794/1  and ST/P004423/1 (JMcG), by the US Department of
Energy under contracts DE-FG02-93ER-40756 (DRP) and DE-SC0015393 (HWG), and by
the Dean's Research Chair programme of the Columbian College of Arts and
Sciences of The George Washington University (HWG).

\newpage

\appendix
\section{Dipole Polarisabilities of the Nucleon} 
\setcounter{equation}{0}
\label{app:polvalues}

For the convenience of the reader, we repeat in table~\ref{tab:pols} the
values and uncertainties of the proton's and neutron's dipole scalar and spin
polarisabilities obtained at $\calO(e^2\delta^4)$ in \ChiEFT with explicit
$\Delta(1232)$ degrees of freedom~\cite{McGovern:2012ew,Myers:2014ace,
  Griesshammer:2015ahu}. The theoretical uncertainties ($68\%$
degree-of-belief intervals) are derived from order-by-order convergence of the
EFT. The scalar polarisabilities are determined from
data~\cite{McGovern:2012ew, Myers:2014ace}, with $\chi^2=113.2$ for $135$
degrees of freedom for the proton, and $45.2$ for $44$ for the neutron. The
respective Baldin sum rules have been used as a constraint. (Note, though,
that the extraction used
$\alphaep+\betamp=13.8\pm0.4$~\cite{OlmosdeLeon:2001zn}. This agrees within
error bars with the more recent value
$\alphaep+\betamp=14.0\pm0.2$~\cite{Gryniuk:2015eza}.) The proton spin
polarisability $\gammammp$ has been fitted as described in
ref.~\cite{McGovern:2012ew}, but $\gammammn$ is then predicted from
that. Scalar polarisabilities are quoted in $10^{-4}~{\rm fm}^3$, spin ones in
$10^{-4}~{\rm fm}^4$. A thorough discussion of all aspects can be found in
ref.~\cite{Griesshammer:2015ahu}.

\begin{table}[!htb]
\begin{center}
\begin{tabular}{|r||l|l|}
\hline
&proton&neutron\\\hline\hline
\alphae&$10.65\pm0.35_\text{stat}\pm0.2_\text{Baldin}\pm0.3_\text{theory}$&
$11.55\pm1.25_\text{stat}\pm0.2_\text{Baldin}\pm0.8_\text{theory}$\\
\betam&$\phantom{0}3.15\mp0.35_\text{stat}\pm0.2_\text{Baldin}
   \mp0.3_\text{theory}$&
   $\phantom{0}3.65\mp1.25_\text{stat}\pm0.2_\text{Baldin}
   \mp0.8_\text{theory}$\\\hline
\gammaee&$-1.1\pm1.9_\text{theory}$&
$-4.0\pm1.9_\text{theory}$\\
\gammamm&$ \phm2.2\pm0.5_\text{stat}\pm0.6_\text{theory}$&
$\phm1.3\pm0.5_\text{stat}\pm0.6_\text{theory}$\\
\gammaem&$-0.4\pm0.6_\text{theory}$&
$-0.1\pm0.6_\text{theory}$\\
\gammame&$\phm 1.9\pm0.5_\text{theory}$&
$\phm2.4\pm0.5_\text{theory}$\\\hline
\end{tabular}
\end{center}
\label{tab:pols}
\caption{The dipole polarisabilities of the proton and neutron in \ChiEFT with explicit \protect$\Delta(1232)$ degrees of freedom at
  \protect$\calO(e^2\delta^4)$~\cite{McGovern:2012ew, Myers:2014ace,
    Griesshammer:2015ahu}.} 
\end{table}

\section{Matrices for Observables} 
\setcounter{equation}{0}
\label{app:matrices}

We  provide here the definition of the Compton scattering amplitude in the
Breit or cm frame and the relations between the $6$ independent amplitudes
$A_i$ and the asymmetries and polarisation-transfer observables.

All the expressions required are given in Babusci
\etal~\cite{Babusci:1998ww}. Those authors use a covariant basis to define $6$
independent amplitudes (which they denote simply $A_i$ but we will call
$A_i^{\text{L}}$ as they are due to L'vov), and construct invariants
$W^{\pm}_{ij}$ from these (the first index, $i=0,1,2,3$, refers to the photon
polarisation, and the second, $j=0,1,2,3$, to the nucleon
polarisation). Linear combinations of these invariants ($13$ of which are
non-vanishing) give cross sections for polarised photons and nucleons
(incoming or outgoing), and from these, in turn, asymmetries (absolute or
relative) can be constructed.

In order to see the influence of the polarisabilities, though, it is more
customary to use a basis of independent tensors constructed from Pauli
matrices as follows:
\begin{equation}
  \label{eq:Tmatrix}
  \begin{array}{rcl}
    T(\w,z)&=& A_1(\w,z)\;(\vec{\epsilon}\,'^*\cdot \vec{\epsilon}) +
    A_2(\w,z)\;(\vec{\epsilon}\,'^*\cdot\hat{\vec{k}})\;(\vec{\epsilon}
    \cdot\hat{\vec{k}}')
    \\&&
    +\ii\,A_3(\w,z)\;\vec{\sigma}\cdot\left(\vec{\epsilon}\,'^*\times\vec{\epsilon}\,\right)
    +\ii\,A_4(\w,z)\;\vec{\sigma}\cdot\left(\hat{\vec{k}}'\times\hat{\vec{k}}\right)
    (\vec{\epsilon}\,'^*\cdot\vec{\epsilon}) \\&&
    +\ii\,A_5(\w,z)\;\vec{\sigma}\cdot
    \left[\left(\vec{\epsilon}\,'^*\times\hat{\vec{k}}
      \right)\,(\vec{\epsilon}\cdot\hat{\vec{k}}')
      -\left(\vec{\epsilon}\times\hat{\vec{k}}'\right)\,
      (\vec{\epsilon}\,'^*\cdot\hat{\vec{k}})\right]
    \\&& +\ii\,A_6(\w,z)\;\vec{\sigma}\cdot
    \left[\left(\vec{\epsilon}\,'^*\times\hat{\vec{k}}'\right)\,
      (\vec{\epsilon}\cdot\hat{\vec{k}}') -\left(\vec{\epsilon}
        \times\hat{\vec{k}} \right)\,
      (\vec{\epsilon}\,'^*\cdot\hat{\vec{k}})\right] \; \;.
  \end{array}
\end{equation}
where $\hat{\vec{k}}$ ($\hat{\vec{k}}'$) is the unit vector in the momentum
direction of the incoming (outgoing) photon with polarisation $\vec{\epsilon}$
($\vec{\epsilon}\,'^*$), $\theta$ is the scattering angle, and $z=\cos\theta$.
This form holds in the Breit and centre-of-mass (cm) frames, and defines amplitudes $A_i^{\text{cm}}(\w_\text{cm}, z_\text{cm})$ or $A_i^{\text{Br}}(\w_\text{Br}, z_\text{Br})$.

The low-energy expansion of the non-Born parts of the Breit-frame amplitudes in terms of the polarisabilities, defined in  eq.~\eqref{eq:H-eff}, is:
\begin{align}\label{eq:polamp}
A_1^\mathrm{Br}&= 4\pi \left(\alphae+z_\text{Br}\betam\right) \w_\text{Br}^2
+{\cal O}(\w_\text{Br}^4)\;\;,\nonumber\\
A_2^\mathrm{Br}&= -4\pi \;\beta_{M1}\,\w_\text{Br}^2+{\cal O}(\w_\text{Br}^4)\;\;,
\nonumber\\
A_3^\mathrm{Br}&=-4\pi\left(\gamma_{E1E1}+ \gamma_{E1M2}+z_\text{Br}( \gamma_{M1M1}+ \gamma_{M1E2})\right)\w_\text{Br}^3
+{\cal O}(\w_\text{Br}^5)\;\;,\nonumber\\
A_4^\mathrm{Br}&= 4 \pi\left(\gamma_{M1E2} -\gamma_{M1M1}\right)\w_\text{Br}^3
+{\cal O}(\w_\text{Br}^5)\;\;,\nonumber\\
A_5^\mathrm{Br}&= 4\pi\;\gamma_{M1M1}\,\w_\text{Br}^3
+{\cal O}(\w_\text{Br}^5)\;\;,\nonumber\\
A_6^\mathrm{Br}&=  4 \pi\;\gamma_{E1M2}\,\w_\text{Br}^3+{\cal O}(\w_\text{Br}^5)\;\;.
\end{align}
The expression for the cm-frame amplitudes in terms of the cm photon energy
and scattering angle is identical, except that the omitted terms all start at
one order lower in $\omega$ because manifest crossing symmetry is lost. These
new terms are all boost corrections and depend only on the same
polarisabilities that already enter above.

When, in sects.~\ref{sec:crosssectionvar} and~\ref{sec:asymmetriesvar}, we
vary the polarisabilities, we add to the chiral amplitudes terms mirroring
eq.~\eqref{eq:polamp} with polarisability shifts $\delta \alphae$,
$\delta \betam$, $\delta\gamma_i$ in place of full polarisabilities.

The relations between the covariant, cm and Breit amplitudes provided in
ref.~\cite{Lensky:2015awa} allow the invariants that are expressed in terms of
the $A_i^{\text{L}}$ in ref.~\cite{Babusci:1998ww} to be written in terms of
the cm or Breit amplitudes. Writing the amplitudes as a vector,
$\vec{A}=(A_1,\dots,A_6)^T$, invariants and observables are conveniently
expressed as bilinears in $\vec{A}$: $\vec{A}^\dagger\,\calM_\alpha\,\vec{A}$.

In the cm frame, these matrices $\calM_\alpha$ for the observables
$\calO_\alpha$ all have a simple form and are independent of the photon
energy, so we give those here. In what follows, $z=\cos\theta_{\text{cm}}$ and
$A_i=A_i^{\text{cm}}$.  As mentioned in sect.~\ref{sec:pols}, the
polarisabilities are most easily defined in the Breit frame, and that is the
frame we use for our numerical work.  However, the expressions for the
matrices are the same in the $\MN\to\infty$ limit, and the simpler forms in
the cm frame allow one to see more clearly where the various amplitudes, and
hence polarisabilities, play a role.
  
The unpolarised differential cross section is given by
\def\TsqUnpol{{\dd\sigma}}
\begin{equation}
  \label{eq:dsigma-domega}
  \left.\frac{\dd\sigma}{\dd\Omega}\right|_{\text{unpol}}=\Phi^2\;|T|^2=\Phi^2\;
  \left(\vec{A}^\dagger\,\calM_\TsqUnpol\,\vec{A}\right)
\end{equation}
where $\Phi$ is the frame-dependent flux factor which tends to $1/4\pi$ at low
energy (see, \eg, the review~\cite{Griesshammer:2012we}). $|T|^2$
is calculated by averaging over the initial photon polarisation and nucleon
spin, and summing over the final states. (It is what Babusci \etal\ call
$W_{00}$, up to a normalisation factor of $4\MN^2$.)

The matrix associated with $|T|^2$ is 
\newcommand{\ms}{}
\begin{equation}
  \label{eq:ampsquared}
  \calM_\TsqUnpol=\frac 1 2 {\ms\arraycolsep=0.3\arraycolsep\begin{pmatrix}
 z^2+1 & z \left(z^2-1\right) & 0 & 0 & 0 & 0 \\
 z \left(z^2-1\right) & \left(z^2-1\right)^2 & 0 & 0 & 0 & 0 \\
 0 & 0 & 3-z^2 & z-z^3 & -2 z \left(z^2-1\right) & 4-4 z^2 \\
 0 & 0 & z-z^3 & 1-z^4 & -2 z^2 \left(z^2-1\right) & -2 z \left(z^2-1\right) \\
 0 & 0 & -2 z \left(z^2-1\right) & -2 z^2 \left(z^2-1\right) & -4 z^4+2 z^2+2 & -6 z \left(z^2-1\right) \\
 0 & 0 & 4-4 z^2 & -2 z \left(z^2-1\right) & -6 z \left(z^2-1\right) & 6-6 z^2
\end{pmatrix}}.
\end{equation}

The asymmetries and polarisation-transfer observables $\Sigma_\alpha$ are then
expressed by matrices $\calM_\alpha$ which can also be derived from the
customary relation between a cross section and the density matrices, see
sect.~\ref{sec:Observables}: $\rho^{(\gamma)}(\vec{\xi})$ for the photon beam
in terms of the Stokes parameters; $\rho(P,\vec{n})$ for the target; and
$\rho(P^\prime\equiv1,\vec{n}^\prime)$ for the recoil. When the recoil
polarisation is not detected, the cross section for a specific polarisation
state is
\begin{equation}
  \label{eq:crossforasym}
  \frac{\dd\sigma}{\dd\Omega}(P,\vec{n};\vec{\xi})=
  \Phi^2\;\tr[T\;\rho^{(\gamma)}(\vec{\xi})\;\rho(P,\vec{n})\;T^\dagger]\;\;.
\end{equation}
For polarisation-transfer observables, one uses
$\rho(P=0,\vec{n})=\frac{1}{2}$ for an unpolarised target:
\begin{equation}
  \label{eq:crossforasym2}
  \frac{\dd\sigma}{\dd\Omega}(\vec{n}^\prime;\vec{\xi})=
  \frac{1}{2}\;\Phi^2\;\tr[T\;\rho^{(\gamma)}(\vec{\xi})\;T^\dagger\;
  \rho(P^\prime\equiv1,\vec{n}^\prime)]\;\;.
\end{equation} 
The matrix $\calM_\alpha$ is then derived by inserting these relations into
the definition of the corresponding observable $\Sigma_\alpha$ in
eqs.~\eqref{eq:asym3} to~\eqref{eq:asym3y} for the asymmetries and their
analogue for the polarisation-transfer coefficients, with the normalisation
such that
\begin{equation}
  \label{eq:obstomatrix}
  \Sigma_\alpha=\frac{\vec{A}^\dagger\,\calM_\alpha\,\vec{A}}
   {2\;\vec{A}^\dagger\,\calM_\TsqUnpol\,\vec{A}}\;\;. 
\end{equation}

A similar derivation yields the difference $\Delta_\alpha$ of the rates for
the different orientations associated with each asymmetry or
polarisation-transfer observable (see figs.~\ref{fig:asymmetries-rates} and
~\ref{fig:asymmetries-rates-neutron} for plots). These are given by the
numerator of eqs.~\eqref{eq:asym3} to~\eqref{eq:asym3y} and the
polarisation-transfer analogues, and are proportional to the numerator
of~\eqref{eq:obstomatrix},
\begin{equation}
  \label{eq:deltamatrix}
  \Delta_\alpha=\frac{g}{2}\;\Phi^2\;\vec{A}^\dagger\,\calM_\alpha\,\vec{A}
  =g\;\Sigma_\alpha\;\left.\frac{\dd\sigma}{\dd\Omega}\right|_{\text{unpol}}\;\;,
\end{equation}
with $g=4$ for $\Delta_{3y}$, $g=2$ for all other asymmetries and for
$\Delta_{3y'}$, and $g=1$ for the other polarisation-transfer observables. The
denominator in the definition of each observable $\Sigma_\alpha$
(eqs.~\eqref{eq:asym3} to~\eqref{eq:asym3y} and analogues) is
$g\;\Phi^2\;\vec{A}^\dagger\,\calM_\TsqUnpol\,\vec{A}$.

For the asymmetries, one finds:
\begin{equation}
  \label{eq:matsig3}
\calM
_3
= (1-z^2){\ms\arraycolsep=0.3\arraycolsep\begin{pmatrix}
 -1 & -z & 0 & 0 & 0 & 0 \\
 -z & 1-z^2 & 0 & 0 & 0 & 0 \\
 0 & 0 & 1 & z & 2 z & 2 \\
 0 & 0 & z & z^2-1 & 2 z^2 & 2 z \\
 0 & 0 & 2 z & 2 z^2 & 4 z^2 & 4 z \\
 0 & 0 & 2 & 2 z & 4 z & 4\end{pmatrix}}
\end{equation}
\begin{equation}
  \label{eq:matsigy}
\calM
_y
= \ii \sqrt{1-z^2}{\ms\arraycolsep=0.3\arraycolsep\begin{pmatrix}
 0 & 0 & -z & -z^2-1 & -2 z^2 & -2 z \\
 0 & 0 & 1-z^2 & z-z^3 & -2 z \left(z^2-1\right) & 2-2 z^2 \\
 z & z^2-1 & 0 & 0 & 0 & 0 \\
 z^2+1 & z \left(z^2-1\right) & 0 & 0 & 0 & 0 \\
 2 z^2 & 2 z \left(z^2-1\right) & 0 & 0 & 0 & 0 \\
 2 z & 2 \left(z^2-1\right) & 0 & 0 & 0 & 0\end{pmatrix}}
\end{equation}
\begin{equation}
  \label{eq:matsig1x}
\calM
_{1x}
= \ii \sqrt{1-z^2}{\ms\arraycolsep=0.3\arraycolsep\begin{pmatrix}
 0 & 0 & z & 0 & z^2+1 & 2 z \\
 0 & 0 & z^2-1 & 0 & z \left(z^2-1\right) & z^2-1 \\
 -z & 1-z^2 & 0 & z^2-1 & z^2+1 & 2 z \\
 0 & 0 & 1-z^2 & 0 & z-z^3 & 1-z^2 \\
 -z^2-1 & z-z^3 & -z^2-1 & z \left(z^2-1\right) & 0 & 2 \left(z^2-1\right) \\
 -2 z & 1-z^2 & -2 z & z^2-1 & 2-2 z^2 & 0\end{pmatrix}}
\end{equation}
\begin{equation}
  \label{eq:matsig1z}
\calM
_{1z}
= \ii (1-z^2){\ms\arraycolsep=0.3\arraycolsep\begin{pmatrix}
 0 & 0 & -1 & 0 & -z & -1 \\
 0 & 0 & -z & 0 & 1-z^2 & 0 \\
 1 & z & 0 & -z & -z & -1 \\
 0 & 0 & z & 0 & z^2+1 & 2 z \\
 z & z^2-1 & z & -z^2-1 & 0 & 0 \\
 1 & 0 & 1 & -2 z & 0 & 0\end{pmatrix}}
\end{equation}
\begin{equation}
  \label{eq:matsig2x}
\calM
_{2x}
= \sqrt{1-z^2}{\ms\arraycolsep=0.3\arraycolsep\begin{pmatrix}
 0 & 0 & -z & 0 & 1-z^2 & 0 \\
 0 & 0 & 1-z^2 & 0 & z-z^3 & 1-z^2 \\
 -z & 1-z^2 & 2 z & z^2+1 & 3 z^2-1 & 2 z \\
 0 & 0 & z^2+1 & 0 & z \left(z^2-1\right) & 1-z^2 \\
 1-z^2 & z-z^3 & 3 z^2-1 & z \left(z^2-1\right) & 4 z \left(z^2-1\right) & 2 \left(z^2-1\right) \\
 0 & 1-z^2 & 2 z & 1-z^2 & 2 \left(z^2-1\right) & 0\end{pmatrix}}
\end{equation}
\begin{equation}
  \label{eq:matsig2z}
\calM
_{2z}
={\ms\arraycolsep=0.3\arraycolsep\begin{pmatrix}
 0 & 0 & -z^2-1 & 0 & z-z^3 & z^2-1 \\
 0 & 0 & z-z^3 & 0 & -\left(z^2-1\right)^2 & 0 \\
 -z^2-1 & z-z^3 & 2 \left(z^2-1\right) & z \left(z^2-1\right) & 3 z \left(z^2-1\right) & 3 \left(z^2-1\right) \\
 0 & 0 & z \left(z^2-1\right) & 0 & \left(z^2-1\right)^2 & 0 \\
 z-z^3 & -\left(z^2-1\right)^2 & 3 z \left(z^2-1\right) & \left(z^2-1\right)^2 & 4 z^2 \left(z^2-1\right) & 4 z \left(z^2-1\right) \\
 z^2-1 & 0 & 3 \left(z^2-1\right) & 0 & 4 z \left(z^2-1\right) & 4 \left(z^2-1\right)\end{pmatrix}}
\end{equation}
\begin{equation}
  \label{eq:matsig3y}
\calM
_{3y}
= \ii \sqrt{1-z^2}{\ms\arraycolsep=0.3\arraycolsep\begin{pmatrix}
 0 & 0 & -z & 1-z^2 & -2 z^2 & -2 z \\
 0 & 0 & 1-z^2 & z-z^3 & -2 z \left(z^2-1\right) & 2-2 z^2 \\
 z & z^2-1 & 0 & 0 & -2 & -2 z \\
 z^2-1 & z \left(z^2-1\right) & 0 & 0 & 0 & 0 \\
 2 z^2 & 2 z \left(z^2-1\right) & 2 & 0 & 0 & 2-2 z^2 \\
 2 z & 2 \left(z^2-1\right) & 2 z & 0 & 2 \left(z^2-1\right) & 0\end{pmatrix}}
\end{equation}
And for the polarisation-transfer observables:
\begin{equation}
  \label{eq:matsig1xp}
\calM
_{1x^\prime}
= \ii \sqrt{1-z^2}{\ms\arraycolsep=0.3\arraycolsep\begin{pmatrix}
 0 & 0 & -1 & 0 & -2 z & -z^2-1 \\
 0 & 0 & 0 & 0 & 1-z^2 & z-z^3 \\
 1 & 0 & 0 & 0 & 2 z & z^2+1 \\
 0 & 0 & 0 & 0 & z^2-1 & z \left(z^2-1\right) \\
 2 z & z^2-1 & -2 z & 1-z^2 & 0 & 2 z \left(z^2-1\right) \\
 z^2+1 & z \left(z^2-1\right) & -z^2-1 & z-z^3 & -2 z \left(z^2-1\right) & 0\end{pmatrix}}
\end{equation}
\begin{equation}
  \label{eq:matsig1zp}
\calM
_{1z^\prime}
= \ii (1-z^2){\ms\arraycolsep=0.3\arraycolsep\begin{pmatrix}
 0 & 0 & 0 & 0 & -1 & -z \\
 0 & 0 & 1 & 0 & 0 & 1-z^2 \\
 0 & -1 & 0 & -1 & 1 & z \\
 0 & 0 & 1 & 0 & 2 z & z^2+1 \\
 1 & 0 & -1 & -2 z & 0 & 2 \left(z^2-1\right) \\
 z & z^2-1 & -z & -z^2-1 & 2-2 z^2 & 0\end{pmatrix}}
\end{equation}
\begin{equation}
  \label{eq:matsig2xp}
\calM
_{2x^\prime}
= \sqrt{1-z^2}{\ms\arraycolsep=0.3\arraycolsep\begin{pmatrix}
 0 & 0 & -1 & 0 & 0 & z^2-1 \\
 0 & 0 & 0 & 0 & z^2-1 & z \left(z^2-1\right) \\
 -1 & 0 & 2 & 2 z & 2 z & 3-z^2 \\
 0 & 0 & 2 z & 0 & z^2-1 & z-z^3 \\
 0 & z^2-1 & 2 z & z^2-1 & 0 & -2 z \left(z^2-1\right) \\
 z^2-1 & z \left(z^2-1\right) & 3-z^2 & z-z^3 & -2 z \left(z^2-1\right) & 4-4 z^2\end{pmatrix}}
\end{equation}
\begin{equation}
  \label{eq:matsig2zp}
\calM
_{2z^\prime}
={\ms\arraycolsep=0.3\arraycolsep\begin{pmatrix}
 0 & 0 & 2 z & 0 & z^2-1 & z-z^3 \\
 0 & 0 & z^2-1 & 0 & 0 & -\left(z^2-1\right)^2 \\
 2 z & z^2-1 & 0 & 1-z^2 & z^2-1 & z \left(z^2-1\right) \\
 0 & 0 & 1-z^2 & 0 & 0 & \left(z^2-1\right)^2 \\
 z^2-1 & 0 & z^2-1 & 0 & 4 z \left(z^2-1\right) & 2 \left(z^4-1\right) \\
 z-z^3 & -\left(z^2-1\right)^2 & z \left(z^2-1\right) & \left(z^2-1\right)^2 & 2 \left(z^4-1\right) & 4 z \left(z^2-1\right)\end{pmatrix}}
\end{equation}
\begin{equation}
  \label{eq:matsig3yp}
\calM
_{3y^\prime}
= \ii \sqrt{1-z^2}{\ms\arraycolsep=0.3\arraycolsep\begin{pmatrix}
 0 & 0 & -z & 1-z^2 & -2 z^2 & -2 z \\
 0 & 0 & 1-z^2 & z-z^3 & -2 z \left(z^2-1\right) & 2-2 z^2 \\
 z & z^2-1 & 0 & 0 & 2 & 2 z \\
 z^2-1 & z \left(z^2-1\right) & 0 & 0 & 0 & 0 \\
 2 z^2 & 2 z \left(z^2-1\right) & -2 & 0 & 0 & 2 \left(z^2-1\right) \\
 2 z & 2 \left(z^2-1\right) & -2 z & 0 & 2-2 z^2 & 0\end{pmatrix}}
\end{equation}

The matrices are either real or imaginary, but are, of course, always
Hermitean.  The matrices for $\Sigma_y$, $\Sigma_{1x}$, $\Sigma_{1z}$,
$\Sigma_{3y}$, $\Sigma_{1x^\prime}$, $\Sigma_{1z^\prime}$ and
$\Sigma_{3y^\prime}$ are pure imaginary but the observables are real because the
observables are nonzero only when the amplitudes have imaginary parts, \ie
above the first inelasticity, namely in our case above the pion-production
threshold.

It is, of course, possible to take linear combinations of amplitudes to form a
new set, each of which depends on a single polarisability or polarisability
combination. The corresponding ``rotation'' matrices can be used to transform
the matrices above, in order to see whether particular combinations of
amplitudes dominate in either basis. The most noticeable result is that both
$\Delta_3$ and $\Delta_y$ turn out to be completely independent of both
$\gammazero$ and $\gammapi$---in the cm frame.  This is particularly
significant for $\Delta_3$, since it raises the possibility that for energies
around the photoproduction threshold, once $\alphae$ and $\betam$ are
well-determined, sensitivity to $\gammaeminus$ and $\gammamminus$ is not
contaminated by lack of knowledge of $\gammapi$.  For higher energies, the
boost corrections which arise in the lab frame reduce the significance of
these observations, see below.

In fact, all matrices except $\calM_{\TsqUnpol}$, $\calM_{3y}$ and
$\calM_{3y'}$ have two zero eigenvalues, which means that there are two linear
combinations of amplitudes to which they are insensitive. However, other than
in the cases of $\calM_3$ and $\calM_y$, the corresponding eigenvectors do not
indicate simple combinations of polarisabilities.  Even more interesting
results emerge if we look at combinations of double polarisation observables
$\calM_{ij}\pm \calM_{ij'}$. $\calM_{3y}+\calM_{3y^\prime}$ has two zero
eigenvalues, and all other combinations of this form have four, and so are
sensitive to only two combinations of amplitudes.  However, only some of these
zero modes correspond to simple combinations of amplitudes. This analysis does
reveal, though, that the sum of $\calM_{3y}$ and $\calM_{3y'}$ is insensitive
to $\gammazero$ and $\gammapi$, while the difference is insensitive to
$\alphae$ and $\betam$.  The combinations $\calM_{1x}-\calM_{1x'}$ and
$\calM_{1z}+\calM_{1z'}$ are completely insensitive to both $\alpha-\beta$ and
$\gammazero$, while the opposite combinations are insensitive to
$\alpha+\beta$ and $\gammapi$. Sums and differences of $\calM_{2x}$ and
$\calM_{2x'}$, or of $\calM_{2z}$ and $\calM_{2z'}$, pair the insensitivities
up the opposite way round.

Though this is an intriguing observation, the asymmetry and
polarisation-transfer reactions are sufficiently different that it may not be
significant for experimental design. And experiments are not conducted in the
cm frame, in which the above decomposition holds, but in the lab frame; this
complicates matters. That complication is least important for observables
which transform easily between frames. 

The incident photon and target nucleon polarisations are identical in all
frames which share the same orientations of the $x$, $y$ and $z$ axes; this
includes the Breit, cm, and lab frames. Therefore, all asymmetries are
form-invariant, \ie to find the lab results, one only has to convert the
scattering angle and photon energy from the cm to the lab frame:
\begin{equation}
  \label{eq:forminv}\Sigma_\alpha^\mathrm{lab}(\omegalab,\theta_\mathrm{lab})=
  \Sigma_\alpha^\mathrm{cm}(\omega_\mathrm{cm}[\omegalab,\theta_\mathrm{lab}],
  \theta_\mathrm{cm}[\omegalab,\theta_\mathrm{lab}])\;\mbox{ for } j\not\in\{x^\prime,z^\prime\}
\end{equation}
The situation is slightly more complicated for polarisation-transfer
coefficients, as the polarisation of the outgoing nucleon is measured relative
to its momentum vector, \ie the $z^\prime$ axis is oriented along
$\pv^{\, \prime}$, and the $x^\prime$ axis is perpendicular to it but in the
scattering plane. In this case, the appropriate coefficients are different for
cm and lab frames, and are given for both in ref.~\cite{Babusci:1998ww}. They
are simply related by a rotation in the reaction plane through $\theta_R$, the
angle in the lab frame from the $z$- to the $z^\prime$-axis, see
fig.~\ref{fig:spinkinematics}. In order to conserve momentum perpendicular to
the incoming photon direction, they must satisfy
\begin{equation} \label{eq:recoil}
  \sin\theta_R=\frac{\omegaprimelab\sin\thetalab}{|\pv'|},\qquad 0\le \theta_R \le \pi/2,
\end{equation}
where $\omegaprimelab$ is the outgoing photon energy. Then, for $i=1,2$,
\begin{align}
  \Sigma_{ix'}^\mathrm{lab}&=\cos\theta_R\Sigma_{ix'}^\mathrm{cm}-
                           \sin\theta_R\Sigma_{iz'}^\mathrm{cm}, \nonumber \\
  \Sigma_{iz'}^\mathrm{lab}&=\sin\theta_R\Sigma_{ix'}^\mathrm{cm}+
                           \cos\theta_R\Sigma_{iz'}^\mathrm{cm}.
\label{eq:rotate}
\end{align}
On the other hand, the orientation of the $y$ axis is unchanged, and so the
polarisation transfer $\Sigma_{3y^\prime}$ is form-invariant and obeys an
equation like \eqref{eq:forminv}.

\section{Comments on Babusci \etal}
\setcounter{equation}{0}
\label{app:readbabusci}

\subsection{Interpreting Coefficients}
\label{sec:coefficients}

For the dedicated student of the invaluable paper of Babusci
\etal~\cite{Babusci:1998ww}, in an aside which is not intended to be read
independently, we note that the lab-frame coefficients defined there as rather
ugly expressions in their eq.~(3.29) can also be written more intuitively as
\begin{equation}
\label{eq:coefficients}
\begin{split}
C_{x'}^K&=-\textfrac 1 2 \left[\omegalab\sin(\theta_R)+\omegaprimelab\sin(\theta+\theta_R)\right] \\
C_{z'}^K&=\textfrac 1 2 \left[\omegalab\cos(\theta_R)+\omegaprimelab\cos(\theta+\theta_R] \right)\\
C_{z'}^Q&=\textfrac 1 2
          \left[\omegalab\cos(\theta_R)-\omegaprimelab\cos(\theta+\theta_R)\right]
          \;\;.
\end{split}
\end{equation}
It may seem odd that, up to signs, these are just the $x'$ and $z'$
coefficients of the \emph{lab-frame} vectors $\vec K$ and $\vec Q$ (defined as
$\textfrac 1 2(\kv'\pm\kv)$), even though the struck nucleon is not at rest in
that frame. The explanation is that a boost into its rest frame just exchanges
$\kv$ and $\kv'$ and changes the sign of the $z'$ components, so that $\vec K$
and $\vec Q$ are trivially related in the rest frames of the target and
recoiling nucleons.

\subsection{Proving Completeness}
\label{app:completeness}

Babusci \etal~\cite{Babusci:1998ww} provide a complete set of observables in
which at most one nucleon and one photon are polarised or analysed, and, at the
beginning of their sect.~III.B, they hint that the amplitudes can be reconstructed 
 from them. We now use the definition of observables in terms of the
quadratic forms with matrices $\calM_\alpha$ in appendix~\ref{app:matrices} to provide an
explicit proof of this, following the presentation in Arenh\"ovel
\etal~\cite{Arenhovel:1998vj}.

We start with the $6$ observables $\dd\sigma/\dd\Omega$, $\Sigma_3$,
$\Sigma_{2x/z}$ and $\Sigma_{2x^\prime/z^\prime}$ which are nonzero below the
first inelasticity (pion-production threshold). We denote these observables $\calO_{\alpha\mathbb{R}}$,
with  $\alpha\mathbb{R}=1,\dots,6$. They are related to the 
 real, independent
amplitudes $A_{1\text{-}6}$ as before,
\begin{equation}
  \label{eq:scalarproduct}
 \calO_{\alpha\mathbb{R}}\propto\vec{A}^\dagger\;\calM_{\alpha\mathbb{R}}\;\vec{A}\;\;, 
\end{equation}
where the six $6\times6$ matrices $\calM_{\alpha\mathbb{R}}$ are real as
well. The constants of proportionality are irrelevant for the present purpose.

To show that at fixed $(\omega,\theta)$, the amplitudes $\vec{A}$ can be
constructed from $6$ observables $\calO_{\alpha\mathbb{R}}$, one follows Arenh\"ovel \etal~\cite{Arenhovel:1998vj} to
combine (any) one column from each of the $\calM_{\alpha\mathbb{R}}$ into a
new $6\times6$ matrix $H_\mathbb{R}^\text{col}$. Here, the superscript reminds us
that there are $6^6$ choices possible. If at least one of these has a nonzero
determinant,
\begin{equation}
\det H_\mathbb{R}^\text{col}\ne0\;\;\mbox{ for some choice of columns,}
\end{equation}
then the column vectors residing in that $H_\mathbb{R}^\text{col}$ are
linearly independent and span the $6$-dimensional space in which the scalar
product eq.~\eqref{eq:scalarproduct} lives. The vector $\vec{A}$ can be
reconstructed, and therefore the observables $\calO_{\alpha\mathbb{R}}$ form a
complete set of observables.  Taking column $1$ from $\dd\sigma/\dd\Omega$,
$\Sigma_3$, $\Sigma_{2x}$ and $\Sigma_{2x^\prime}$, and column $3$ from
$\Sigma_{2z}$ and $\Sigma_{2z^\prime}$ produces one of many
$H_\mathbb{R}^\text{col}$ of non-zero determinant, and so shows that these $6$
observables suffice to reconstruct the $6$ amplitudes below the pion-production threshold.  Possible (discrete) sign ambiguities can be removed by
requiring that amplitudes evolve continuously in $\omega$ and $\theta$ from
the Thomson limit.

We now apply this method to the situation above the first inelasticity. There,
the amplitudes are complex, $\vec{A}=\Re\vec{A}+\ii\,\Im\vec{A}$, and the matrices
$\calM_{\alpha\mathbb{I}}$ ($\alpha\mathbb{I}=7,\dots,13$) of the $7$ additional
observables are all purely imaginary. Therefore, the scalar product in
eq.~\eqref{eq:scalarproduct} takes on two different forms for the two types of
observables (Einstein's Summation Convention is understood):
\newcommand{\Imag}{-\ii}
\begin{equation}
  \label{eq:scalarprodabove}
  \calO_{\alpha\mathbb{R}}=
  (\Re[A_i]\Re[A_j]+\Im[A_i]\Im[A_j])\;\calM_{\alpha\mathbb{R}}^{ij}
  \;\;\mbox{ and }\;\;
 \calO_{\alpha\mathbb{I}}=-2\ii\,\;
 \Re[A_i]\Im[A_j]\,\calM_{\alpha\mathbb{I}}^{ij}
\end{equation}
We turn this quadratic form into a scalar product between real vectors by
defining a $12$-dimensional vector space via
$\vec{u}^T=(\Re[\vec{A}]^T,\Im[\vec{A}]^T
)$
in which the observables are found from real and symmetric $12\times12$
matrices
\begin{equation}
	\begin{pmatrix}
		\calM_{\alpha\mathbb{R}}&0\\0&\calM_{\alpha\mathbb{R}}
	\end{pmatrix}\;\;\mbox{ and }\;\;
	\begin{pmatrix}
		0&(\Imag\calM_{\alpha\mathbb{I}})^T\\\Imag\calM_{\alpha\mathbb{I}}&0
	\end{pmatrix}	\;\;.
\end{equation}
However, one can only determine $11$ real parameters of the amplitudes, with
the overall phase free. Therefore, one may without loss of generality choose
one imaginary part $y_{j_0}$, $j_0\in\{7,\dots,12\}$, and eliminate it from
the vector space by deleting the $j_0$th row of the vector $\vec{u}$. We
denote the resulting $11$-dimensional vector by $\vec{u}^{\noj}$. The matrices
then turn into $11\times11$ objects
\begin{equation}
\label{eq:eleven}
	\begin{pmatrix}
		\calM_{\alpha\mathbb{R}}&0\\0&\calM_{\alpha\mathbb{R}}^{\noj\noj}
	\end{pmatrix}\;\;\mbox{ and }\;\;
	\begin{pmatrix}
		0&(\Imag\calM_{\alpha\mathbb{I}}^{\noj})^T\\\Imag\calM_{\alpha\mathbb{I}}^{\noj}&0
	\end{pmatrix}	\;\;,
\end{equation}
where the superscripts indicate that the lower $\calM_{\alpha\mathbb{R}}$
block has become a $5\times5$ matrix by elimination of the $j_0$th row
and column, and $\calM_{\alpha\mathbb{I}}$ an object with $6$ columns and
$5$ rows by elimination of its $j_0$th row.

We now construct matrices analogous to $H_\mathbb{R}^\text{col}$. First, pick
any $11$ of the available $13$ observables, then assemble one column of each
into an $11\times11$ matrix $H_\mathbb{C}^\text{col}$. If
\begin{equation}
  \det H_\mathbb{C}^\text{col}\ne0\;\;\mbox{ for some choice of columns,}
\end{equation}
then this set of observables suffices to determine $\vec{u}^{\noj}$ and hence
the amplitudes $A_{1\text{-}6}$, up to the overall phase.  This method
described by Arenh\"ovel \etal~\cite{Arenhovel:1998vj} for complex quadratic
forms is more general, but we will now take advantage of the fact that our
matrices $\calM_\alpha$ are either real or imaginary.

Our search simplifies if we start by choosing the $6$ observables which were
already used to find $\vec{A}$ below threshold. In this way, we exploit the
block-diagonal structure of the matrices in Eq.~(\ref{eq:eleven}), by writing
\begin{equation}
	H_\mathbb{C}^\text{col} = \begin{pmatrix}A&0\\0&B\end{pmatrix}\;\;.
\end{equation}
and choosing
$A=H_\mathbb{R}^\text{col}$. We then only need to augment these
$6$ observables by
$5$ linearly-independent columns from the set of matrices of the additional
above-threshold observables, ${\cal M}_{\alpha
  \mathbb{I}}$, which will then make up the matrix
$B=H_\mathbb{I}^\text{col}$, and finally show that
\begin{equation}
  \det H_\mathbb{I}^\text{col}\ne0\;\;\mbox{ for some choice of columns.}
\end{equation}
This can be done by picking one column apiece from any $5$ of the $7$ new
observables, with any one row $(\noj-6)$ eliminated. For example, one can
construct $B$ by dropping $\Sigma_{3y}$ and $\Sigma_{3y^\prime}$, combining
column $1$ of $\Sigma_y$, $\Sigma_{1x}$ and $\Sigma_{1x^\prime}$ with column
$3$ of $\Sigma_{1z}$ and $\Sigma_{1z^\prime}$, and eliminating the $1$st row.

Therefore, this set of observables suffices to determine the amplitudes $A_i$
up to discrete ambiguities and an overall phase. This is not the most general
set of observables above threshold, since we used the fact that the
observables $\calO_{\alpha\mathbb{R}}$ already can be turned into matrices
$H_\mathbb{R}^\text{col}$ with nonzero determinants, but it suffices.

It should be mentioned that the imaginary parts of the amplitudes may also be
reconstructed as multipoles of the total photoproduction cross
sections~\cite{Hildebrandt:2003fm}. This may provide alternative input or
valuable cross checks, in particular in order to reduce error bars or sign
ambiguities. The sign ambiguity $\vec{A}\to-\vec{A}$ and the overall phase may
be resolvable by requiring that amplitudes evolve continuously in $\omega$ and
$\theta$ from low-energy theorems (like the Thomson limit), whose phases are
determined by choice.

Finally, our proof resorted to the observables expressed in terms of the
matrices $\calM_\alpha$ in the cm frame.  The Lorentz boost into the lab frame
only involves kinematic factors and cannot generate new linear or quadratic
dependencies. Unsurprisingly then, the lab frame Compton amplitudes for
spin-$\half$ targets can also be reconstructed from the $6$ observables
$\dd\sigma/\dd\Omega$, $\Sigma_3$, $\Sigma_{2x/z}$ and
$\Sigma_{2x^\prime/z^\prime}$, supplemented above the pion-production
threshold with \emph{any} $5$ of the $7$ observables $\Sigma_y$,
$\Sigma_{1x/z}$, $\Sigma_{1x^\prime/z^\prime}$, $\Sigma_{3y}$ and
$\Sigma_{3y^\prime}$. To do a ``complete experiment" below the first
inelasticity, it is therefore mandatory to perform at least two polarisation
transfer experiments, and at least one more is necessary above threshold.

\section{Online Supplement: Sensitivity of Neutron Observables} 
\setcounter{equation}{0}
\label{app:moreplots}

For ease of comparison, each plot uses the same contour map as the
corresponding proton plot. The values of the neutron polarisabilities,
including theoretical and experimental uncertainties, are summarised in
ref.~\cite{Griesshammer:2015ahu} and appendix~\ref{app:polvalues}.

\begin{figure}[!htbp]
\begin{center}
     \includegraphics[width=\textwidth]{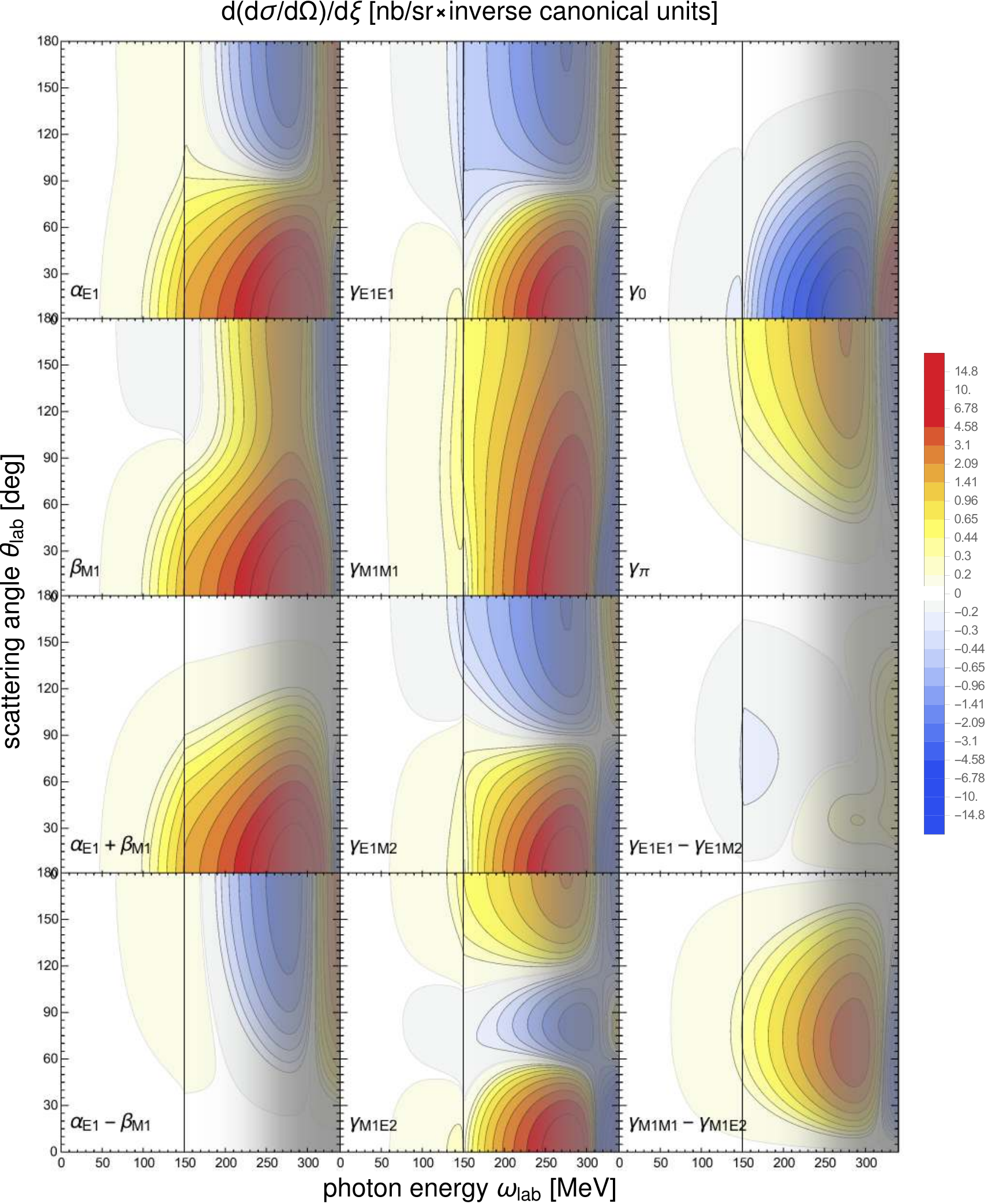}
     \caption{(Colour online) Sensitivity of the neutron cross section to varying the
       polarisabilities.}
\label{fig:neutron-crosssection-polsvar}
\end{center}
\end{figure}

\begin{figure}[!htbp]
\begin{center}
     \includegraphics[width=\textwidth]{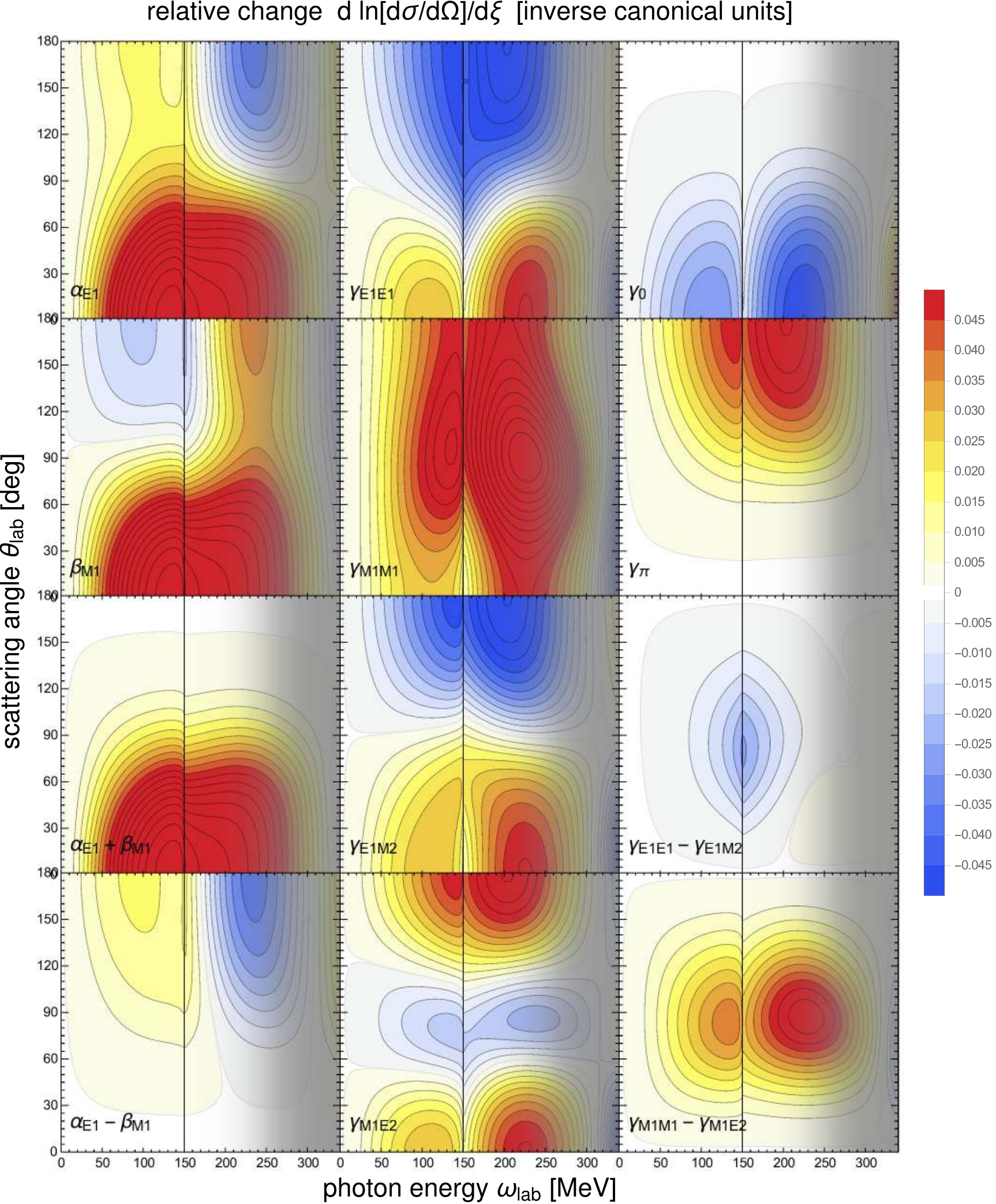}
     \caption{(Colour online) Sensitivity of the neutron  cross
       section to varying the polarisabilities, normalised to the cross
       section.}
     \label{fig:neutron-crosssection-relative-change-polsvar}
\end{center}
\end{figure}

\begin{figure}[!htbp]
\begin{center}
     \includegraphics[width=\textwidth]{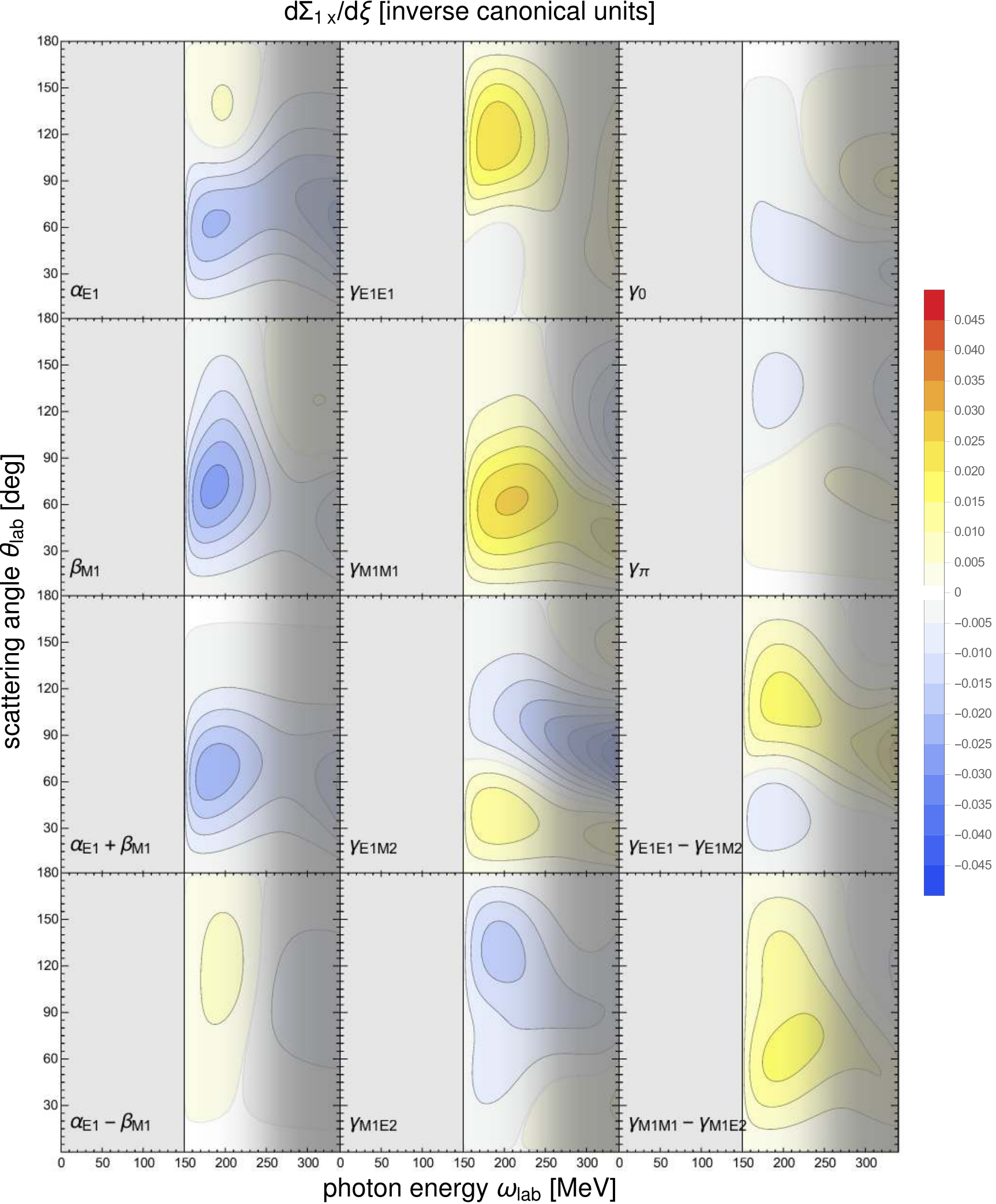}
     \caption{(Colour online) Sensitivity of the double asymmetry $\Sigma_{1x}$
       (linearly polarised photons on a neutron target polarised along the
       $x$ axis) to varying the polarisabilities.}
     \label{fig:neutron-polsvar-1X}
\end{center}
\end{figure}

\begin{figure}[!htbp]
\begin{center}
     \includegraphics[width=\textwidth]{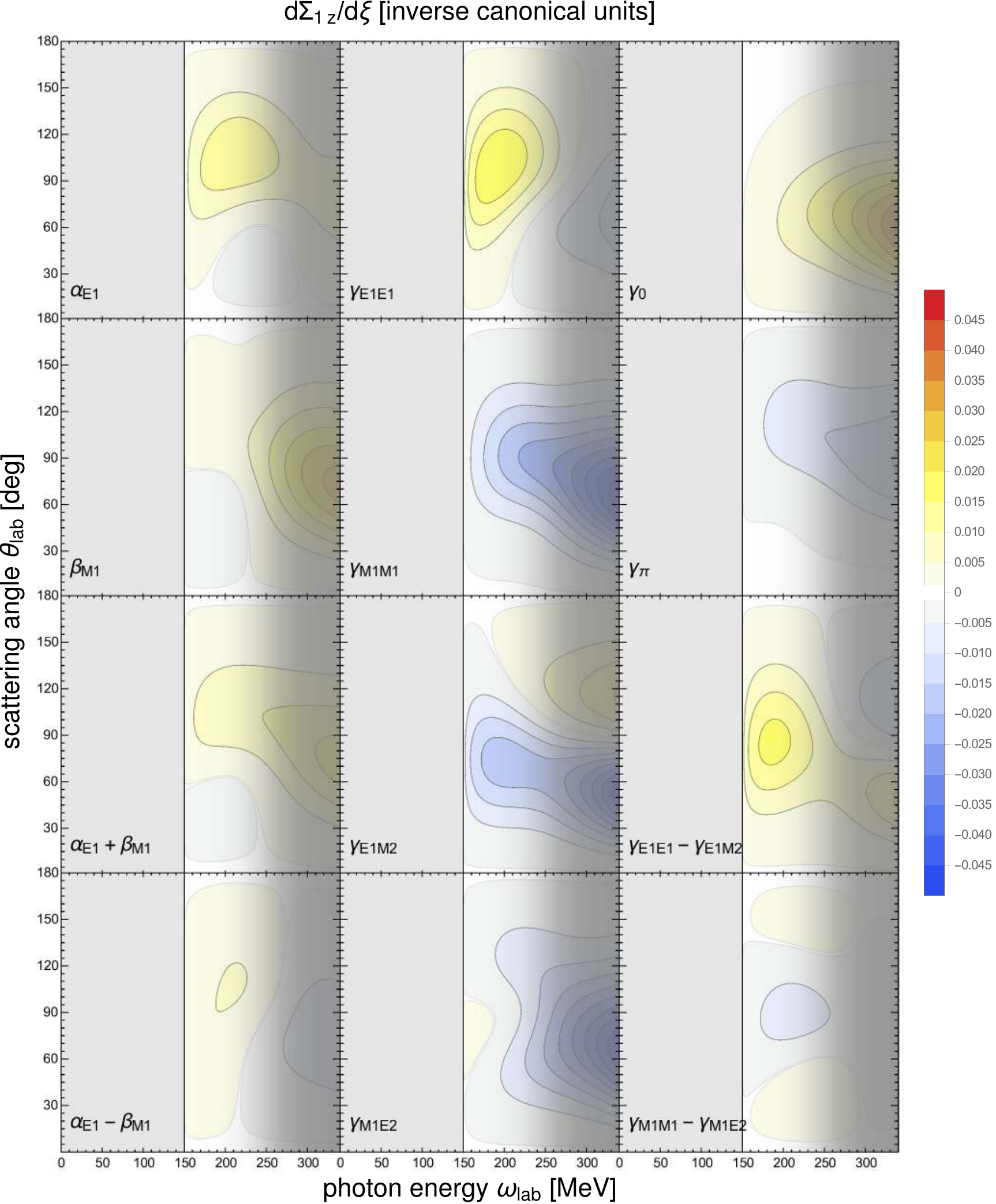}
     \caption{(Colour online) Sensitivity of the double asymmetry $\Sigma_{1z}$
       (linearly polarised photons on a neutron target polarised along the
       $z$ axis) to varying the polarisabilities.}
     \label{fig:neutron-polsvar-1Z}
\end{center}
\end{figure}

\begin{figure}[!htbp]
\begin{center}
     \includegraphics[width=\textwidth]{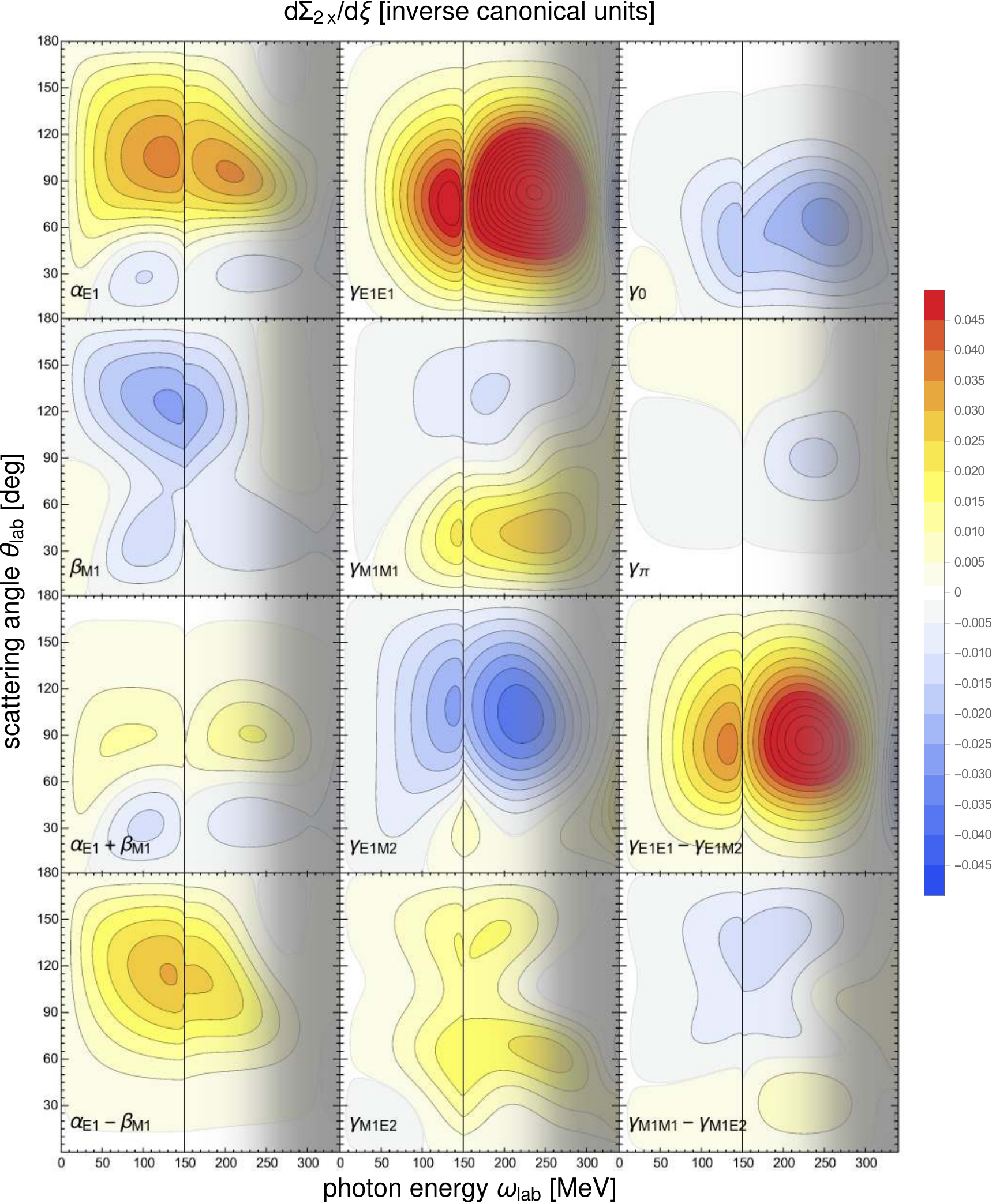}
     \caption{(Colour online) Sensitivity of the double asymmetry
       $\Sigma_{2x}$ (circularly polarised photons on a neutron target
       polarised along the $x$ axis) to varying the polarisabilities; see text
       for details.}
     \label{fig:neutron-polsvar-2X}
\end{center}
\end{figure}

\begin{figure}[!htbp]
\begin{center}
     \includegraphics[width=\textwidth]{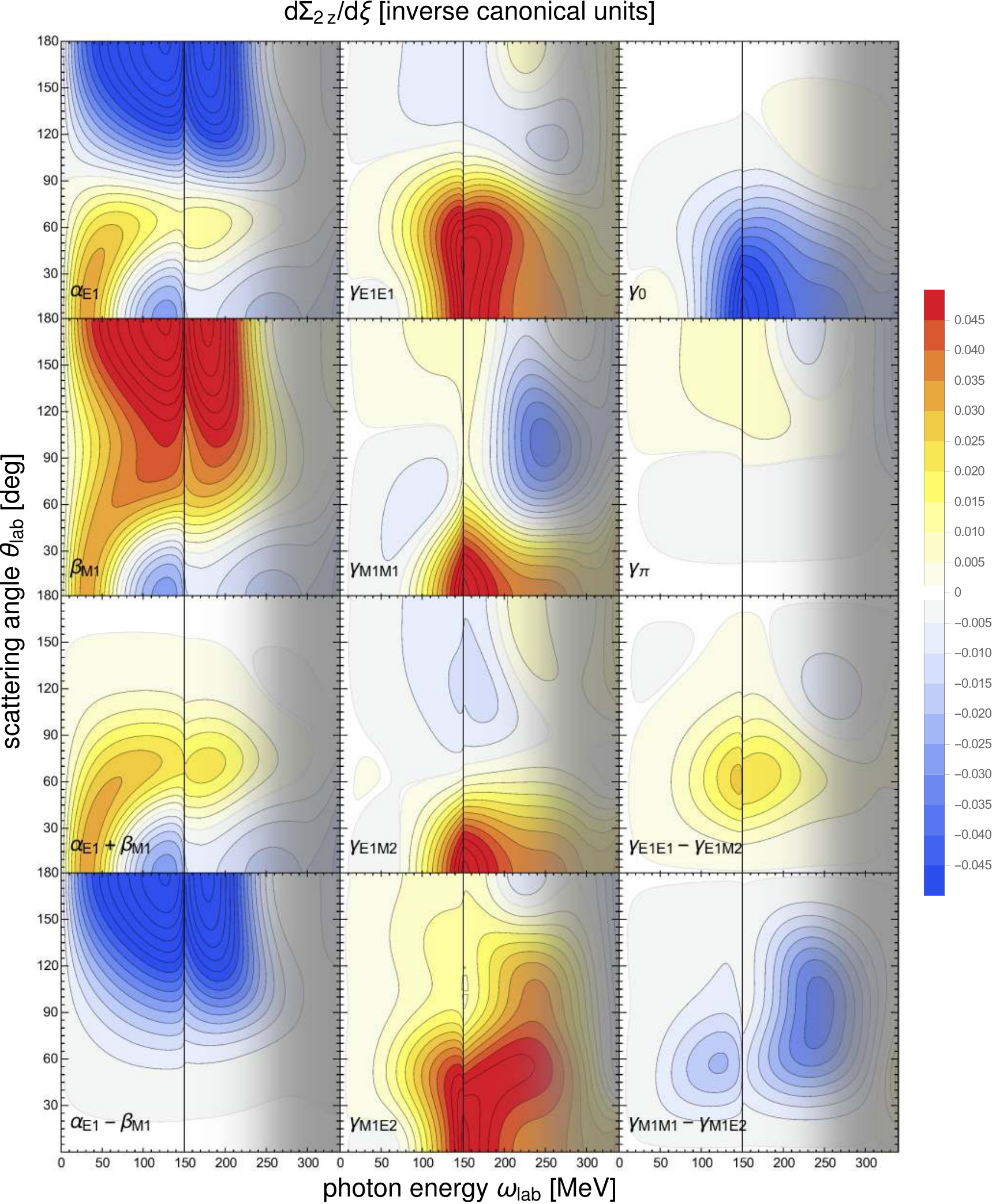}
     \caption{(Colour online) Sensitivity of the double asymmetry $\Sigma_{2z}$
       (circularly polarised photons on a neutron target polarised along the
       $z$ axis) to varying the polarisabilities.}
     \label{fig:neutron-polsvar-2Z}
\end{center}
\end{figure}

\begin{figure}[!htbp]
\begin{center}
     \includegraphics[width=\textwidth]{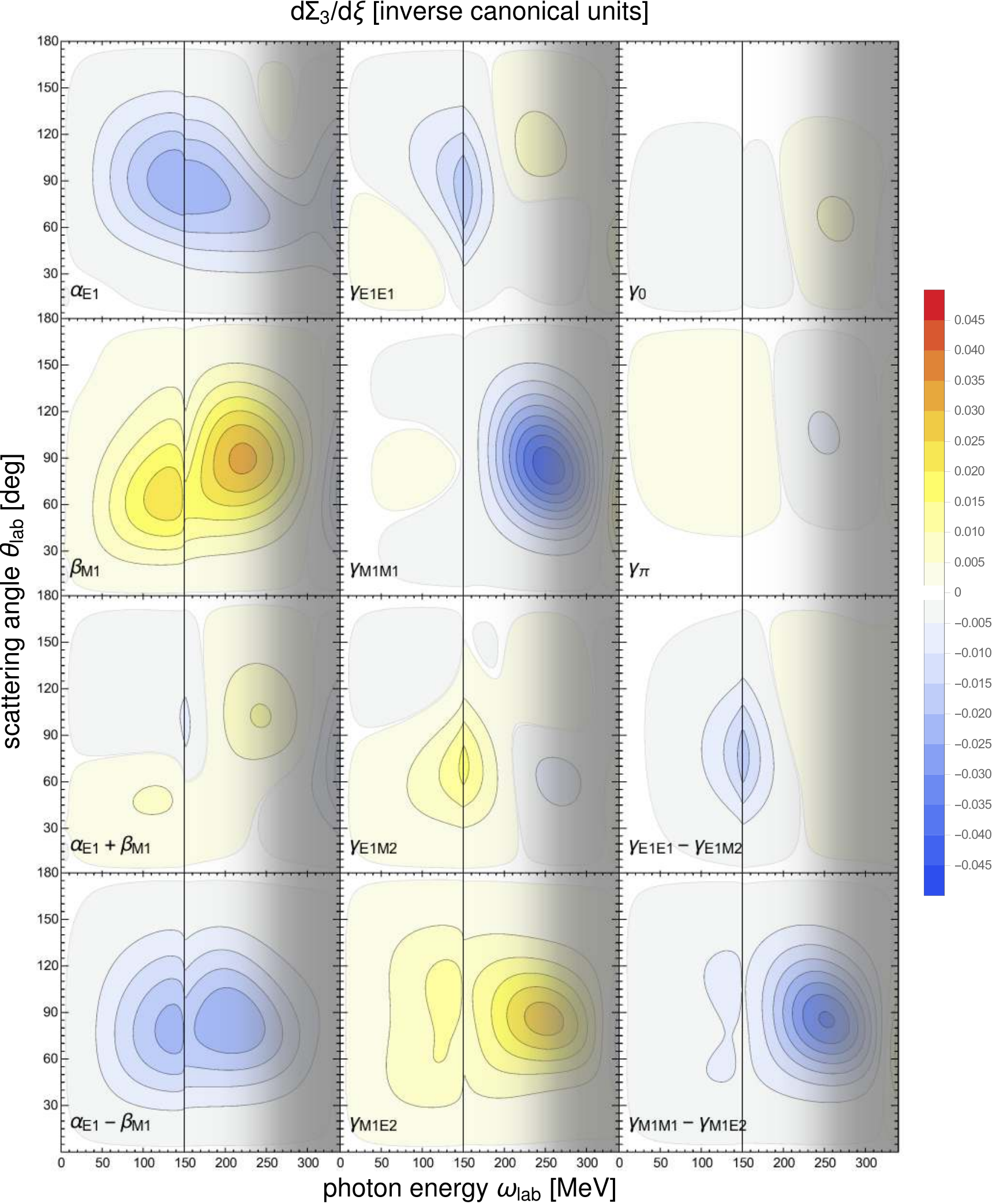}
     \caption{(Colour online) Sensitivity of the beam asymmetry $\Sigma_{3}$
       (linearly polarised photons on an unpolarised neutron target) to varying
       the polarisabilities.}
     \label{fig:neutron-polsvar-3}
\end{center}
\end{figure}

\begin{figure}[!htbp]
\begin{center}
     \includegraphics[width=\textwidth]{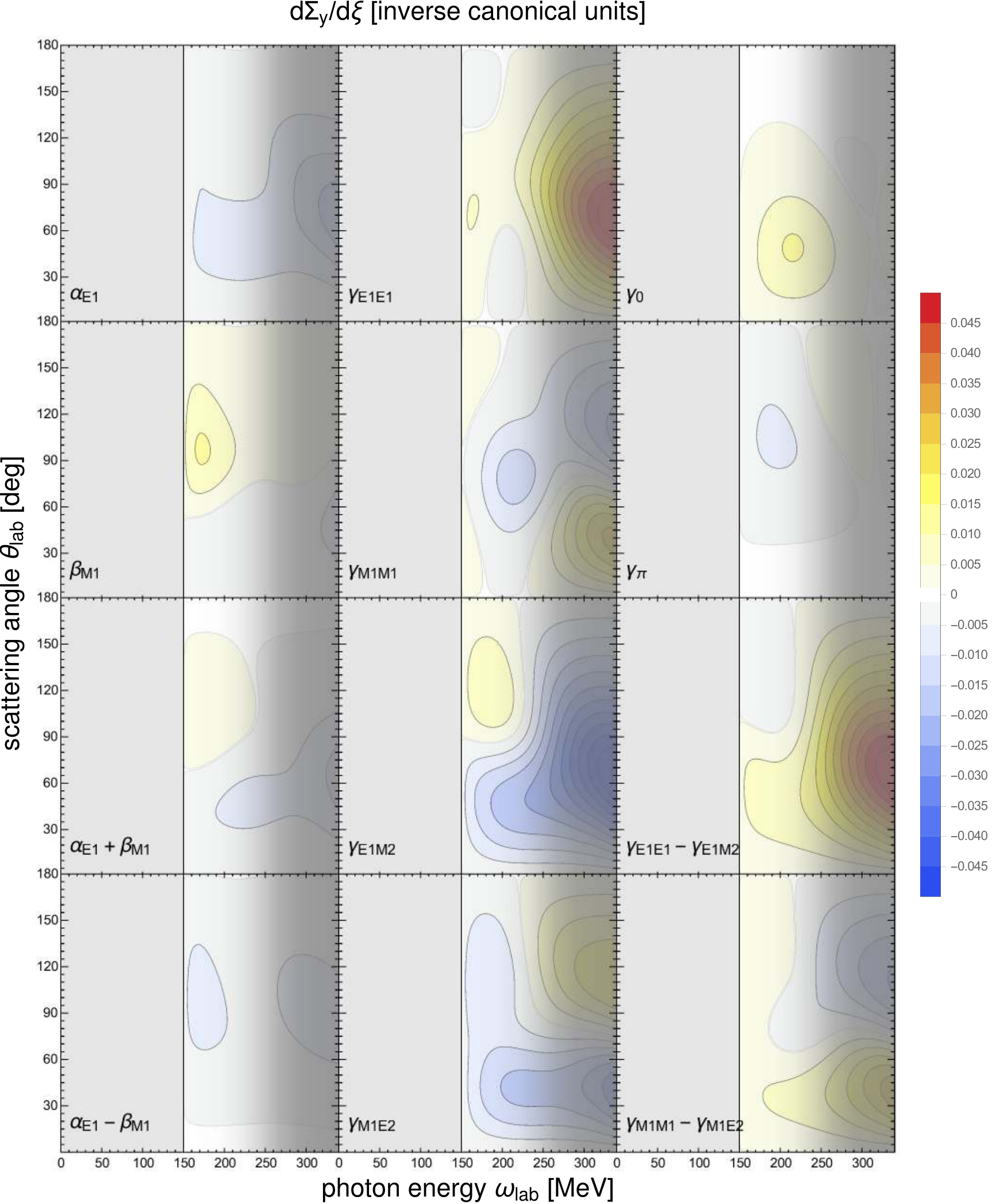}
     \caption{(Colour online) Sensitivity of the target asymmetry $\Sigma_{y}$
       (unpolarised photons on a neutron target along the $y$ axis) to varying
       the polarisabilities.}
     \label{fig:neutron-polsvar-Y}
\end{center}
\end{figure}

\begin{figure}[!htbp]
\begin{center}
     \includegraphics[width=\textwidth]{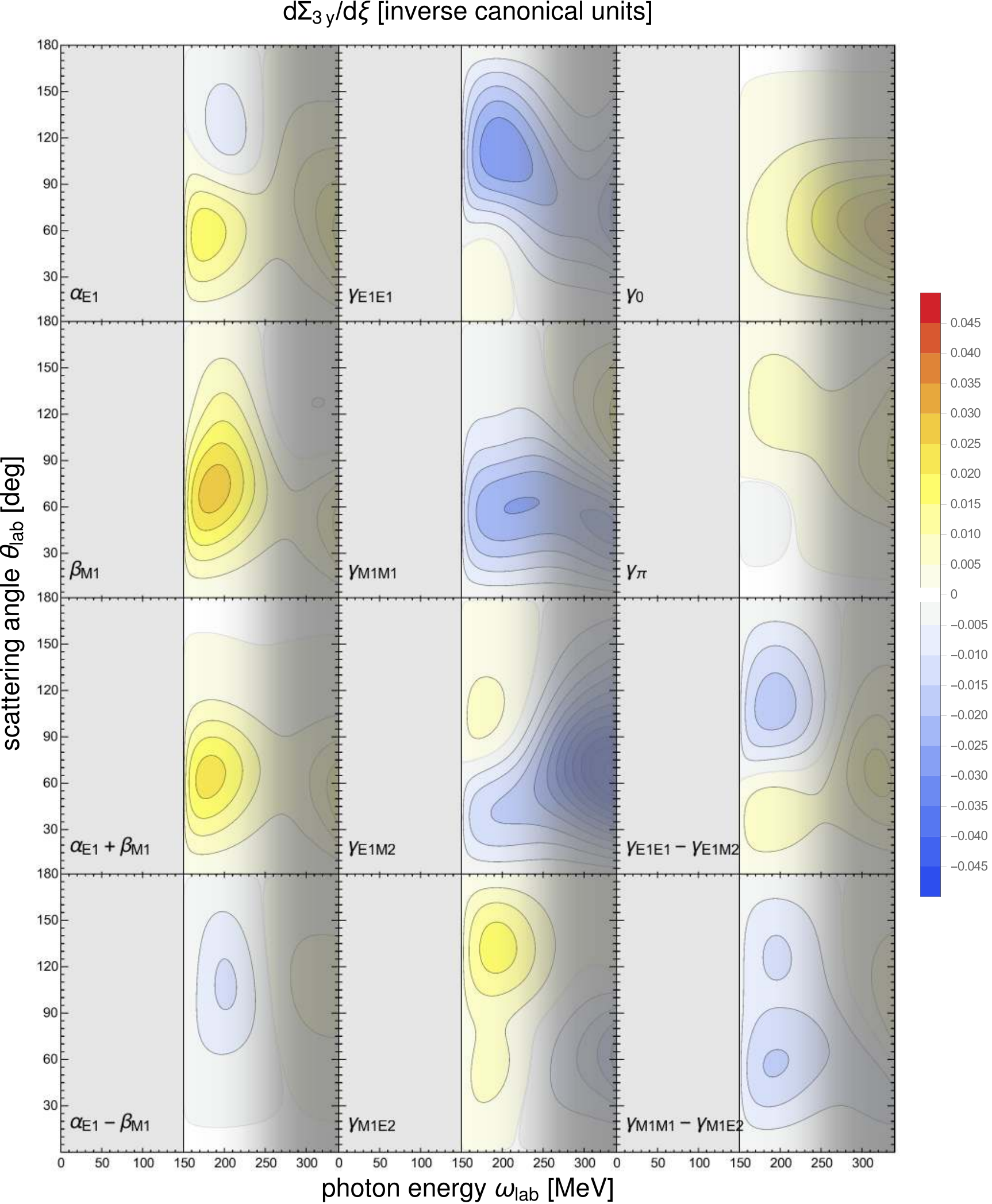}
     \caption{(Colour online) Sensitivity of the double asymmetry $\Sigma_{3y}$
       (linearly polarised photons on a neutron target polarised along the
       $y$ axis) to varying the polarisabilities.}
     \label{fig:neutron-polsvar-3Y}
\end{center}
\end{figure}


\begin{figure}[!htbp]
\begin{center}
     \includegraphics[width=\textwidth]{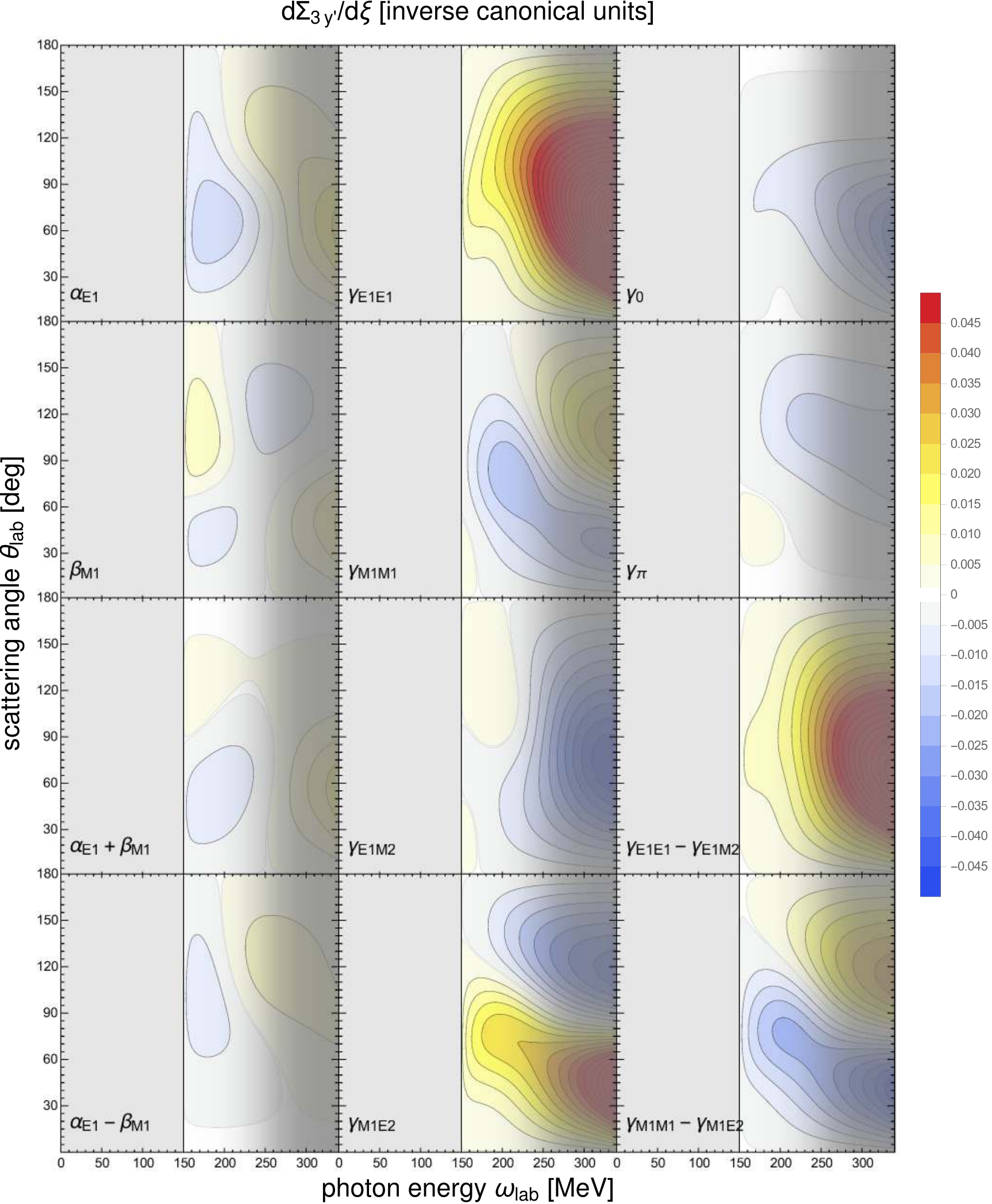}
     \caption{(Colour online) Sensitivity of the polarisation-transfer
       observable $\Sigma_{3y^\prime}$
       (linearly polarised photons on an unpolarised neutron target, recoil polarised along the $y^\prime$ axis) to varying the polarisabilities.}
     \label{fig:neutron-polsvar-3Yp}
\end{center}
\end{figure}

\begin{figure}[!htbp]
\begin{center}
     \includegraphics[width=\textwidth]{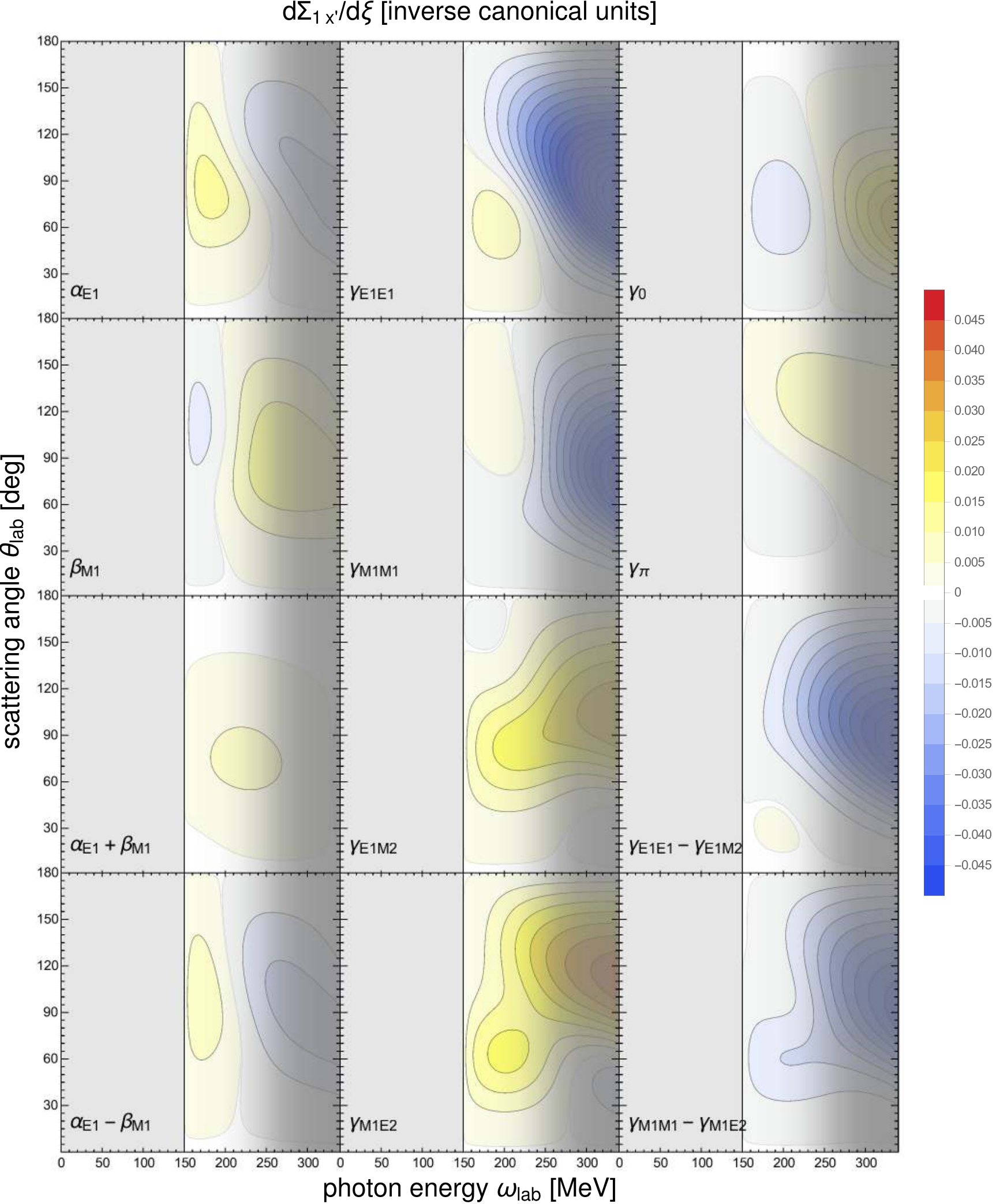}
     \caption{(Colour online) Sensitivity of the polarisation-transfer
       observable $\Sigma_{1x^\prime}$ (linearly polarised photons on an
       unpolarised neutron target, recoil polarised along the $x^\prime$ axis) to
       varying the polarisabilities.}
     \label{fig:neutron-polsvar-1Xp}
\end{center}
\end{figure}

\begin{figure}[!htbp]
\begin{center}
     \includegraphics[width=\textwidth]{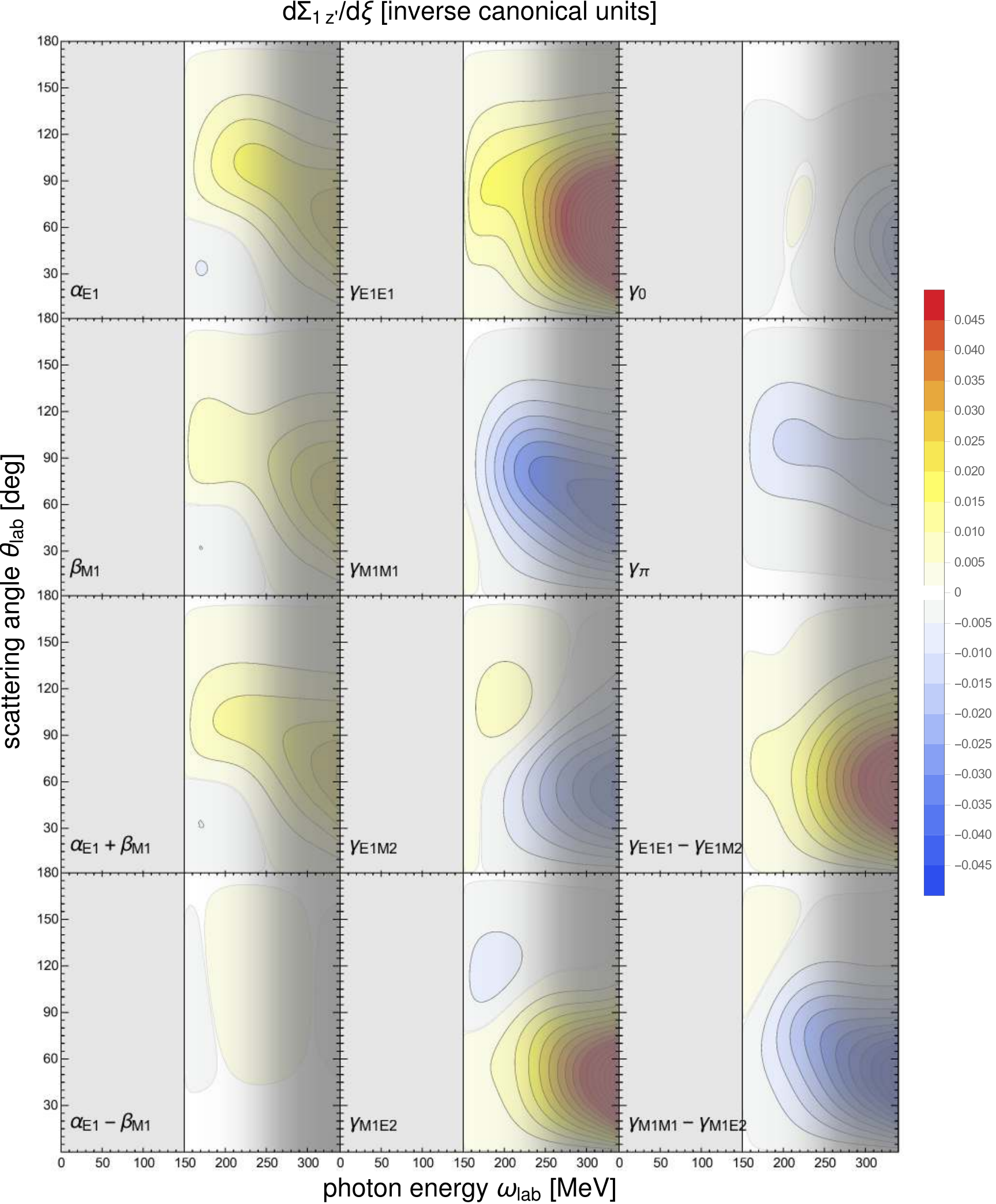}
     \caption{(Colour online) Sensitivity of the polarisation-transfer
       observable $\Sigma_{1z^\prime}$ (linearly polarised photons on an
       unpolarised neutron target, recoil polarised along the $z^\prime$ axis) to
       varying the polarisabilities.}
     \label{fig:neutron-polsvar-1Zp}
\end{center}
\end{figure}

\begin{figure}[!htbp]
\begin{center}
     \includegraphics[width=\textwidth]{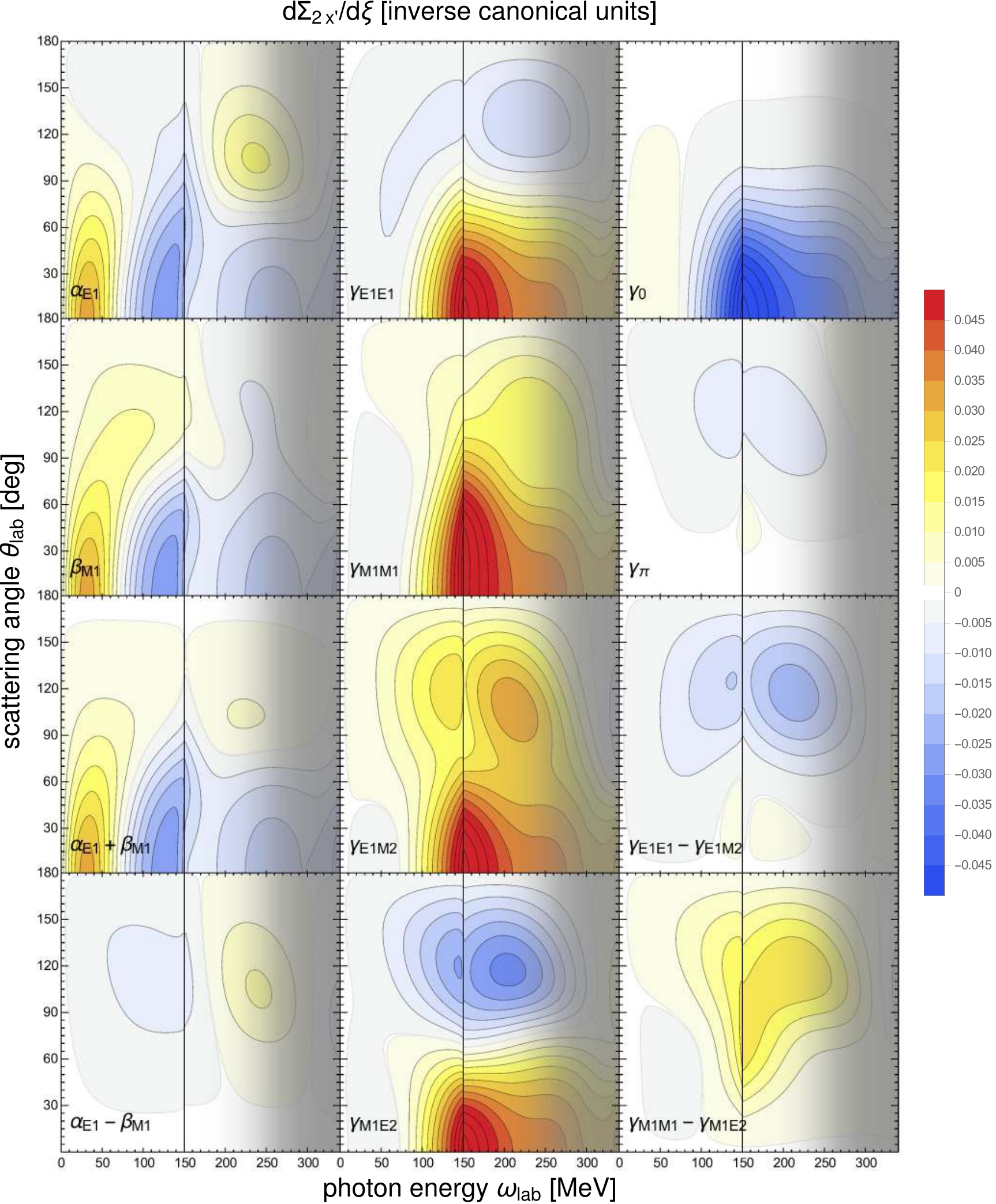}
     \caption{(Colour online) Sensitivity of the polarisation-transfer
       observable $\Sigma_{2x^\prime}$ (circularly polarised photons on an
       unpolarised neutron target, recoil polarised along the $x^\prime$ axis) to
       varying the polarisabilities.}
     \label{fig:neutron-polsvar-2Xp}
\end{center}
\end{figure}

\begin{figure}[!htbp]
\begin{center}
     \includegraphics[width=\textwidth]{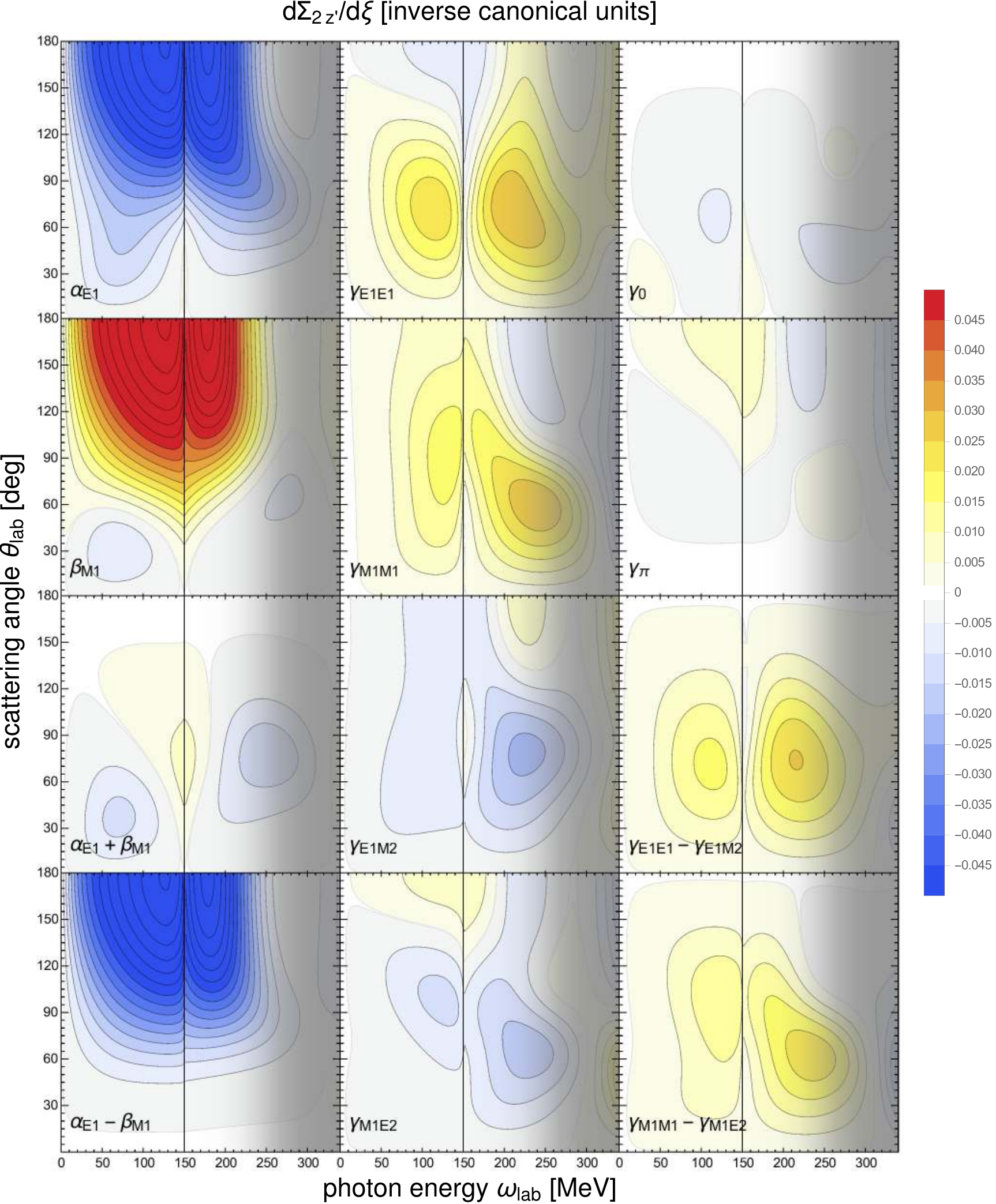}
     \caption{(Colour online) Sensitivity of the polarisation-transfer
       observable $\Sigma_{2z^\prime}$ (circularly polarised photons on an
       unpolarised neutron target, recoil polarised along the $z^\prime$ axis) to
       varying the polarisabilities.}
     \label{fig:neutron-polsvar-2Zp}
\end{center}
\end{figure}
\clearpage
\end{document}